\documentclass[pre,eqsecnum,tighten]{revtex4}

\usepackage{amssymb,amsmath}

\usepackage{epsfig}

\usepackage{graphicx}

\usepackage{bm}

\begin{document}

\title{Enskog Theory for Polydisperse Granular Mixtures.
I. Navier-Stokes order Transport}
\author{Vicente Garz\'{o}\footnote[1]{Electronic address: vicenteg@unex.es}}
\address{Departamento de F\'{\i}sica, Universidad de Extremadura, E-06071
Badajoz, Spain}
\author{James W. Dufty\footnote[2]{Electronic address: dufty@phys.ufl.edu}}
\address{Department of Physics, University of Florida, Gainesville, Florida 32611}
\author{Christine M. Hrenya
\footnote[3]{Electronic address: hrenya@colorado.edu}}
\address{Department of Chemical and
Biological Engineering, University of Colorado, Boulder, CO 80309}

\begin{abstract}
A hydrodynamic description for an $s$-component mixture of inelastic, smooth hard disks (two dimensions) or
spheres (three dimensions) is derived based on the revised Enskog theory for the single-particle velocity
distribution functions. In this first portion of the two-part series, the macroscopic balance equations for
mass, momentum, and energy are derived. Constitutive equations are calculated from exact expressions for the
fluxes by a Chapman--Enskog expansion carried out to first order in spatial gradients, thereby resulting in a
Navier-Stokes order theory. Within this context of small gradients, the theory is applicable to a wide range of
restitution coefficients and densities. The resulting integral-differential equations for the zeroth- and
first-order approximations of the distribution functions are given in exact form. An approximate solution to
these equations is required for practical purposes in order to cast the constitutive quantities as algebraic
functions of the macroscopic variables; this task is described in the companion paper.

\end{abstract}


\date{\today}

\maketitle

\draft

\section{ Introduction}
\label{sec1}

Flows of polydisperse particles (mixtures) are ubiquitous in nature and industry alike. Examples of the former
include pyroclastic flows, landslides, pollutant transport, and planetary rings. Examples of the latter include
pneumatic conveying of grains, ores, and chemicals; fluidized-bed operation for power production and catalytic
cracking; mixing of pharmaceutical powders (medication and binder) and poultry feedstock (grains and vitamins).
A non-uniform particle distribution may be a property of the starting material itself, or it may be
intentionally utilized in order to improve process performance. For an example of the latter, the addition of
fines to a relatively monodisperse material has been shown to (i) decrease attrition in high-speed conveying
lines \cite{Knowlton94}, (ii) increase conversion in high-velocity, fluidized-bed reactors \cite{Pell88} and
(iii) improve heat transfer efficiency in a circulating fluidized bed (CFB) combustor \cite{Lee97}. Polydisperse
materials are also known to exhibit counter-intuitive behaviors that have no monodisperse counterpart. For
example, agitation of polydisperse materials via vibration, free-fall, or flow down an incline leads to
segregation among unlike particles (de-mixing). Enhancing or suppressing this segregation tendency may be
critical to process performance, depending on whether the desired outcome is a separated or well-mixed state,
respectively.

In the current effort, attention is restricted to rapid flows, in which particle collisions are assumed to be
both binary and instantaneous in nature. For monodisperse systems, kinetic-theory-based treatments have been
successful at predicting not only rapid granular flows (in which the role of the interstitial fluid is assumed
negligible), but have also been incorporated into models of high-velocity, gas-solid systems. In particular,
kinetic-theory-based descriptions are now standard components in both commercial and open-source CFD
(computational fluid dynamics) software packages for multiphase flows such as Fluent$^{\rm R}$ and MFIX
(http://www.mfix.org/), respectively. Nonetheless, the development and application of kinetic-theory-based
descriptions for polydisperse systems is in its infancy relative to their monodisperse counterparts, as has been
highlighted in several recent review articles and perspectives \cite{Ottino00,Sundaresan01,Muzzio02,Curtis04}.
The main challenge associated with the derivation of kinetic-theory-based descriptions for mixtures is the
increased complexity associated with the additional hydrodynamic fields and associated transport coefficients,
and in particular with the accurate evaluation of the collisions integrals. Correspondingly, many of the early
contributions resorted to assumptions which are only strictly true in the limit of perfectly elastic spheres in
a uniform steady state: a Maxwellian (single-particle) velocity distribution \cite
{Jenkins87,Huilin01,Rahaman03} or an equipartition of energy \cite {Jenkins89,Zamankhan95,Arnarson98,Willits99}.
However, the presence of a non-Maxwellian velocity distribution in granular flows is well-documented
\cite{Campbell90,Goldshtein95,Goldhirsch96,Esipov97,vanNoije98,Brey99,Losert99,Kudrolli00}, and has been shown
to have a significant impact on some transport coefficients \cite{GMD06}. Moreover, a non-equipartition of
energy between unlike particles is widely established \cite
{Garzo99,MG02bis,Barrat02,Wildman02,Feitosa02,Clelland02,Dahl02,Alam03,Paolotti03} , and has been shown to
significantly contribute to the driving force for segregation \cite{Galvin05} and to lead to a reversal of the
segregation direction \cite{Brey05,Yoon06,G06} in certain systems. A more recent theory \cite{Iddir05} involves
the lifting of both of these assumptions, except in the evaluation of collision integrals involving two unlike
particles, in which case a Maxwellian velocity distribution is assumed for each particle type. Two current
theories exist that do not involve either of these assumptions \cite{Garzo02,Serero06}, though both are based on
the Boltzmann equation and thus are limited to dilute flows. Another key difference between existing
polydisperse theories is the base state used in the Chapman--Enskog (CE) expansion. Some theories \cite
{Jenkins87,Huilin01,Rahaman03,Jenkins89,Zamankhan95,Arnarson98,Willits99,Iddir05,Serero06} assume an expansion
about a perfectly elastic (molecular equilibrium) base state, and thus are restricted to nearly-elastic systems.
However, in the CE method the base state must not be chosen a priori, but rather it is determined as the
solution to the kinetic equation to zeroth order in the gradient expansion. This solution is found to correspond
to the {\em local} homogeneous cooling state (HCS) and was used in Ref.\ \cite {Garzo02} as the reference state
to determine the Navier-Stokes transport coefficients of a dilute mixture, without any restriction on the level
of inelasticity.

The objective of the current effort is twofold. First, a kinetic-theory-based description for the flow of an
$s$-component mixture in $d$ dimensions is derived which (i) incorporates non-Maxwellian and non-equipartition
effects, (ii) is applicable to a wide range of restitution coefficients, and (iii) is applicable to both dilute
and (moderately) dense flows. In particular, a CE expansion of the revised Enskog theory for inelastic, hard
spheres is carried out for both disks ($d=2$) and spheres ($d=3$) up to the Navier-Stokes order. Second, the
derivation of the resulting theory is critically compared and contrasted to that of existing theories, in an
effort to clearly reveal the implications of various treatments on both the governing equations and constitutive
relations. For this reason, the derivation is presented in a detailed and somewhat pedagogical fashion. This
work takes the form of two self-contained, companion papers. In this first paper, the results of the exact
analysis are given. The follow-on paper details the leading order approximations needed for the explicit
evaluation of all properties derived here: the distribution functions, the ``equations of state'' (cooling rate
and pressure), and the transport coefficients.  In addition, the methodology used to obtain these results is
critically compared there to that of previous theories.

A confusing issue in the granular community is the context of the Navier-Stokes hydrodynamic equations in freely
cooling granular gases derived in this paper. The expressions for the Navier-Stokes transport coefficients are
not limited to weak inelasticity and so the calculations provided here apply even for strong dissipation. The
Navier-Stokes hydrodynamic equations may or may not be limited with respect to inelasticity, depending on the
particular states analyzed. The CE method assumes that the relative changes of the hydrodynamic fields over
distances of the order of the mean free path are small. For ordinary (elastic) gases this can be controlled by
the initial or boundary conditions. However, in the case of granular fluids the situation is more complicated
since in some cases (e.g., steady states such as the simple shear flow problem \cite{Santos04}) the boundary
conditions imply a relationship between dissipation and gradients so that both cannot be chosen independently.
In these cases, the Navier-Stokes approximation only holds for nearly elastic particles \cite{Santos04}.
However, the transport coefficients characterizing the Navier-Stokes hydrodynamic equations are nonlinear
functions of the coefficients of restitution, regardless the applicability of those equations.

In spite of the above cautions, the Navier-Stokes approximation is relevant to describe a wide class of granular
flows. One of them corresponds to spatial perturbations of the HCS for an isolated system. Computer simulations
have confirmed the accuracy of the Navier-Stokes hydrodynamic equations with their associated transport
coefficients to quantitatively describe cluster formation \cite{BRC99}. The same kinetic theory results apply to
driven systems as well. This is so since the reference state is a {\em local} HCS whose parameters change
throughout the system to match the physical values in each cell. Another examples of good agreement between
theory and simulation \cite{BRMG01} and experiments \cite{YHCMW02,HYCMW04} include the application of the
Navier-Stokes hydrodynamics to describe density/temperature profiles in vertical vibrated gases, supersonic flow
past at wedge in real experiments \cite{RBSS02}, and nonequipartition and size segregation in agitated granular
mixtures \cite{Feitosa02,Wildman02,SUKSS06}. In summary, the Navier-Stokes hydrodynamics with the constitutive
equations obtained in this paper constitute an important and useful description for many different physical
situations, although more limited than for elastic gases.

\section{ Overview of Derivation}

The theoretical basis for a hydrodynamic description of molecular gases is most completely established at low
density using the Boltzmann kinetic equation. There, the CE solution and its prediction of transport
coefficients is well-established from both computer simulation and experiment \cite{Ferziger72}. For a
moderately dense gas there is no accurate and practical generalization of the Boltzmann equation except for the
idealized hard sphere fluid. In that case, the Enskog kinetic equation describes the dominant positional
corrections to the Boltzmann equation due to excluded volume effects of other particles on a colliding pair
\cite{Ferziger72}. The neglected velocity correlations are important only at much higher densities. The
derivation of hydrodynamics and evaluation of transport coefficients based on the Enskog kinetic equation leads
to an accurate and unique description of moderately dense gases. The generalization to mixtures requires a
revision of the original Enskog theory for thermodynamic consistency (revised Enskog theory, or RET)
\cite{vanBeijeren73}, and its application to hydrodynamics and mixture transport coefficients was accomplished
twenty years ago \cite{deHaro83}. As noted above, for granular (dissipative) gases, there remains an open
problem of predicting transport properties at moderate densities, as occur in current experiments and
simulations. This problem is addressed here in its full generality using the extension of this revised Enskog
theory to inelastic collisions without limits on the number of components, densities, temperature, or degree of
dissipation. This subsumes all previous analyses for both molecular and granular gases, which are recovered in
the appropriate limits.

Due to the extreme length of the derivation, an outline of the steps involved is given here for easy reference.
\bigskip

\begin{itemize}

\item \hspace*{.2in}{\it Section III.}  The starting point of the
derivation process is the revised Enskog kinetic equations for
mixtures of inelastic, hard spheres. These equations for the
single-particle, position
and velocity distribution functions of each species, $\left\{ f_{i}\right\} $%
, take the form of nonlinear, integro-differential equations,
where the integral portion arises from the collision operator.

\item \hspace*{.2in}{\it Section IV.} The macroscopic variables of
interest (number density $\left\{ n_{i}\right\} $, etc.) are defined
exactly in terms of moments of $\left\{ f_{i}\right\} $ (e.g.,
$n_{i}\left( \mathbf{r}\right) =\int d\mathbf{v}f_{i}\left(
\mathbf{r} ,\mathbf{v,}t\right) $, where $\mathbf{v}$ is the
velocity of species $i$). Thus, the macroscopic balance equations
can be obtained by appropriate manipulation of the Enskog kinetic
equations (e.g., multiplication by $d\mathbf{v}$ followed by
integration over the velocity to obtain the species mass balance).
At this stage, all of the constitutive quantities (cooling rate,
stress tensor, conduction, and mass flux) appearing in the
macroscopic balances are integral functionals of $\left\{
f_{i}\right\} $, which depend explicitly on space and time only
through their dependence on $\left\{ f_{i}\right\}$.

\item \hspace*{.2in}{\it Section V.} \hspace*{.1in} In order to
obtain a hydrodynamic description (one in which the constitutive
quantities are determined entirely by the macroscopic or
hydrodynamic variables), the concept of a normal solution is
introduced. These are special solutions to the Enskog equations for
which the $\left\{ f_{i}\right\}$ depend on space and time only
through an implicit {\em functional} dependence on the macroscopic
fields (or equivalently as explicit {\em functions} of these local
fields and their gradients at the spatial point of interest).

\item\hspace*{.2in}{\it Section VI.} An exact analytical solution
for $\left\{ f_{i}\right\} $ is not a practical objective in the
most general case, and thus attention is restricted to states with
small spatial gradients. In this case the gradients provide a small
parameter, allowing a small spatial gradients, or small Knudsen
number, expansion (i.e., the CE expansion). The analysis is carried
out here to first (Navier-Stokes) order: $
f_{i}=f_{i}^{(0)}+f_{i}^{(1)}$, where $f_{i}^{(0)}$ is the zeroth
order solution and $f_{i}^{(1)}$ is the first-order correction
(zero- and first-order in gradients, respectively). The kinetic
equations then become integral-differential equations for the
determination of $f_{i}^{(0)}$ and $ f_{i}^{(1)}$.

\item\hspace*{.2in} {\it Section VII.} Correspondingly, the
constitutive equations are identified as functions of the
hydrodynamic variables and their gradients through their dependence
on $\left\{ f_{i}\right\} $, with coefficients of the gradients
defining the transport coefficients. Hence, all equations of state
(pressure and reference state cooling rate) and all transport
coefficients, which are integrals involving $f_{i}^{(0)}$, inherit
this dependence on the hydrodynamic variables and their gradients.
The coefficients are determined from solutions to the integral
equations.
\end{itemize}

This completes the derivation reported in this manuscript. Up until this point, the results are exact for
Navier-Stokes order hydrodynamics (first order in spatial gradients) of the RET. This determines the form of the
Navier-Stokes hydrodynamics, but more explicit dependence of the transport coefficients on the macroscopic
variables requires a corresponding explicit solution to the integral equations for $f_{i}^{(0)}$ and
$f_{i}^{(1)}$. One approximate method, known to be accurate for ordinary fluids, is detailed in the follow-on
paper \cite{GDH06}, resulting in constitutive quantities that are algebraic functions of the macroscopic
variables.

\section{Revised Enskog Kinetic Theory}
\label{sec2}

The system considered is a mixture of $\{N_i\}$ smooth hard disks ($d=2$) or spheres ($d=3$) of masses $\left\{
m_{i}\right\}$ and diameters $\left\{ \sigma _{i}\right\}$, where the subscript $i$ labels one of the $s$
mechanically different species and $d$ is the dimension. In general, collisions among all pairs are inelastic
and are characterized by independent constant normal restitution coefficients $\left\{ \alpha _{ij}=\alpha
_{ji}\right\} $, where $\alpha _{ij}$ is the restitution coefficient for collisions between particles of species
$i$ and $ j$, $0<\alpha _{ij}\leq 1$. The macroscopic (or hydrodynamic) properties of interest (number
densities, flow velocity, and energy density) are determined from the single particle position and velocity
distribution functions $f_{i}(\mathbf{r}_{1},\mathbf{v}_{1};t)$, for $i=1,..s$, where $
f_{i}(\mathbf{r}_{1},\mathbf{v}_{1};t)d\mathbf{r}_{1}d\mathbf{v}_{1}$ is proportional to the probability to find
a particle of species $i$ in the position and velocity element $d\mathbf{r}_{1}d\mathbf{v}_{1}$ at time $t$. The
fundamental description of any system is based on the probability density for \textit{all} constituent particles
and the Liouville equation for its time evolution; this is equivalent to solving the collective equations of
motion for all particles in the system and becomes computationally prohibitive for a large number of particles.
However, for the macroscopic fields only the reduced distribution functions $\left\{ f_{i}\right\} $, obtained
from the integration of the probability density over all except one particle's position and velocity for each of
the species, are required for calculation of the macroscopic properties. The equations for these reduced
distribution functions resulting from the partial integrations of the Liouville equation, give rise to the BBGKY
hierarchy equations. The first level of this hierarchy gives the time dependence of $\left\{ f_{i}\right\}$
\cite{Brey97,vanN01}
\begin{equation}
\left( \partial _{t}+\mathbf{v}_{1}\cdot \mathbf{\nabla
}_{\mathbf{r}_{1}}
\mathbf{+}m_{i}^{-1}\mathbf{F}_{i}(\mathbf{r}_{1})\cdot
\mathbf{\nabla }_{ \mathbf{v}_{1}}\right)
f_{i}(\mathbf{r}_{1},\mathbf{v}_{1};t)=C_{i}(\mathbf{r
}_{1},\mathbf{v}_{1};t),  \label{2.1}
\end{equation}
where
\begin{eqnarray}
C_{i}(\mathbf{r}_{1},\mathbf{v}_{1};t) &=&\sum_{j=1}^{s}\sigma
_{ij}^{d-1}\int d\mathbf{v}_{2}\int
d\widehat{{\boldsymbol\sigma}}\Theta (
\widehat{{\boldsymbol\sigma}}\cdot
\mathbf{g}_{12})(\widehat{{\boldsymbol\sigma}}
\cdot \mathbf{g}_{12})  \label{2.1.1} \\
&&\times \left( \alpha _{ij}^{-2}f_{ij}(\mathbf{r}_{1},\mathbf{v}
_{1}^{\prime \prime
},\mathbf{r}_{1}-\boldsymbol{\sigma}_{ij},\mathbf{v} _{2}^{\prime
\prime };t)-f_{ij}(\mathbf{r}_{1},\mathbf{v}_{1},\mathbf{r
}_{1}+\boldsymbol{\sigma }_{ij},\mathbf{v}_{2};t)\right).  \notag
\end{eqnarray}
The left sides of these equations describe changes in the
distribution functions due to motion in the presence of external
conservative forces $ \mathbf{F}_{i}(\mathbf{r}_{1})$. The right
side describes changes due to collisions among the particles. The
function $f_{ij}(\mathbf{r}_{1},
\mathbf{v}_{1},\mathbf{r}_{2},\mathbf{v}_{2};t)d\mathbf{r}_{1}d\mathbf{v}
_{1}d\mathbf{r}_{2}d\mathbf{v}_{2}$ is proportional to the joint
probability of finding a particle of species $i$ in
$d\mathbf{r}_{1}d\mathbf{v}_{1}$ and one of species $j$ in
$d\mathbf{r}_{2}d\mathbf{v}_{2}$. The position $ \mathbf{r}_{2}$
in these functions appears only for $\mathbf{r}_{2}=\mathbf{r
}_{1}\pm {\boldsymbol{\sigma}}_{ij}$, where
${\boldsymbol\sigma}_{ij}=\widehat{{\boldsymbol\sigma}}\sigma_{ij}$
and $\sigma_{ij}\equiv \left(\sigma_{i}+\sigma_{j}\right) /2$;
this means that the two particles are at contact. The vector
$\widehat{{\boldsymbol\sigma}}$ is a unit vector directed along
the line of centers from the sphere of species $j$ to that of
species $i$ at contact and the integration
$d\widehat{{\boldsymbol\sigma}}$ is over a solid angle for this
contact sphere. The Heaviside step function $\Theta $ assures that
the relative velocities $\mathbf{g}_{12}=
\mathbf{v}_{1}-\mathbf{v}_{2}$ are such that a collision takes
place, and the ``restituting'' (pre-collisional) velocities
$\mathbf{v}_{1}^{\prime \prime }$ and $\mathbf{v}_{2}^{\prime
\prime }$ are related to the post-collisional velocities by
\begin{equation}
\mathbf{v}_{1}^{\prime \prime }=\mathbf{v}_{1}-\mu _{ji}\left(
1+\alpha _{ij}^{-1}\right) (\widehat{{\boldsymbol\sigma}}\cdot
\mathbf{g}_{12})\widehat{
{\boldsymbol\sigma}},\hspace{0.3in}\mathbf{v}_{2}^{\prime \prime
}=\mathbf{v} _{2}+\mu _{ij}\left( 1+\alpha _{ij}^{-1}\right)
(\widehat{{\boldsymbol\sigma}} \cdot
\mathbf{g}_{12})\widehat{{\boldsymbol\sigma}} \label{2.2}
\end{equation}
where $\mu _{ij}=m_{i}/\left( m_{i}+m_{j}\right) $. It is
convenient for the discussion here to write that equation in a
more symbolic form by introducing the notation
\begin{equation}
X(\mathbf{v}_{1}^{\prime \prime },\mathbf{v}_{2}^{\prime \prime
})=b_{ij}^{-1}X(\mathbf{v}_{1},\mathbf{v}_{2}),  \label{2.2a}
\end{equation}
so that $b_{ij}^{-1}$ is a general substitution operator that
changes the argument of a function to its precollision velocities
given by (\ref{2.2}). Then, changing variables
$\widehat{{\boldsymbol\sigma}}\rightarrow -\widehat{
{\boldsymbol\sigma}}$ in the second term on the right side of
(\ref{2.1}) and noting that
$b_{ij}^{-1}\widehat{{\boldsymbol\sigma}}\cdot
\mathbf{g}_{12}=-\alpha
_{ij}^{-1}\widehat{{\boldsymbol\sigma}}\cdot \mathbf{g}_{12}$
gives the equivalent form \cite{Brey97,vanN01}
\begin{eqnarray}
\left( \partial _{t}+\mathbf{v}_{1}\cdot \mathbf{\nabla }_{\mathbf{r}_{1}}
\mathbf{+}m_{i}^{-1}\mathbf{F}_{i}(\mathbf{r}_{1})\cdot \mathbf{\nabla }_{ \mathbf{v}_{1}}\right)
f_{i}(\mathbf{r}_{1},\mathbf{v}_{1};t) &=&-\sum_{j=1}^{s}\sigma _{ij}^{d-1}\int d\mathbf{v}_{2}\int
d\widehat{{\boldsymbol\sigma}}(\alpha _{ij}^{-1}b_{ij}^{-1}+1)(\widehat{{\boldsymbol\sigma}}\cdot
\mathbf{g}_{12})  \notag \\
&&\times \Theta (-\widehat{{\boldsymbol\sigma}}\cdot
\mathbf{g}_{12})f_{ij}(
\mathbf{r}_{1},\mathbf{v}_{1},\mathbf{r}_{1}-{\boldsymbol{\sigma
}}_{ij},\mathbf{v} _{2};t).  \label{2.2b}
\end{eqnarray}
This demonstrates that the two particle distributions
$f_{ij}$ appear only on the contact hemisphere given by
$\Theta (-\widehat{{\boldsymbol\sigma}} \cdot \mathbf{g}_{12})$,
correponding to particles that are directed toward each other and
hence have a change in their velocities.

Equation (\ref{2.1}) becomes a kinetic theory (i.e., closed
equations for the set of $f_{i}$) only after specifying $f_{ij}$
on the right side as a functional of the set of $f_{i}$ (the
alternative of making approximations at higher levels of the BBGKY
hierarchy has not been productive in general for molecular gases).
As indicated above, this is required for $f_{ij}$ only when the
particles are at contact and on that hemisphere for which the
relative velocities are directed toward each other. In this
restricted context, the Enskog kinetic theory results from a
neglect of velocity correlations, i.e. the Enskog approximation
\begin{equation}
f_{ij}(\mathbf{r}_{1},\mathbf{v}_{1},\mathbf{r}_{2},\mathbf{v}
_{2};t)\rightarrow \chi _{ij}\left(
\mathbf{r}_{1},\mathbf{r}_{2}\mid \left\{ n_{i}\right\} \right)
f_{i}(\mathbf{r}_{1},\mathbf{v}_{1};t)f_{j}(
\mathbf{r}_{2},\mathbf{v}_{2};t).  \label{2.3}
\end{equation}
Spatial correlations arising from volume exclusion effects are
retained through the factor $\chi_{ij}\left(
\mathbf{r}_{1},\mathbf{r}_{2}\mid \left\{ n_{i}\right\} \right) $.
In the special case of a uniform system, it is simply related to
the nonequilibrium pair correlation function $g_{ij}\left( \left|
\mathbf{r}_{1}-\mathbf{r}_{2}\right| ;\left\{ n_{i}\right\}
\right) $ (probability density to find a particle of species $i$
at $\mathbf{r}_{1}$ and $j$ at $\mathbf{r}_{2}$) by
\cite{Lutsko01}
\begin{equation}
g_{ij}\left( \sigma _{ij};\left\{ n_{k}\right\} \right)
=\frac{1+\alpha _{ij}}{2\alpha _{ij}}\chi _{ij}\left( \sigma
_{ij};\left\{ n_{k}\right\} \right)  \label{2.3a}
\end{equation}
This relationship is proved in Appendix \ref{appA} and provides some partial interpretation for $\chi_{ij}$. It
is important to note that these correlation functions are \textit{functionals} of the actual species densities
$\left\{ n_{i}\right\} $ (defined below in Eq.\ (\ref{3.1})). This functional dependence is what distinguishes
the RET from the original ``standard'' Enskog theory (SET), where the $g_{ij}$ are \textit{functions} of the
species densities at the single position of interest, $\mathbf{r}_{1}$. Some partial justification for the
approximation (\ref{2.3}) for ordinary atomic fluids is given in Appendix \ref{appA}, where it is known to
provide accurate results for moderately dense gases, and reasonable estimates even for dense gases. Its use for
granular gases is justified largely from expectations based on these results for ordinary fluids.

Substitution of the Enskog approximation (\ref{2.3}) into the exact first level hierarchy equations (\ref{2.1})
defines the RET for the distribution functions $\left\{ f_{i}\right\} $
\begin{equation}
\left( \partial _{t}+\mathbf{v}_{1}\cdot \mathbf{\nabla
+}m_{i}^{-1}\mathbf{F }_{i}(\mathbf{r}_{1})\cdot \mathbf{\nabla
}_{\mathbf{v}_{1}}\right) f_{i}(
\mathbf{r}_{1},\mathbf{v}_{1};t)=\sum_{j=1}^{s}J_{ij}\left[
\mathbf{r}_{1}, \mathbf{v}_{1}\mid f(t)\right] \;.  \label{2.5}
\end{equation}
The collision operators $\left\{ J_{ij}\left[
\mathbf{r}_{1},\mathbf{v} _{1}\mid f(t)\right] \right\} $ are
given by
\begin{eqnarray}
J_{ij}\left[ \mathbf{r}_{1},\mathbf{v}_{1}\mid f(t)\right] &\equiv
&\sigma _{ij}^{d-1}\int d\mathbf{v}_{2}\int
d\widehat{{\boldsymbol\sigma}}\Theta (
\widehat{{\boldsymbol\sigma}}\cdot
\mathbf{g}_{12})(\widehat{{\boldsymbol\sigma}}
\cdot \mathbf{g}_{12})  \notag \\
&&\times \left[ \alpha_{ij}^{-2}\chi_{ij}\left(
\mathbf{r}_{1},\mathbf{r} _{1}-{\boldsymbol\sigma}_{ij}\mid
\left\{ n_{i}\right\} \right) f_{i}(\mathbf{r}
_{1},\mathbf{v}_{1}^{\prime \prime
};t)f_{j}(\mathbf{r}_{1}-{\boldsymbol\sigma}
_{ij},\mathbf{v}_{2}^{\prime \prime };t)\right.  \notag \\
&&\left. -\chi_{ij}\left(
\mathbf{r}_{1},\mathbf{r}_{1}+{\boldsymbol\sigma} _{ij}\mid
\left\{ n_{i}\right\} \right) f_{i}(\mathbf{r}_{1},\mathbf{v}
_{1};t)f_{j}(\mathbf{r}_{1}+{\boldsymbol\sigma}_{ij},\mathbf{v}_{2};t)\right]
. \label{2.6}
\end{eqnarray}
The corresponding Boltzmann equations for a dilute mixture follow from this result since $\chi_{ij}\left(
\mathbf{r}_{1},\mathbf{r}_{1}-{\boldsymbol\sigma}_{ij}\mid \left\{ n_{i}\right\} \right) \rightarrow 1$ at low
density. Furthermore, on length scales of the order of the mean free path or greater, the different centers
$\left( \mathbf{r}_{1},\mathbf{r}_{2}=\mathbf{ r}_{1}\pm {\boldsymbol\sigma}_{ij}\right) $ of the colliding pair
in Eq.\ (\ref{2.6}) can be neglected $\left( \mathbf{r}_{1}\approx \mathbf{r}_{2}\right) $ since the diameters
of the particles are small compared to the mean free path at low density. As will be shown below, a nonzero
distance between the particle centers gives rise to the collisional contributions to the transport coefficients,
which are not present in dilute systems. These two modifications to $f_{ij}$ result in the usual Boltzmann
description for a granular mixture. The results obtained here therefore encompass earlier work on granular
mixtures at low density \cite{Garzo02}. In the elastic limit, $\alpha _{ij}\rightarrow 1$, these equations
become the Enskog theory for mixtures of dense molecular gases studied in Ref.\ \cite{deHaro83}.

 As happens for elastic collisions, the inelastic Enskog equation provides a semiquantitative
description of the hard sphere system that neglects the velocity correlations between the particles that are
about to collide (molecular chaos assumption). The Enskog approximation is expected to be valid for short times
since as the system evolves corrections to the Enskog equation due to multiparticle collisions, including
recollision events (``ring'' collisions) should be incorporated. The latter are expected to be stronger for
fluids with inelastic collisions where the colliding pairs tend to be more focused. Therefore, some deviations
from molecular chaos have been observed in molecular dynamics (MD) simulations \cite{ML98,SM01,PTNE02} of
granular fluids as the density increases. Although the existence of these correlations restricts the range of
validity of the Enskog equation, there is substantial evidence in the literature for the validity of the Enskog
theory at moderate densities and higher restitution coefficients especially at the level of macroscopic
properties (such as transport coefficients). In the case of molecular dynamics (MD) simulations, the Enskog
theory compares quite well with simulations for the radial distribution function \cite{Lutsko01}, the
self-diffusion coefficient \cite{BRCG00,LBD02}, the kinetic temperatures of a binary mixture in homogeneous
cooling state \cite{Dahl2-02}, and the rheological properties of a mixture under simple shear flow
\cite{L01,MGAL06}. The agreement between MD and Enskog equation is good for moderate densities (solid volume
fraction up to 0.15) and even conditions of strong dissipation (restitution coefficients $\alpha_{ij}>0.7$). For
higher densities the $\alpha$ range is more limited but the Enskog theory still captures the relevant
qualitative features. The Enskog transport coefficients for a monocomponent gas \cite{Garzo99a} have also been
tested against NMR experiments of a system of mustard seeds vibrated vertically \cite{YHCMW02,HYCMW04}. The
average value of the coefficient of restitution of the grains used in this experiment is $\alpha=0.87$, which
lies outside of the quasielastic limit ($\alpha \approx 0.99$). Comparison between theory and experiments shows
that the Enskog kinetic theory successfully models the density and granular temperature profiles away from the
vibrating container bottom and quantitatively explains the temperature inversion observed in experiments
\cite{Martin05}. All these results clearly show the applicability of the Enskog theory for densities outside the
Boltzmann limit and values of dissipation beyond the quasielastic limit. In this context, one can conclude that
the Enskog equation provides a unique basis for the description of dynamics across a wide range of densities,
length scales, and degrees of dissipation. No other theory with such generality exists.

\section{Macroscopic Balance Equations}
\label{sec3}

In the previous section, the Enskog assumption (\ref{2.3}) was used to
obtain a closed set of kinetic equations (\ref{2.5}) for a moderately dense
mixture of inelastic hard spheres.  The result takes the form of
nonlinear, integral-differential equations for the distribution function
$f_i$, which contains information on a {\it microscopic} scale.  In this section,
this theory will be used to obtain the corresponding description on the
{\it macroscopic} (or hydrodynamic) scale.  First, the relevant macroscopic
variables will be identified and defined.  Next, the corresponding balance
equations will be derived.  Finally, expressions for the equations of
state (pressure and cooling rate) and fluxes will be
presented as integral expressions containing $f_i$.

The variables of interest for a macroscopic description of the mixture are the number densities for all species,
$ \left\{ n_{i}\left( \mathbf{r},t\right) \right\}$ (or equivalently, the mass densities $\left\{ \rho_i\left(
\mathbf{r},t\right)=m_in_{i}\left( \mathbf{r},t\right) \right\}$), the total energy density, $e\left(
\mathbf{r},t\right) $, and the total momentum, $\mathbf{p} \left( \mathbf{r},t\right) $. These are expected to
be the $s+1+d$ slow variables that dominate the dynamics for long times through a closed autonomous set of
equations, the hydrodynamic equations. The reasoning behind this is that these are the densities for global
conserved quantities in molecular fluids, and therefore have decay times set by the wavelength of the
excitations. Long wavelength (space scales large compared to the mean free path) phenomena therefore persist at
long times (compared to a mean free time) after which the complex transient microscopic dynamics has become
negligible. For granular fluids, the energy is not conserved but is characterized by a cooling rate at long
wavelengths. Still, this cooling rate may be slow compared to the transient dynamics and thus the energy remains
a relevant slow variable. This is confirmed by MD simulations showing a rapid approach to this cooling law after
only a few collisions \cite{Dahl2-02}.

These macroscopic variables will be referred to collectively as
the hydrodynamic fields. They are defined without approximation in
terms of moments of the distribution functions
\begin{equation}
n_i\left( \mathbf{r},t\right)\equiv \int d\mathbf{v}f_{i}(\mathbf{r}, \mathbf{v};t),\hspace{0.3in}i=1,..s,
\label{3.1}
\end{equation}
\begin{equation}
e\left( \mathbf{r},t\right) \equiv \sum_{i=1}^{s}\int
d\mathbf{v}\frac{1}{2} m_{i}v^{2}f_{i}(\mathbf{r},\mathbf{v};t)
\label{3.2}
\end{equation}
\begin{equation}
\mathbf{p}\left( \mathbf{r},t\right) \equiv \sum_{i=1}^{s}\int
d\mathbf{v} m_{i}\mathbf{v}f_{i}(\mathbf{r},\mathbf{v};t)
\label{3.3}
\end{equation}
The time dependence occurs entirely through the distribution function and hence is determined from the Enskog
kinetic equations (\ref{2.5}). However, rather than solving the kinetic equation to determine this complete time
dependence it is useful for the purposes of deriving the simpler hydrodynamic description to first obtain the
balance equations. These equations express the time derivative of the hydrodynamic fields in terms of local
fluxes and sources due to collisions or the external force. These equations and the identification of the fluxes
follow in detail from the form of the collision operators in (\ref{2.1.1}) as shown in Appendix \ref{appB} (in
fact they are obtained there exactly from the first hierarchy equation (\ref{2.1}) without the Enskog
approximation (\ref{2.3}) and hence are exact). The results for the balance equations are
\begin{equation}
\partial _{t}n_{i}\left( \mathbf{r},t\right) +m_i^{-1}\nabla \cdot \mathbf{j}
_{i}\left( \mathbf{r},t\right) =0,  \label{3.4}
\end{equation}
\begin{equation}
\partial _{t}e\left( \mathbf{r},t\right) +\nabla \cdot \mathbf{s}\left(
\mathbf{r},t\right) =-w\left( \mathbf{r},t\right) +\sum_{i=1}^{s}m_i^{-1}\mathbf{F} _{i}\left( \mathbf{r}\right)
\cdot \mathbf{j}_{i}\left( \mathbf{r},t\right) , \label{3.5}
\end{equation}
\begin{equation}
\partial _{t}p_{\beta }\left( \mathbf{r},t\right) +\partial _{r_{\gamma
}}t_{\gamma \beta }\left( \mathbf{r},t\right)
=\sum_{i=1}^{s}n_{i}\left( \mathbf{r},t\right) F_{i\beta }\left(
\mathbf{r}\right) .  \label{3.6}
\end{equation}
The explicit expressions for $\mathbf{j}_i$, $\mathbf{s}$, $w$ and
$t_{\gamma\beta}$ are contained in Appendix \ref{appB} and not
shown here since they are cast in a more convenient form below.

The mass fluxes $\left\{ \mathbf{j}_{i}\left( \mathbf{r},t\right)
\right\} ,$ energy flux $\mathbf{s}\left( \mathbf{r},t\right) $,
and momentum flux $ t_{\beta\gamma }\left( \mathbf{r},t\right) $
describe the rate of transport of the hydrodynamic fields through
a given cross sectional area. They consist of parts due to pure
convection and parts due to collision. To identify the convective
(kinetic) parts, the local flow field $\mathbf{U}\left(
\mathbf{r},t\right) $ is defined in terms of the momentum density
by
\begin{equation}
\mathbf{p}\left( \mathbf{r},t\right) \equiv \rho \left(
\mathbf{r},t\right) \mathbf{U}\left( \mathbf{r},t\right)
,\hspace{0.3in}\rho \left( \mathbf{r} ,t\right)
=\sum_{i=1}^{k}m_{i}n_{i}\left( \mathbf{r},t\right) ,
\label{3.11}
\end{equation}
where the second equation defines the mass density. Also, the
energy density is written in terms of the internal energy density
$e_{0}\left( \mathbf{r} ,t\right) $ in the local rest frame, plus
the energy due to flow
\begin{equation}
e\left( \mathbf{r},t\right) =e_{0}\left( \mathbf{r},t\right)
+\frac{1}{2} \rho \left( \mathbf{r},t\right) U^{2}\left(
\mathbf{r},t\right). \label{3.11a}
\end{equation}
In terms of $\mathbf{U}\left( \mathbf{r},t\right)$ the fluxes
become
\begin{equation}
\mathbf{j}_{i}\left( \mathbf{r},t\right) =\rho_{i}\left( \mathbf{r},t\right) \mathbf{U}\left(
\mathbf{r},t\right) +\mathbf{j}_{0i}\left( \mathbf{r} ,t\right),  \label{3.13}
\end{equation}
\begin{equation}
s_{\beta }\left( \mathbf{r},t\right) =\left( e_{0}\left(
\mathbf{r},t\right) +\frac{1}{2}\rho \left( \mathbf{r},t\right)
U^{2}\left( \mathbf{r},t\right) \right) U_{\beta }\left(
\mathbf{r},t\right) +P_{\beta \gamma }\left( \mathbf{r},t\right)
U_{\gamma }\left( \mathbf{r},t\right) +q_{\beta }\left(
\mathbf{r},t\right),  \label{3.14}
\end{equation}
\begin{equation}
t_{\beta \gamma }\left( \mathbf{r},t\right) =\rho \left(
\mathbf{r},t\right) U_{\beta }\left( \mathbf{r},t\right) U_{\gamma
}\left( \mathbf{r},t\right) +P_{\beta \gamma }\left(
\mathbf{r},t\right). \label{3.15}
\end{equation}
The first terms on the right sides describe convective transport,
while the diffusion fluxes $\mathbf{j}_{0i}\left(
\mathbf{r},t\right) ,$ heat flux $ \mathbf{q}\left(
\mathbf{r},t\right) $, and pressure tensor $P_{\beta \gamma
}\left( \mathbf{r}_{1},t\right) $ describe the residual transport
for each fluid element in its local rest frame. Before giving
their forms more explicitly, it is instructive to insert
(\ref{3.13})--(\ref{3.15}) into ( \ref{3.4})--(\ref{3.6}) to get
the equivalent form for the balance equations
\begin{equation}
D_{t}n_{i}+n_{i}\nabla \cdot \mathbf{U}+m_i^{-1}\nabla \cdot \mathbf{j}_{0i}=0, \label{3.20}
\end{equation}
\begin{equation}
D_{t}e_{0}+\left( e_{0}\delta _{\gamma \beta }+P_{\gamma \beta }\right)
\partial _{r_{\gamma }}U_{\beta }+\nabla \cdot \mathbf{q}=-w\left( \mathbf{r}
,t\right) +\sum_{i=1}^{s}m_i^{-1}\mathbf{F}_{i}\left(
\mathbf{r}\right) \cdot \mathbf{j}_{0i}\left( \mathbf{r},t\right)
,  \label{3.21}
\end{equation}
\begin{equation}
\rho D_{t}U_{\beta }+\partial _{r_{\gamma }}P_{\gamma \beta }=\sum_{i=1}^{s}n_{i}\left( \mathbf{r},t\right)
F_{i\beta }\left( \mathbf{r} \right),  \label{3.22}
\end{equation}
where $D_{t}=\partial _{t}+\mathbf{U}\cdot \nabla $ is the material
derivative.

The independent hydrodynamic fields are now $\left\{ n_{i}\left( \mathbf{r} ,t\right) \right\}$, $e_{0}\left(
\mathbf{r},t\right) $, and $\mathbf{U} \left( \mathbf{r},t\right)$. The remaining quantities in the balance
equations are the energy loss rate $w\left( \mathbf{r},t\right) $, the mass fluxes $\left\{
\mathbf{j}_{0i}\left( \mathbf{r},t\right) \right\} $, the heat flux $\mathbf{q}\left( \mathbf{r},t\right) $, and
the pressure tensor $P_{\beta \gamma }\left( \mathbf{r}_{1},t\right) $. These quantities, which are defined in
terms of the distribution functions, are obtained by the explicit forms for $\mathbf{j}_i$, $\mathbf{s}$, $w$,
and $t_{\gamma\beta}$ given in Appendix \ref{appB} together with Eqs.\ (\ref{3.13})--(\ref{3.15}).

Specifically, the energy loss rate is due to inelastic collisions
\begin{eqnarray}
w\left( \mathbf{r},t\right) &\equiv
&\frac{1}{4}\sum_{i,j=1}^{s}\left( 1-\alpha _{ij}^{2}\right)
m_{i}\mu _{ji}\sigma _{ij}^{d-1}\int d\mathbf{v}
_{1}\int d\mathbf{v}_{2}\int d\widehat{\boldsymbol {\sigma }}\,  \notag \\
&&\times \Theta (\widehat{\boldsymbol {\sigma }}\cdot
\mathbf{g}_{12})(\widehat{ \boldsymbol {\sigma }}\cdot
\mathbf{g}_{12})^{3}f_{ij}(\mathbf{r}_{1},
\mathbf{v}_{1},\mathbf{r}_{1}+{\boldsymbol {\sigma
}}_{ij},\mathbf{v}_{2};t), \label{3.26}
\end{eqnarray}
whereas the diffusion flux arises from convective (kinetic) transport
\begin{equation}
\mathbf{j}_{0i}\left( \mathbf{r}_{1},t\right) \equiv m_i\int d\mathbf{v}_{1}
\mathbf{V}_{1}f_{i}(\mathbf{r}_{1},\mathbf{v}_{1};t), \label{3.27}
\end{equation}
where  ${\bf V}_1={\bf v}_1-{\bf U}({\bf r},t)$ is the velocity in the local rest frame. The heat flux has both
``kinetic'' and ``collisional'' transfer contributions
\begin{equation}
\mathbf{q}\left( \mathbf{r}_{1},t\right) \equiv
\mathbf{q}^{k}\left( \mathbf{ r}_{1},t\right)
+\mathbf{q}^{c}\left( \mathbf{r}_{1},t\right) ,  \label{3.28}
\end{equation}
with
\begin{equation}
\mathbf{q}^{k}\left( \mathbf{r}_{1},t\right) =\sum_{i=1}^{s}\int
d\mathbf{v}
_{1}\frac{1}{2}m_{i}V_{1}^{2}\mathbf{V}_{1}f_{i}(\mathbf{r}_{1},\mathbf{v}
_{1};t),  \label{3.29}
\end{equation}
\begin{eqnarray}
\mathbf{q}^{c}\left( \mathbf{r}_{1},t\right)
&=&\sum_{i,j=1}^{k}\frac{1}{8} \left( 1+\alpha _{ij}\right)
m_{j}\mu _{ij}\sigma _{ij}^{d}\int d\mathbf{v} _{1}\int
d\mathbf{v}_{2}\int d\widehat{\boldsymbol {\sigma }}\,\Theta
(\widehat{\boldsymbol {\sigma }}\cdot \mathbf{g}_{12})  \notag \\
&&\times (\widehat{\boldsymbol {\sigma }}\cdot
\mathbf{g}_{12})^{2}\left[ \left( 1-\alpha _{ij}\right) \left( \mu
_{ji}-\mu _{ij}\right)( \widehat{\boldsymbol {\sigma}}\cdot
\mathbf{g}_{12})+4\widehat{\boldsymbol {\sigma }}
\cdot \mathbf{G}_{ij}\right)  \notag \\
&&\times \widehat{\boldsymbol {\sigma
}}\int_{0}^{1}dxf_{ij}(\mathbf{r} _{1}-x\boldsymbol {\sigma
}_{ij},\mathbf{v}_{1},\mathbf{r}_{1}+\left( 1-x\right) \boldsymbol
{\sigma}_{ij},\mathbf{v}_{2};t), \label{3.29a}
\end{eqnarray}
where ${\bf G}_{ij}=\mu_{ij} {\bf V}_1+\mu_{ji} {\bf V}_2$ is the
center-of-mass velocity.

Similarly, the pressure tensor has both kinetic and collisional
contributions
\begin{equation}
P_{\gamma \beta }\left( \mathbf{r}_{1},t\right) \equiv P_{\gamma
\beta }^{k}\left( \mathbf{r}_{1},t\right) +P_{\gamma \beta
}^{c}\left( \mathbf{r} _{1},t\right) ,  \label{3.30}
\end{equation}
where
\begin{equation}
P_{\gamma \beta }^{k}\left( \mathbf{r}_{1},t\right) =\sum_{i=1}^{s}\int d \mathbf{v}_{1}m_{i}V_{1\beta
}V_{1\gamma }f_{i}(\mathbf{r}_{1},\mathbf{v} _{1};t),  \label{3.31}
\end{equation}
\begin{eqnarray}
P_{\gamma \beta }^{c}\left( \mathbf{r}_{1},t\right) &=&\frac{1}{2}
\sum_{i,j=1}^{s}m_{j}\mu _{ij}\left( 1+\alpha _{ij}\right) \sigma
_{ij}^{d}\int d\mathbf{v}_{1}\int d\mathbf{v}_{2}\int
d\widehat{\boldsymbol {\sigma }}\,\Theta (\widehat{\boldsymbol
{\sigma }}\cdot
\mathbf{g}_{12})(\widehat{\boldsymbol {\sigma }}\cdot \mathbf{g}_{12})^{2}  \notag \\
& &\times \widehat{\sigma}_{\beta}\widehat{\sigma}_{\gamma} \int_{0}^{1}dxf_{ij}(\mathbf{r}_{1}-x{\boldsymbol
{\sigma }}_{ij},\mathbf{v}_{1}, \mathbf{r}_{1}+\left( 1-x\right) {\boldsymbol {\sigma }}_{ij},\mathbf{v}_{2};t).
\label{3.32}
\end{eqnarray}
Equations (\ref{3.20})--(\ref{3.22}) together with the definitions (\ref{3.26})--(\ref{3.32}) represent the
macroscopic balance equations for a granular mixture, without restrictions on the densities or degrees of
dissipation. In the case of a three-dimensional system ($d=3$), the above equations reduce to previous results
\cite{GM03} derived for hard spheres. When the approximate form (\ref{2.3}) is used in the first hierarchy
equation and in these expressions for the cooling rate and fluxes, the Enskog theory results.

For historical consistency with the usual constitutive equations for a ordinary fluid, the temperature $T\left(
\mathbf{r},t\right) $ is used in the following instead of the internal energy density $e_{0}\left( \mathbf{r}
,t\right) $, with the definition
\begin{equation}
e_{0}\left( \mathbf{r},t\right) \equiv \frac{d}{2}n\left(
\mathbf{r} ,t\right) T\left( \mathbf{r},t\right) .  \label{3.33}
\end{equation}
As a definition, this amounts only to a change of variables and
there are no thermodynamic implications involved in the use of
this temperature for a granular fluid. The corresponding
hydrodynamic equation for $T\left( \mathbf{ r},t\right) $ follows
directly from (\ref{3.21})
\begin{equation}
\frac{d}{2}n\left( D_{t}+\zeta \right) T+P_{\gamma \beta }\partial _{r_{\gamma }}U_{\beta }+\nabla \cdot
\mathbf{q}-\frac{d}{2} T\sum_{i=1}^{s}m_i^{-1}\nabla \cdot \mathbf{j}_{0i}=\sum_{i=1}^{s}m_i^{-1}\mathbf{F}
_{i}\cdot \mathbf{j}_{0i}.  \label{3.34}
\end{equation}
To obtain these results the continuity equation has been used
\begin{equation}
D_{t}\rho +\rho \nabla \cdot \mathbf{U}=0.  \label{3.35}
\end{equation}
This follows from the definitions of $\rho $ and $\mathbf{U}$ and
the conservation laws for the $\left\{ n_{i}\left(
\mathbf{r},t\right) \right\} $ . A related consequence is
\begin{equation}
\sum_{i=1}^{s}\mathbf{j}_{0i}=0,  \label{3.36}
\end{equation}
so that only $s-1$ dissipative mass fluxes are independent.
Finally, the ``cooling rate''\ $\zeta $ has been introduced in
(\ref{3.34}) by the definition
\begin{eqnarray}
\zeta &=&\frac{2}{dnT}w=\frac{1}{2dnT}\sum_{i,j=1}^{s}\left( 1-\alpha _{ij}^{2}\right) m_{i}\mu _{ji}\sigma
_{ij}^{d-1}\int d\mathbf{v}
_{1}\int d\mathbf{v}_{2}\int d\widehat{\boldsymbol {\sigma }},  \notag \\
&&\times \Theta (\widehat{\boldsymbol {\sigma }}\cdot
\mathbf{g}_{12})(\widehat{ \boldsymbol {\sigma }}\cdot
\mathbf{g}_{12})^{3}f_{ij}(\mathbf{r}_{1},
\mathbf{v}_{1},\mathbf{r}_{1}+\boldsymbol {\sigma
}_{ij},\mathbf{v}_{2};t). \label{3.37}
\end{eqnarray}

\section{Concept of a Normal Solution and Hydrodynamics}
\label{sec4}

The form of the equations of state and fluxes given in the previous section, (\ref{3.27})--(\ref{3.32}) and
(\ref{3.37}), are cast as functionals of the distributions $\left\{ f_{i}\right\}$, which depend explicitly on
space and time.  As a result, the macroscopic balance equations are not entirely expressed in terms of the
hydrodynamic fields, and thus do not comprise a closed set of equations. If these distributions can instead be
expressed as functionals of the hydrodynamic fields ({\em normal} solution), then $\zeta \left(
\mathbf{r},t\right) $, $\left\{ \mathbf{j}_{0i}\left( \mathbf{r},t\right) \right\} $, $\mathbf{q}\left(
\mathbf{r},t\right) $, and $P_{\beta \gamma }\left( \mathbf{r}_{1},t\right) $ also will become functionals of
the hydrodynamic fields through (\ref{3.27})--(\ref{3.32}) and (\ref{3.37}). Such expressions are called
``constitutive relations''. They provide the missing link between the  balance equations and a closed set of
equations for the hydrodynamic fields alone. Such a closed set of equations defines ``hydrodynamics'' in its
most general sense.

It is seen, therefore, that any derivation of hydrodynamics
proceeds first by construction of normal solutions to the kinetic
equations. More precisely, a normal solution is one whose space
and time dependence occurs entirely through the hydrodynamic
fields, denoted
\begin{equation}
f_{i}(\mathbf{r}_{1},\mathbf{v}_{1};t)=f_{i}(\mathbf{v}_{1}\mid \left\{
y_{\beta }\left( \mathbf{r}_{1},t\right) \right\} ),  \label{4.1}
\end{equation}
where $\left\{y_{\beta }\left( \mathbf{r}_{1},t\right) \right\}$
denotes generically the set of hydrodynamic fields
\begin{equation}
y_{\beta }\mathbf{\Leftrightarrow }\left\{ T,\mathbf{U},\left\{
n_{i}\left( \mathbf{r},t\right) \right\} \right\}.  \label{4.1a}
\end{equation}
Therefore, the space and time derivatives of the kinetic
equation are given by
\begin{equation}
\left( \partial _{t}+\mathbf{v}_{1}\cdot \mathbf{\nabla
}_{\mathbf{r} }\right) f_{i}(\mathbf{v}_{1}\mid \left\{ y_{\beta
}\left( \mathbf{r} _{1},t\right) \right\} )=\int
d\mathbf{r}\frac{\delta f_{i}(\mathbf{v} _{1}\mid \left\{ y_{\beta
}\left( t\right) \right\} )}{\delta y_{\eta }\left(
\mathbf{r};t\right) }\left( \partial _{t}+\mathbf{v}_{1}\cdot
\mathbf{\nabla }_{\mathbf{r}}\right) y_{\eta }\left(
\mathbf{r};t\right) . \label{4.2}
\end{equation}
Furthermore, the balance equations for the hydrodynamic fields (\ref{3.20})--(\ref{3.22}) can be used to express
$
\partial _{t} y_{\eta} \left( \mathbf{r}; t \right)$ in
(\ref{4.2}) in terms of space derivatives of the hydrodynamic fields. For such a solution for $f_i$, Eqs.\
(\ref{3.27})--(\ref{3.32}) and (\ref{3.37}) give directly by integration the desired constitutive relations.

The determination of $f_{i}(\mathbf{v}_{1}\mid \left\{ y_{\beta
}\left( \mathbf{r}_{1},t\right) \right\} )$ from the kinetic
equations (\ref{2.5}) is a very difficult task in general, and
further restriction on the class of problems considered is
required at this point to make progress. Any \emph{functional} of
the fields can be represented equivalently as a local
\emph{function} of the fields \emph{and} all of their gradients.
In many cases, gradients of high degree are small and may be
negligible so that the normal distribution becomes
\begin{equation}
f_{i}(\mathbf{v}_{1}\mid \left\{ y_{\beta }\left(
\mathbf{r}_{1},t\right) \right\} )\rightarrow
f_{i}(\mathbf{v}_{1};\left\{ y_{\beta }\left( \mathbf{r
}_{1},t\right) ,\mathbf{\nabla }_{\mathbf{r}_{1}}y_{\beta }\left(
\mathbf{r} _{1},t\right) ,\cdots\right\} )  \label{4.3}
\end{equation}
This representation does not imply that the low degree gradients
are small, and $f_{i}$ may be a non-linear function of the
relevant gradients. This occurs in many important applications for
granular fluids \cite {Santos04}. In the limiting case where the
low-degree gradients can be controlled by boundary or initial
conditions and made small, a further Taylor series expansion can
be given
\begin{eqnarray}
f_{i}(\mathbf{v}_{1}\mid \left\{ y_{\beta }\left(
\mathbf{r}_{1},t\right) \right\} )&\rightarrow&
f_{i}^{(0)}(\mathbf{v}_{1};\left\{ y_{\beta }\left(
\mathbf{r}_{1},t\right) \right\}
)+f_{i}^{(1)}(\mathbf{v}_{1};\left\{ y_{\beta }\left(
\mathbf{r}_{1},t\right) \right\} )+\cdots \nonumber\\
&\rightarrow&f_{i}^{(0)}(\mathbf{v}_{1};\left\{ y_{\beta }\left(
\mathbf{r}_{1},t\right) \right\} )+ \mathbf{Y}_{i\alpha
}(\mathbf{v} _{1};\left\{ y_{\beta }\left( \mathbf{r}_{1},t\right)
\right\} )\cdot \mathbf{\nabla }_{\mathbf{r}_{1}}y_{\alpha }\left(
\mathbf{r}_{1},t\right) +\cdots \nonumber\\
\label{4.4}
\end{eqnarray}
It follows that the leading order distributions have the exact
properties
\begin{equation}
n_{i}\left( \mathbf{r},t\right) \equiv \int
d\mathbf{v}f_{i}^{(0)}(\mathbf{v} ;\left\{ y_{\beta }\left(
\mathbf{r},t\right) \right\} ),\hspace{0.3in} i=1,..s,
\label{4.4a}
\end{equation}
\begin{equation}
\frac{d}{2}n\left( \mathbf{r},t\right) T\left( \mathbf{r},t\right) \equiv \sum_{i=1}^{s}\int
d\mathbf{v}\frac{1}{2}m_{i}V^2 f_{i}^{(0)}(\mathbf{v};\left\{ y_{\beta }\left( \mathbf{r} ,t\right) \right\} ),
\label{4.4b}
\end{equation}
\begin{equation}
\rho \left( \mathbf{r},t\right) \mathbf{U}\left(
\mathbf{r},t\right) \equiv \sum_{i=1}^{s}\int
d\mathbf{v}m_{i}\mathbf{v}f_{i}^{(0)}(\mathbf{v};\left\{ y_{\beta
}\left( \mathbf{r},t\right) \right\} ),  \label{4.4c}
\end{equation}
and the corresponding moments of all higher order terms in (\ref{4.4}) must vanish. Generalization of this type
of gradient expansion for the normal solution to include a class of nonlinear gradients in the reference state
has been discussed recently \cite{Lutsko06,G06bis}.

As is standard for molecular gases, the gradient expansion will be
taken with respect to the reference local HCS, i.e. that resulting
from the neglect of all gradients in the functional but evaluated at
the value of the fields at the chosen point and time $\left\{
y_{\beta }\left( \mathbf{r}_{1},t\right) \right\} $. This point is
crucial in our analysis since most of the previous results have
taken elastic Maxwell distributions as the base state. Note that in
the CE method the form of $f_{i}^{(0)}$ comes from the solution to
the kinetic equation to zeroth order in gradients and cannot be
chosen a priori. Accordingly, $f_{i}^{(0)}(\mathbf{v}_{1};\left\{
y_{\beta }\left( \mathbf{r}_{1},t\right) \right\} )\rightarrow
f_{i}^{(0)}(V_{1};\left\{ y_{\beta }\left( \mathbf{r}_{1},t\right)
\right\} )$, where $V\mathbf{=}\left|
\mathbf{v-U(r},t\mathbf{)}\right| $ is homogeneous and isotropic
with respect to its velocity dependence. This symmetry implies that
the leading (zero) order contributions to (\ref{3.27}) and
(\ref{3.28}) for the vector fluxes $\left\{ \mathbf{j}_{0i}\left(
\mathbf{r},t\right) \right\} $ and $\mathbf{q}\left(
\mathbf{r},t\right) $ must vanish, and this contribution to the
pressure tensor $P_{\gamma \beta}$ must be isotropic (proportional
to $ \delta _{\gamma \beta })$. Similar symmetry considerations to
the first order contribution (linear in the gradients) determines
the exact structure of the constitutive equation to this order.
Based on these symmetry considerations, the constitutive quantities
are known to take the forms
\begin{equation}
\zeta \left( \mathbf{r},t\right) \rightarrow \zeta ^{(0)}\left(
\left\{ y_{\beta }\left( \mathbf{r},t\right) \right\} \right)
+\zeta _{U}\left( \left\{ y_{\beta }\left( \mathbf{r},t\right)
\right\} \right) \mathbf{\nabla }\cdot \mathbf{U}\left(
\mathbf{r},t\right) ,  \label{4.5}
\end{equation}
\begin{eqnarray}
\mathbf{j}_{0i}\left( \mathbf{r},t\right) &\rightarrow
&-\sum_{j=1}^{s}m_im_{j} \frac{n_{j}\left( \mathbf{r},t\right)
}{\rho \left( \mathbf{r},t\right) } D_{ij}\left( \left\{ y_{\beta
}\left( \mathbf{r},t\right) \right\} \right)
\mathbf{\nabla }\ln n_{j}\left( \mathbf{r},t\right)  \notag \\
&&-\rho \left( \mathbf{r},t\right) D_{i}^{T}\left( \left\{
y_{\beta }\left( \mathbf{r},t\right) \right\} \right)
\mathbf{\nabla }\ln T\left( \mathbf{r},t\right)
-\sum_{j=1}^{s}D_{ij}^{F}\left( \left\{ y_{\beta }\left(
\mathbf{r},t\right) \right\} \right) \mathbf{F}_{j}\left(
\mathbf{r} \right),  \label{4.6}
\end{eqnarray}
\begin{eqnarray}
\mathbf{q}\left( \mathbf{r},t\right) &\rightarrow &-\lambda \left(
\left\{ y_{\beta }\left( \mathbf{r},t\right) \right\} \right)
\mathbf{\nabla }T\left( \mathbf{r},t\right)  \notag \\
&&-\sum_{i,j=1}^{s}\left( T^2\left( \mathbf{r},t\right) D_{q,ij}\left( \left\{ y_{\beta }\left(
\mathbf{r},t\right) \right\} \right) \mathbf{\nabla }\ln n_{j}\left( \mathbf{r},t\right) +L_{ij}\left( \left\{
y_{\beta }\left( \mathbf{r},t\right) \right\} \right) \mathbf{F}_{j}\left( \mathbf{r}\right) \right),
\nonumber\\ \label{4.7}
\end{eqnarray}
\begin{eqnarray}
P_{\gamma \lambda }\left( \mathbf{r},t\right) &=&p\left( \left\{
y_{\beta }\left( \mathbf{r},t\right) \right\} \right) \delta
_{\gamma \lambda }-\eta \left( \left\{ y_{\beta }\left(
\mathbf{r},t\right) \right\} \right) \left(
\partial _{r_{\gamma }}U_{\lambda }\left( \mathbf{r},t\right) +\partial
_{r_{\lambda}}U_{\gamma }\left( \mathbf{r},t\right)
-\frac{2}{d}\mathbf{ \nabla }\cdot \mathbf{U}\left(
\mathbf{r},t\right) \right)
\notag \\
&&-\kappa \left( \left\{ y_{\beta }\left( \mathbf{r},t\right)
\right\} \right) \mathbf{\nabla }\cdot \mathbf{U}\left( \mathbf{r}
,t\right).  \label{4.8}
\end{eqnarray}
The unknown quantities in these constitutive equations (\ref{4.5})--(\ref{4.8}) include the cooling rate $\zeta
^{(0)}\left( \left\{ y_{\beta }\left( \mathbf{r},t\right) \right\} \right) $, the hydrostatic granular pressure
$p\left( \left\{ y_{\beta }\left( \mathbf{r},t\right) \right\} \right) $, and the transport coefficients $\zeta
_{U}$, $D_{ij}$, $D_{i}^{T}$, $D_{ij}^{F}$, $D_{q,ij}$, $L_{ij}$, $\eta$, and $\kappa$. These quantities can be
expressed as explicit functions of the hydrodynamic variables once $f_{i}^{(0)}$ and $f_{i}^{(1)}$ are known.
The equations governing the solution of $f_{i}^{(0)}$ and $f_{i}^{(1)}$ are found using the CE method, as
described below.

\section{Chapman--Enskog Normal Solution}
\label{sec5}

The CE method is a procedure for constructing an approximate normal solution. It is perturbative, using the
spatial gradients as the small expansion parameter. More precisely, the small parameter is Knudsen number
($\mathrm{Kn}$), defined as the gradient of the hydrodynamic fields relative to their local value times the mean
free path. This means that the conditions for the solution are restricted to small variations of the
hydrodynamic fields over distances of the order of the mean free path. In the presence of an external force it
is necessary to characterize the magnitude of this force relative to the gradients as well. Here, it is assumed
that the magnitude of the force is first order in perturbation expansion. This allows comparison with the
results of Ref.\ \cite{deHaro83} for the elastic case.

The perturbation is carried out by considering the Enskog kinetic
equations successively at each order in the gradients. As
described below, the zeroth order equation is first obtained for
$f_{i}^{(0)}$.  Next, the first order equation for $f_{i}^{(1)}$
is obtained.  This expansion leads to integral-differential
equations for the determination of $f_{i}^{(0)}$ and
$f_{i}^{(1)}$, which are solved explicitly in the follow-on paper
\cite{GDH06}.

As detailed in Appendix \ref{appC}, to zeroth order in the gradients,
the kinetic equation (\ref{2.5}) becomes
\begin{equation}
\left( \partial _{t}T\right) \partial
_{T}f_{i}^{(0)}(\mathbf{v}_{1};\left\{ y_{\beta }\left(
\mathbf{r}_{1},t\right) \right\}
)=\sum_{j=1}^{s}J_{ij}^{(0)}\left[ \mathbf{v}_{1}\mid
f^{(0)}(t)\right] \;, \label{5.1}
\end{equation}
where
\begin{eqnarray}
J_{ij}^{(0)}\left[ \mathbf{v}_{1}\mid f^{(0)}(t)\right] &\equiv
&\chi _{ij}^{(0)}\left( \sigma _{ij};\left\{ n_{i}\left(
\mathbf{r}_{1},t\right) \right\} \right) \sigma _{ij}^{d-1}\int
d\mathbf{v}_{2}\int d\widehat{\boldsymbol {\sigma }}\Theta
(\widehat{\boldsymbol {\sigma }}\cdot \mathbf{g}_{12})(\widehat{
\boldsymbol {\sigma }}\cdot \mathbf{g}_{12})  \notag \\
&&\times \left[ \alpha _{ij}^{-2}f_{i}^{(0)}(V_{1}^{\prime \prime
};\left\{ y_{\beta }\left( \mathbf{r}_{1},t\right) \right\}
)f_{j}^{(0)}(V_{2}^{\prime \prime };\left\{ y_{\beta }\left(
\mathbf{r}_{1},t\right) \right\} )\right.\nonumber\\
& & \left. -f_{i}^{(0)}(V_{1};\left\{ y_{\beta }\left(
\mathbf{r}_{1},t\right) \right\} )f_{j}^{(0)}(V_{2};\left\{
y_{\beta }\left( \mathbf{r}_{1},t\right) \right\} )\right] .
\label{5.2}
\end{eqnarray}
All spatial gradients are neglected at this lowest order. Equation
(\ref{5.1}) determines the velocity dependence of
$f_{i}^{(0)}(V_{1};\left\{ y_{\beta }\left(
\mathbf{r}_{1},t\right) \right\} );$ the space and time dependence
is local and entirely through the fields $\left\{ y_{\beta }\left(
\mathbf{r} _{1},t\right) \right\} $ at the space and time point of
interest. This has been exploited by writing the time dependence
of $f_{i}^{(0)}$ in terms of the time dependence of the fields,
and recognizing that all time derivatives of the latter are
proportional to space gradients, except the temperature, through
the balance equations
\begin{equation}
\partial _{t}n_{i}=0,\quad \partial _{t}T=-\zeta ^{(0)}T,
\quad \partial _{t}\mathbf{U=0}.  \label{5.2a}
\end{equation}
Here, $\zeta ^{(0)}$ is the cooling rate (\ref{3.37}) to zeroth
order in the gradients
\begin{eqnarray}
\zeta ^{(0)} &=&\frac{1}{2dnT}\sum_{i,j=1}^{s}\left( 1-\alpha
_{ij}^{2}\right) m_{i}\mu _{ji}\chi _{ij}^{(0)}\left( \sigma
_{ij};\left\{ n_{i}\left( \mathbf{r}_{1},t\right) \right\} \right)
\sigma _{ij}^{d-1}\int d
\mathbf{v}_{1}\int d\mathbf{v}_{2}\int d\widehat{\boldsymbol {\sigma }}\,  \notag \\
&&\times \Theta (\widehat{\boldsymbol {\sigma }}\cdot \mathbf{g}_{12})(\widehat{\boldsymbol {\sigma }}\cdot
\mathbf{g}_{12})^{3}f_{i}^{(0)}(V_{1};\left\{ y_{\beta }\left( \mathbf{r}_{1},t\right) \right\}
)f_{j}^{(0)}(V_{2};\left\{ y_{\beta }\left( \mathbf{r}_{1},t\right) \right\} ).  \label{5.2b}
\end{eqnarray}
Similarly, the functional dependence of $\chi _{ij}^{(0)}\left( \mathbf{r}_{1}, \mathbf{r}_{2}\mid \left\{
n_{i}\right\} \right) $ on the compositions to zeroth order in the gradients has the functional dependence on
the densities replaced by $\left\{ n_{i}\right\} \rightarrow \left\{ n_{i}\left( \mathbf{r} _{1},t\right)
\right\} $, at the point of interest. The result is translational and rotational invariant $\chi
_{ij}^{(0)}\left( \mathbf{r}_{1}, \mathbf{r}_{2}\mid \left\{ n_{i}\right\} \right) \rightarrow \chi
_{ij}^{(0)}\left( \left| \mathbf{r}_{1}-\mathbf{r}_{2}\right| ;\left\{ n_{i}\left( \mathbf{r}_{1},t\right)
\right\} \right) ,$ a \emph{function} of the densities. Finally, gradients in the distribution functions of the
collision operators must be neglected, e.g. $f_{i}^{(0)}(\mathbf{v} _{1};\left\{ y_{\beta }\left(
\mathbf{r}_{1}+\boldsymbol{\sigma }_{ij},t\right) \right\} )\rightarrow $ $f_{i}^{(0)}(\mathbf{v}_{1};\left\{
y_{\beta }\left( \mathbf{r}_{1},t\right) \right\} )$.

A further simplification of these equations for the lowest order
distribution functions occurs when they are written in terms of
the corresponding dimensionless forms $\left\{ \phi _{i}\right\}$
\begin{equation}
f_{i}^{(0)}(V;\left\{ y_{\beta }\right\} )=n_{i}v_{0}^{-d}(T)\phi
_{i}\left( V^{\ast };\left\{ n_{i}^{\ast }\right\} \right) ,
\label{5.3}
\end{equation}
with the definitions
\begin{equation}
\mathbf{V}^{\ast }=\frac{\mathbf{V}}{v_{0}(T)},\quad v_{0}(T)= \sqrt{\frac{2T}{m}},\quad n_{i}^{\ast
}=n_{i}\sigma _{i}^{d}, \quad m=\frac{1}{s}\sum_{i=1}^{s}m_{i} \label{5.4}.
\end{equation}
The solution depends on the flow field only through the relative
velocity $ \mathbf{V}$. Furthermore, since there is no external
energy scale the temperature can occur only through the scaling of
the dimensionless velocity through the thermal velocity
$v_{0}(T)$. Equation (\ref{5.1}) now takes the dimensionless form
\begin{equation}
-\frac{1}{2}\zeta ^{\ast }\nabla _{\mathbf{V}^*}\cdot \left( \mathbf{V}^{\ast }\phi _{i}\right)
=\sum_jJ_{ij}^{(0)\ast}\left( V^{\ast}\mid \phi _{i}\right) , \label{5.5}
\end{equation}
where $\nabla_{{\bf V}^*}\equiv \partial/\partial {\bf V}^*$ and
\begin{equation}
\zeta ^{\ast }=\frac{\ell }{v_{0}}\zeta ^{(0)},\quad J_{ij}^{(0)\ast }= \frac{\ell
}{n}v_{0}^{d-1}J_{ij}^{(0)},\quad \ell =\frac{1}{n\sigma ^{d-1}},\quad n=\sum_{i=1}^sn_i, \quad \sigma
=\frac{1}{s}\sum_{i=1}^{s}\sigma _{i}. \label{5.5a}
\end{equation}
The solution to this equation is a universal function of the
magnitude of the velocity $V^{\ast }$ and is otherwise independent
of the temperature and flow field. For a one component fluid it is
independent of the density as well. However, for mixtures it is
parameterized by the dimensionless species densities through the
factors $\chi _{ij}^{(0)}\left( \sigma _{ij};\left\{ n_{i}^{\ast
}\right\} \right) .$ Equation (\ref{5.1}) has the same form as the
corresponding dimensionless Enskog equations for a strictly \emph{
homogeneous} state. The latter is called the HCS. Here, however,
the state is not homogeneous because of the requirements
(\ref{4.4a})--(\ref{4.4c}). Instead it is a \emph{local} \textit{\
}HCS. As said before, an important point to recognize is that the
occurrence of this local HCS as the reference state is not an
assumption of the CE expansion but rather a consequence of the
kinetic equations to lowest order in the gradient expansion.

The analysis to first order in the gradients is similar and the
details are given in Appendix \ref{appC}. The result has the form
(\ref{4.4})
\begin{eqnarray}
f_{i}^{(1)} &\rightarrow &\boldsymbol{\mathcal{A}}_{i}\left(
\mathbf{V}\right)\cdot  \nabla \ln
T+\sum_{j=1}^{s}\boldsymbol{\mathcal{B}}_{i}^j\left(
\mathbf{V}\right) \cdot \nabla \ln n_{j}  \nonumber\\
&&+\mathcal{C}_{i,\gamma \eta }\left( \mathbf{V} \right)
\frac{1}{2}\left( \partial _{\gamma }U_{\eta }+\partial _{\eta
}U_{\gamma }-\frac{2}{d}\delta _{\gamma \eta }\nabla \cdot
\mathbf{U} \right)  \nonumber\\
&&+\mathcal{D}_{i}\left( \mathbf{V} \right) \nabla \cdot
\mathbf{U}+\sum_{j=1}^{s}\boldsymbol{\mathcal{E}}_{i}^j\left(
\mathbf{V} \right) \cdot \mathbf{F}_{j}. \label{5.6}
\end{eqnarray}
The contributions from the flow field gradients have been
separated into independent traceless and diagonal components, as
follows from fluid symmetry. The velocity dependence of the
gradient contributions is contained in the functions
$\boldsymbol{\mathcal{A}}_{i}\left( \mathbf{V}\right)
,\boldsymbol{\mathcal{B}} _{i}^j\left( \mathbf{V}\right)
,\mathcal{C}_{i,\gamma \eta }\left( \mathbf{V} \right)
,\mathcal{D}_{i}\left( \mathbf{V}\right) $, and
$\boldsymbol{\mathcal{E}}_{i}^j\left( \mathbf{V}\right) $. The
kinetic equations determine these functions as the solutions to
the integral equations
\begin{equation}
\left( \left( \mathcal{L}-\frac{1}{2}\zeta ^{(0)}\right) \boldsymbol{\mathcal{A}}\right) _{i}=\mathbf{A}_{i},
\label{5.7}
\end{equation}
\begin{equation}
\left( \mathcal{L}\boldsymbol{\mathcal{B}}^j\right)_{i}-n_{j}\frac{\partial \zeta ^{(0)}}{\partial
n_{j}}\boldsymbol{\mathcal{A}}_{i}=\mathbf{B}_{i}^{j}, \label{5.8}
\end{equation}
\begin{equation}
\left( \left( \mathcal{L}+\frac{1}{2}\zeta ^{(0)}\right) \mathcal{C}_{\gamma \eta }\right) _{i}=C_{i,\gamma \eta
}, \label{5.9}
\end{equation}
\begin{equation}
\left( \left( \mathcal{L}+\frac{1}{2}\zeta ^{(0)}\right) \mathcal{D}\right) _{i}=D_{i},  \label{5.10}
\end{equation}
\begin{equation}
\left(
\left(\mathcal{L}+\zeta^{(0)}\right)\boldsymbol{\mathcal{E}}^j\right)_{i}=\mathbf{E}_{i}^{j}.
\label{5.11}
\end{equation}

The linear operator $\mathcal{L}$ is given by
\begin{equation}
\left( \mathcal{L}X\right) _{i}=\frac{1}{2}\zeta ^{(0)}\nabla
_{\mathbf{V} }\cdot \left( \mathbf{V}X_{i}\right) +\left(
LX\right) _{i},  \label{5.12}
\end{equation}
\begin{equation}
\left( LX\right) _{i}=-\sum_{j=1}^{s}\left( J_{ij}^{(0)}\left[ \mathbf{v} _{1}\mid X_{i},f_{j}^{(0)}\right]
+J_{ij}^{(0)}\left[ \mathbf{v}_{1}\mid f_{i}^{(0)},X_{j}\right] \right), \label{5.12bis}
\end{equation}
and the inhomogeneous terms are defined by
\begin{equation}
A_{i,\gamma}\left( \mathbf{V}\right)=\frac{1}{2} V_{\gamma}\nabla
_{\mathbf{V}}\cdot \left( \mathbf{V}f_{i}^{(0)}\right)
-\frac{p}{\rho }\partial_{V_{\gamma}}
f_{i}^{(0)}+\frac{1}{2}\sum_{j=1}^{k}\mathcal{ K}_{ij,\gamma
}\left[\nabla_{\mathbf{V}}\cdot \left( \mathbf{V}
f_{j}^{(0)}\right) \right] ,  \label{5.13}
\end{equation}
\begin{eqnarray}
B_{i,\gamma}^{j}\left( \mathbf{V}\right) &=& -V_\gamma n_{j}\partial _{n_{j}}f_{i}^{(0)}-\rho^{-1}
(\partial_{V_\gamma}f_{i}^{(0)})n_{j}(\partial _{n_{j}}p)
\nonumber\\
 &&-\sum_{\ell
=1}^{s}\mathcal{K}_{i\ell,\gamma}\left[\left( n_{j}\partial
_{n_{j}}+\frac{1}{2}\left( n_{\ell }\frac{\partial \ln \chi
_{i\ell }^{(0)}}{\partial n_{j}}+I_{i\ell j}\right) \right)
f_{\ell }^{(0)} \right],  \label{5.14}
\end{eqnarray}
\begin{eqnarray}
\label{5.15} C_{i,\gamma \beta }\left(
\mathbf{V}\right)&=&\frac{1}{2} \left(V_\gamma
\partial_{V_\beta}f_{i}^{(0)}+V_\beta
\partial_{V_\gamma}f_{i}^{(0)}-\frac{2}{d}\delta_{\beta
\gamma}{\bf V}\cdot \nabla_{\bf V}f_{i}^{(0)}\right)\nonumber\\
& & +\frac{1}{2}\sum_{j
=1}^{s}\left(\mathcal{K}_{ij,\gamma}[\partial_{V_\beta}f_{j}^{(0)}]+
\mathcal{K}_{ij,\beta}[\partial_{V_\gamma}f_{j}^{(0)}]-\frac{2}{d}
\delta_{\beta
\gamma}\mathcal{K}_{ij,\lambda}[\partial_{V_\lambda}f_{j}^{(0)}]\right),
\end{eqnarray}
\begin{eqnarray}
D_{i}({\bf V})&=&\frac{1}{d}\mathbf{V}\cdot \nabla _{\mathbf{V}} f_{i}^{(0)}-\frac{1}{2} \left( \zeta
_{U}+\frac{2}{n T d}p\right) \nabla _{\mathbf{V}}\cdot \left( \mathbf{V}
f_{i}^{(0)}\right)\nonumber\\
& &  +\sum_{j=1}^s\left( n_{j}\partial _{n_{j}}f_{i}^{(0)}+\frac{1
}{d}\mathcal{K}_{ij,\gamma}\left[\partial
_{V_{\gamma}}f_{j}^{(0)}\right]\right),   \label{5.16}
\end{eqnarray}
\begin{equation}
\mathbf{E}_{i}^{j}({\bf V})=-\left(\nabla_{\mathbf{V}} f_{i}^{(0)}\right)\frac{1}{m_j}\left(\delta
_{ij}-\frac{n_{j}m_{j}}{\rho}\right),  \label{5.17}
\end{equation}
where the operator $\mathcal{K}_{ij,\gamma}[X]$ is defined by Eq.\
(\ref{c.12}).

This completes the construction of the normal solution to the
revised Enskog equations up through first order in the gradients.
Equation\ (\ref{5.5}) determines the $f_{i}^{(0)}$ through the
definition (\ref{5.3}); solution to the linear integral equations
(\ref{5.7})--(\ref {5.11}) determines the $f_{i}^{(1)}$ through
the definition (\ref{5.6}). The unknown fluxes and cooling rate of
the hydrodynamic equations can then be calculated with these
solutions. This is made explicit in the next section.

\section{Constitutive Equations and Transport Coefficients}
\label{sec7}

The forms for the constitutive equations to first order in the gradients are given by Eqs.\
(\ref{4.5})--(\ref{4.8}). The explicit representations for the coefficients in these equations are given in
terms of the solutions to the integral equations for $f_{i}^{(0)}$ and $f_{i}^{(1)}$ of the previous section.
Details of the simplification of these expressions in terms of $f_{i}^{(0)}$ and $f_{i}^{(1)}$ are given in
Appendix \ref{app4} and only the final results are presented here.

Recall that the results below are based on the assumption that the
external force is of the same magnitude as $f_{i}^{(1)}$;  a force of
different magnitude would result in different constitutive relations.

\subsection{Cooling Rate}

The cooling rate is calculated from Eq.\ (\ref{3.37}), resulting
in the form (\ref {4.5})
\begin{equation}
\zeta \rightarrow \zeta ^{(0)}+\zeta _{U}\mathbf{ \nabla }\cdot
\mathbf{U} , \label{6.6a}
\end{equation}
with
\begin{equation}
\zeta ^{(0)}=\frac{B_{3}}{2dnT}\sum_{i,j=1}^{s}\left( 1-\alpha
_{ij}^{2}\right) \frac{m_{i}m_{j}}{m_{i}+m_{j}}\chi
_{ij}^{(0)}\sigma _{ij}^{d-1}\int d\mathbf{v}_{1}\int d\mathbf{v}
_{2}f_{i}^{(0)}(V_{1})f_{j}^{(0)}(V_{2})\,g_{12}^{3},
\label{6.6b}
\end{equation}
and
\begin{eqnarray}
\label{6.6c}
 \zeta _{U}
&=&-\frac{d+2}{dnT}B_{4}\sum_{i,j=1}^{s}\left( 1-\alpha _{ij}^{2}\right) \mu_{ji}\chi _{ij}^{(0)}\sigma
_{ij}^{d} n_{i}n_{j} T_{i}^{(0)}
\nonumber\\
&&+\frac{B_{3}}{d n T}\sum_{i,j=1}^{s}\left( 1-\alpha
_{ij}^{2}\right) \frac{m_{i}m_{j}}{m_{i}+m_{j}}\chi
_{ij}^{(0)}\sigma _{ij}^{d-1}\int d\mathbf{v}_{1}\int
d\mathbf{v}_{2}\;g_{12}^3f_{i}^{(0)}({\bf
V}_{1})\mathcal{D}_{j}({\bf V}_{2}).
\end{eqnarray}
The constant $B_{n}$ is defined by
\begin{equation}
B_{n}\equiv \pi ^{\left( d-1\right) /2}\frac{\Gamma \left(
\frac{n+1}{2} \right) }{\Gamma \left( \frac{n+d}{2}\right) }.
\label{6.6d}
\end{equation}
Also in (\ref{6.6c}) the species temperatures $\left\{
T_{i}^{(0)}\right\} $ have been defined by
\begin{equation}
\frac{d}{2}n_{i}T_{i}^{(0)}=\int d\mathbf{v}\frac{1}{2}
m_{i}V^{2}f_{i}^{(0)}(\left\{ n_{i}\right\} ,T,V).  \label{6.4}
\end{equation}
There is no special significance to these quantities other than naming the
integral on the right side, which is a specified function of the
hydrodynamic fields $\left\{ n_{i}\right\} $ and the global temperature $T$
through $f_{i}^{(0)}$.

\subsection{Mass Fluxes}

The mass fluxes are determined from the definition of (\ref{3.27})
leading to the form (\ref{4.6}) to first order in the gradients
\begin{equation}
\mathbf{j}_{0i}\rightarrow -\sum_{j=1}^{s}m_im_{j}\frac{n_{j}}{\rho }D_{ij} \mathbf{\nabla }\ln n_{j}-\rho
D_{i}^{T}\mathbf{\nabla }\ln T-\sum_{j=1}^{s}D_{ij}^{F}\mathbf{F}_{j}.
\end{equation}
The transport coefficients are identified as
\begin{equation}
D_{i}^{T}=-\frac{m_i}{\rho d}\int d\mathbf{v}\mathbf{V}\cdot \boldsymbol{\mathcal{A}}_{i}\left(
\mathbf{V}\right) , \label{6.8}
\end{equation}
\begin{equation}
D_{ij}=-\frac{\rho }{m_{j}n_{j}d}\int d\mathbf{v}\mathbf{V}\cdot \boldsymbol{\mathcal{B}}_{i}^{j}\left(
\mathbf{V}\right) ,\label{6.9}
\end{equation}
\begin{equation}
D_{ij}^{F}=-\frac{m_i}{d}\int d\mathbf{v}\mathbf{V}\cdot
\boldsymbol{\mathcal{E}}_{i}^{j}\left( \mathbf{V}\right).
\label{6.10}
\end{equation}

\subsection{Energy Flux}

The energy flux to first order in the gradients is given by
(\ref{4.7})
\begin{equation}
\mathbf{q}\rightarrow -\lambda \nabla  T-\sum_{i,j=1}^{s}\left(
T^2D_{q,ij}\nabla\ln n_{j}+L_{ij}\mathbf{F}_{j}\right).
\end{equation}
There are both kinetic and collisional transfer contributions according to Eq.\ (\ref{3.28}), $\mathbf{q}\equiv
\mathbf{q}^{k}+\mathbf{q}^{c}$. The kinetic contributions to the transport coefficients are identified as
\begin{equation}
\lambda ^{k}=\sum_{i=1}^{s}\lambda _{i}^{k}=-\frac{1}{dT}\sum_{i=1}^{s}\int d
\mathbf{v}\frac{1}{2}m_iV^{2}\mathbf{V}\cdot \boldsymbol{\mathcal{A}}_i\left( \mathbf{V}\right), \label{6.15}
\end{equation}
\begin{equation}
D_{q,ij}^{k}=-\frac{1}{dT^2}\int d\mathbf{v}\frac{1}{2}
m_iV^{2}\mathbf{V}\cdot \boldsymbol{\mathcal{B}}_{i}^{j}\left(
\mathbf{V}\right) , \label{6.13}
\end{equation}
\begin{equation}
L_{ij}^{k}=-\frac{1}{d}\int d\mathbf{v}\frac{1}{2}
m_iV^{2}\mathbf{V}\cdot \boldsymbol{\mathcal{E}}_{i}^{j}\left(
\mathbf{V}\right). \label{6.14}
\end{equation}
For convenience below, the partial thermal conductivities $\lambda
_{i}^{k}$ have been introduced in Eq.\ (\ref{6.15}). The collision
transfer contributions are obtained from (\ref{3.29a}) to first
order in the gradients. These are calculated in Appendix
\ref{app4} with the results
\begin{eqnarray}
\lambda ^{c}&=&\sum_{i,j=1}^{s}\frac{1}{8}\left( 1+\alpha
_{ij}\right) m_{j}\mu _{ij}\sigma _{ij}^{d}\chi _{ij}^{(0)}\left\{
2B_{4}\left( 1-\alpha _{ij}\right)\left( \mu _{ij}-\mu
_{ji}\right)n_{i} \left[ \frac{2}{m_j}\lambda
_{j}^{k}+(d+2)\frac{T_{i}^{(0)}}{m_im_jT}\rho
D_{j}^{T}\right] \right.   \nonumber\\
&&\left. +\frac{8B_{2}}{2+d}n_{i}\left[ \frac{2\mu
_{ij}}{m_j}\lambda _{j}^{k}-(d+2)\frac{T_i^{(0)}}{m_im_jT}\left(
2\mu _{ij}-\mu _{ji}\right)\rho D_{j}^{T}\right]
-T^{-1}C_{ij}^{T}\right\} , \label{f.36}
\end{eqnarray}
\begin{eqnarray}
D_{q,ij}^{c}&=&\sum_{p=1}^{s}\frac{1}{8}\left( 1+\alpha
_{ip}\right) m_{p}\mu _{ip}\sigma _{ip}^{d}\chi _{ip}^{(0)}\left\{
2B_{4}\left( 1-\alpha _{ip}\right)\left( \mu _{ip}-\mu
_{pi}\right)\right.  \nonumber\\
& &\times  n_{i}\left[ \frac{ 2}{m_{p}}D_{q,pj}^{k}+(d+2)
\frac{T_{i}^{(0)}}{T^2}\frac{m_{j}n_{j}
}{\rho m_{i}}D_{pj}\right]    \nonumber\\
&&\left. +\frac{8B_{2}}{d+2}n_{i}\left[ \frac{2\mu
_{pi}}{m_{p}}D_{q,pj}^k-(d+2)\left(2\mu _{ip}-\mu
_{pi}\right)\frac{T_i^{(0)}}{T^2} \frac{n_jm_{j}}{ m_{i}\rho
}D_{pj}\right] -T^{-2}C_{ipj}^{T}\right\} ,
\nonumber\\
\label{f.37}
\end{eqnarray}
\begin{eqnarray}
L_{ij}^{c}&=&\sum_{p=1}^{s}\frac{1}{8}\left( 1+\alpha _{ip}\right)
m_{p}\mu _{ip}\sigma _{ip}^{d}\chi _{ip}^{(0)}\left\{ 2B_{4}\left(
1-\alpha _{ip}\right)\left( \mu _{ip}-\mu
_{pi}\right)\right.  \nonumber\\
& &\times  n_{i}\left[ \frac{2}{m_{p}}L_{pj}^{k}+(d+2)
\frac{T_{i}^{(0)}}{m_im_{p}}D_{pj}^F\right]    \nonumber\\
&&\left. +\frac{8B_{2}}{d+2}n_{i}\left[ \frac{2\mu
_{pi}}{m_{p}}L_{pj}^k-(d+2)\left(2\mu _{ip}-\mu
_{pi}\right)\frac{T_i^{(0)}}{m_im_p}D_{pj}^F\right]\right\},
\label{f.38}
\end{eqnarray}
where the coefficients $C_{ij}^{T}$ and $C_{ipj}^{T}$ are given by Eqs.\ (\ref {f.28}) and (\ref{f.29}). These
expressions also depend on the transport coefficients of the mass fluxes, $D_{i}^{T}$, $D_{ij}$, and
$D_{ij}^{F}$ given by Eqs.\ (\ref{6.8}), (\ref{6.9}), and (\ref{6.10}), respectively, and on the kinetic
contributions $\lambda_i^k$, $D_{q,ij}^k$, and $L_{ij}^k $.

\subsection{Momentum Flux}

The pressure tensor is evaluated from Eqs.\
(\ref{3.30})--(\ref{3.32}). To zeroth order in the gradients, one
gets the pressure $p$ as
\begin{equation}
p\left( \left\{ n_{i}\right\} ,T\right) =\frac{1}{d}P_{\gamma
\gamma }^{(0)}\equiv p^{k}\left( \left\{ n_{i}\right\} ,T\right)
+p^{c}\left( \left\{ n_{i}\right\} ,T\right) =\frac{1}{d}P_{\gamma
\gamma }^{(0)k}+\frac{1 }{d}P_{\gamma \gamma }^{(0)c},
\label{6.1}
\end{equation}
where
\begin{equation}
p=p^{k}+p^{c},  \label{f.17}
\end{equation}
\begin{equation}
p^{k}=nT,\hspace{0.3in}p^{c}=B_{2}\sum_{i,j=1}^{s}\mu _{ji}\left( 1+\alpha _{ij}\right) \sigma _{ij}^{d}\chi
_{ij}^{(0)}n_{i}n_{j}T_{i}^{(0)}.\label{f.18}
\end{equation}
Similarly the shear viscosity is $\eta =\eta ^{k}+\eta ^{c}$ where
\begin{equation}
\eta ^{k}=\sum_{i=1}^{s}\eta _{i}^{k},\quad \eta
_{i}^{k}=-\frac{1}{(d+2)(d-1)}\sum_{i=1}^{s}\int d\mathbf{v}
m_{i}V_{\lambda }V_{\gamma}\mathcal{C}_{i,\lambda \gamma}\left(
\mathbf{V}\right),   \label{f.19}
\end{equation}
\begin{equation}
\eta ^{c} =\frac{2B_{2}}{\left( d+2\right) }\sum_{i,j=1}^{s}\mu
_{ij}\left( 1+\alpha _{ij}\right) \chi _{ij}^{(0)}n_{i}\sigma
_{ij}^{d}\eta _{j}^{k} +\frac{d}{d+2}\kappa^c. \label{f.20}
\end{equation}
Finally, the bulk viscosity is $\kappa =\kappa ^{k}+\kappa ^{c}$ where
\begin{equation}
\kappa ^{k}=0,\quad \kappa ^{c}=\frac{B_{3}\left( d+1\right)
}{2d^{2} }\sum_{i,j=1}^{s}m_{j}\mu _{ij}\left( 1+\alpha
_{ij}\right) \chi _{ij}^{(0)}\sigma _{ij}^{d+1}\int
d\mathbf{v}_{1}\int d\mathbf{v} _{2}f_{i}^{(0)}({\bf
V}_{1})f_{j}^{(0)}({\bf V}_{2})g_{12}. \label{f.21}
\end{equation}

\section{Discussion}
\label{sec9}

The most complete and accurate description of mixtures for ordinary fluids is based on the revised Enskog
kinetic equations for hard spheres. The explicit construction of solutions to those equations by the CE
expansion to first order in the gradients was given more that twenty years ago in Ref.\ \cite{deHaro83}. These
solutions, together with the macroscopic balance equations obtained from the kinetic equations, provide a
self-consistent derivation of Navier-Stokes hydrodynamics for mixtures and the identification of expressions for
all the Navier-Stokes parameters (equations of state, transport coefficients). In the context of the chosen
kinetic theory, the analysis and the expressions for these parameters are exact. At this formal level questions
of principle could be addressed, prior to the introduction of subsequent approximations for practical
evaluations. For example, it was shown that application of the analysis to the original and revised Enskog
theories leads to qualitatively different Navier-Stokes hydrodynamics, only one of which is consistent with
irreversible thermodynamics. Since no approximations were involved this was sufficient to reject the Enskog
kinetic theory in favor of its revised version \cite{vanBeijeren73}.

The present work is simply an extension of that in Ref.\ \cite{deHaro83} to {\em inelastic} hard sphere granular
mixtures. Modification of the collisions to account for inelasticity leads to significant differences from
ordinary fluids in detail, but the formal structure of the CE expansion remains the same. Similarly, granular
Navier-Stokes hydrodynamics results exactly from the CE solution to first order in the gradients and the
corresponding modified balance equations. The form of these hydrodynamic equations and expressions for the
transport coefficients are exact, as in the ordinary fluid case. The primary motivation for this analysis is to
provide the basis for practical applications, as noted in the Introduction, and described in the following
paper. However, at the formal level, important fundamental questions can be addressed and clarified as well.

The existence of hydrodynamics for granular fluids has been questioned, due to the many known differences from
ordinary fluids: there is no equilibrium or even stationary reference state; the temperature is not a
hydrodynamic field (failure of energy conservation), or conversely, multiple temperature fields could be
required for mixtures (failure of equilibrium state equipartition for the corresponding granular HCS). In the
end, qualitative discussions must be resolved by controlled analysis. Here, the validity of the RET for some
range of densities and degree of dissipation has been assumed as a mesoscopic basis for possible macroscopic
dynamics in a granular mixture. As shown in the text, sufficient conditions are the macroscopic balance
equations (verified) and a normal solution to the kinetic equations. The normal solution is defined in terms of
a chosen set of hydrodynamic fields, and the question of hydrodynamics reduces to its existence. The details of
the Appendices give the explicit construction of this solution to first order in the gradients, together with a
proof of the existence of solutions to the associated integral equations. It can be concluded from this that a
closed set of hydrodynamic equations for the species densities, flow velocity, and a single temperature exist
for sufficiently small gradients.

This conclusion is consistent with the observations that the reference state is not equilibrium, depends on the
cooling temperature, energy loss can be large at strong dissipation, and the kinetic temperatures of species are
different. None of these facts compromises implementation of the CE expansion for solution to the kinetic
equation. The parameters of the resulting Navier-Stokes equations incorporate such effects through the integral
equations that determine them, and their dependence on the time dependent fields. This in turn affects the
solutions to the Navier-Stokes equations under different physical conditions, and is responsible for some of the
observed peculiarities of granular fluids. Clearly, it is important to get the details of the Navier-Stokes
equations accurately before concluding that any observed experimental phenomenon is hydrodynamic or not. This is
another primary motivation for the present work.

These details entail solution to the equation for the reference state and solution to the integral equations for
the transport coefficients, to determine them as functions of the hydrodynamic fields (temperature, flow field,
and species densities) and the system parameters (restitution coefficients, masses, particle sizes). There has
been considerable study of the reference state, as an expansion about a Gaussian for relatively small velocities
(asymptotic forms for large velocities are known as well). The integral equations can be solved approximately as
truncated expansions in a complete set of polynomials with Gaussian weight factors. For ordinary fluids the
leading approximation is generally quite accurate, and the following paper gives its extension to the granular
mixture. Still, there are open questions about this approximation for strong dissipation and large mechanical
disparity (e.g., mass ratio). Previous results obtained for granular mixtures at low-density
\cite{Montanero03,Garzo04} and for the shear viscosity of a heated granular mixture at moderate density
\cite{GM03} have shown the accuracy of the above approximation, even for strong dissipation.

An accurate solution to the integral equations will predict the transport coefficients as functions of the
dissipation. There is only one correct result for this dependence, given by the formulas obtained here. However,
its measurement in a given experiment or simulation can entangle and affect this dependence of the transport
coefficients due to higher order gradients beyond the Navier-Stokes limit. It may be tempting to compare
experimental or simulation data to a corresponding Navier-Stokes solution, adjusting the transport coefficients
for a best fit and reporting these as the ``measured'' values. This can be misleading for granular fluids under
conditions where the size of the gradients increase with the degree of dissipation. For such states, strong
dissipation can require additional terms in the constitutive equations beyond those of Navier-Stokes order
\cite{Sela98,Santos04,Hrenya06}. This does not mean that the results obtained here are not correct at strong
dissipation, only that they must be distinguished carefully from other effects of the same order. A careful
tabulation of the Navier-Stokes results given here (e.g., via Monte Carlo simulation) is required for an
accurate analysis of experiments of current interest. It is an interesting new feature of granular fluids that
hydrodynamic states beyond Navier-Stokes order may be the norm rather than the exception.

\section{Acknowledgments}

 V. G. acknowledges partial support from the Ministerio de Ciencia y Tecnolog\'{\i}a
(Spain) through Grant No. FIS2007--60977. C.M.H. is grateful to the National Science Foundation for providing
financial support of this project through grant CTS-0318999 with additional support provided by the American
Chemical Society Petroleum Research Fund (Grant 43393-AC9) and the Engineering and Physical Sciences Research
Council (Grant EP/DO30676/1). C. M. H. and J. W. D. are also grateful to the organizers and participants of the
Granular Physics Workshop at the Kavli Institute of Theoretical Physics (with partial support from the National
Science Foundation under grant PHY99-07949), which provided a starting forum for much of this work.

\appendix

\section{RET and Spatial Correlations at Contact}
\label{appA}

In the case of ordinary fluids, the Enskog approximation can be understood as a short time, or Markovian
approximation. This follows if the initial distribution has the Enskog form
\begin{equation}
f_{ij}(\mathbf{r}_{1},\mathbf{v}_{1},\mathbf{r}_{1}-\boldsymbol
{\sigma
}_{ij},\mathbf{v}_{2};t=0)=f_{i}(\mathbf{r}_{1},\mathbf{v}_{1},t=0)f_{j}(
\mathbf{r}_{1}-\boldsymbol {\sigma }_{ij},\mathbf{v}_{2};t=0)\chi
_{ij}\left( \mathbf{r}_{1},\mathbf{r}_{1}-\boldsymbol {\sigma
}_{ij}\mid \left\{ n_{k}\right\} \right) .  \label{a.1.1}
\end{equation}
In fact, this is a quite plausible class of initial conditions since correlations in that case are generally
induced by the interparticle structure that is independent of the velocities. Then at finite times, it is
assumed that $f_{ij}(\mathbf{r}_{1},\mathbf{v}_{1},\mathbf{r}_{1}- \boldsymbol {\sigma }_{ij},\mathbf{v}_{2};t)$
becomes a functional of $f_{i}$
\begin{equation}
f_{ij}(\mathbf{r}_{1},\mathbf{v}_{1},\mathbf{r}_{1}-\boldsymbol
{\sigma
}_{ij},\mathbf{v}_{2};t)=F_{ij}(\mathbf{r}_{1},\mathbf{v}_{1},\mathbf{r}
_{1}-\boldsymbol {\sigma }_{ij},\mathbf{v}_{2};t\mid f_{i}(t)).
\label{a.1.2}
\end{equation}
The Enskog approximation corresponds to evaluating this functional
at $t=0$
\begin{equation}
f_{ij}(\mathbf{r}_{1},\mathbf{v}_{1},\mathbf{r}_{1}-\boldsymbol
{\sigma }_{ij},\mathbf{v}_{2};t)\rightarrow
F_{ij}(\mathbf{r}_{1},\mathbf{v} _{1},\mathbf{r}_{1}-\boldsymbol
{\sigma }_{ij},\mathbf{v}_{2};t=0\mid f_{i}(t)). \label{a.1.3}
\end{equation}
Thus for the special class of initial conditions the Enskog approximation is asymptotically exact at short
times, and assumes that the generator for dynamics at later times is the same as that initially. This idea
provides a simple mean field theory for particles with continuous potentials of interaction, but is more
realistic for hard spheres where there is instantaneous momentum transport at the initial time. The presence of
inherent velocity correlations for granular fluids suggests that the form (\ref{a.1.1}) is less justified than
in the ordinary fluid case. However, it is noted that velocity correlations are present for any nonequilibrium
state even with elastic collisions and it is known that the Enskog equation still provides a good approximation
in these latter cases.

An important exact boundary condition for hard spheres is given by
\cite {Lutsko01}
\begin{equation}
\Theta (\widehat{\boldsymbol {\sigma }}\cdot
\mathbf{g}_{12})f_{ij}(\mathbf{r}
_{1},\mathbf{v}_{1},\mathbf{r}_{1}-\boldsymbol {\sigma
}_{ij},\mathbf{v} _{2};t)=\alpha _{ij}^{-2}b_{ij}^{-1}\Theta
(-\widehat{\boldsymbol {\sigma }}\cdot
\mathbf{g}_{12})f_{ij}(\mathbf{r}_{1},\mathbf{v}_{1},\mathbf{r}_{1}-
\mathbf{\sigma }_{ij},\mathbf{v}_{2};t).  \label{a.2}
\end{equation}
This equation implies that the distribution of particles that have collided is the same as those about to
collide, but with their velocities changed according to the collision rules. In general the two particle
distribution function can be written as
\begin{eqnarray}
f_{ij}(\mathbf{r}_{1},\mathbf{v}_{1},\mathbf{r}_{1}-\boldsymbol
{\sigma }_{ij},\mathbf{v}_{2};t) &=&\Theta (-\widehat{\boldsymbol
{\sigma }}\cdot \mathbf{g}
_{12})f_{ij}(\mathbf{r}_{1},\mathbf{v}_{1},\mathbf{r}_{1}-\boldsymbol {\sigma }_{ij}
,\mathbf{v}_{2};t)  \notag \\
&&+\Theta (\widehat{\boldsymbol {\sigma}}\cdot
\mathbf{g}_{12})f_{ij}(
\mathbf{r}_{1},\mathbf{v}_{1},\mathbf{r}_{1}-\boldsymbol {\sigma
}_{ij},\mathbf{v} _{2};t).  \label{a.3}
\end{eqnarray}
If the Enskog approximation (\ref{a.1.3}) is introduced in the
first term on the right side of (\ref{a.3}), then the
corresponding approximation on the right side of Eq.\ (\ref{a.3})
gives the approximate two particle distribution function at
contact as
\begin{eqnarray}
f_{ij}(\mathbf{r}_{1},\mathbf{v}_{1},\mathbf{r}_{1}-\boldsymbol
{\sigma }_{ij},\mathbf{v}_{2};t) &\rightarrow &\Theta
(-\widehat{\boldsymbol {\sigma }}\cdot \mathbf{g}_{12})\chi
_{ij}\left( \mathbf{r}_{1},\mathbf{r}_{1}-\boldsymbol {\sigma
}_{ij}\mid \left\{ n_{k}\right\} \right)
f_{i}(\mathbf{r}_{1},\mathbf{
v}_{1};t)f_{j}(\mathbf{r}_{1}-\boldsymbol {\sigma
}_{ij},\mathbf{v}_{2};t) \nonumber\\
 &+&\alpha
_{ij}^{-2}b_{ij}^{-1}\Theta (-\widehat{\boldsymbol {\sigma}}\cdot
\mathbf{g}_{12})\chi _{ij}\left(
\mathbf{r}_{1},\mathbf{r}_{1}-\boldsymbol {\sigma }_{ij}\mid
\left\{ n_{k}\right\} \right) f_{i}(\mathbf{r}_{1},\mathbf{
v}_{1};t)\nonumber\\
& & \times f_{j}(\mathbf{r}_{1}-\boldsymbol
{\sigma}_{ij},\mathbf{v}_{2};t)
\nonumber\\
&=&\Theta (-\widehat{\boldsymbol {\sigma}}\cdot
\mathbf{g}_{12})\chi _{ij}\left(
\mathbf{r}_{1},\mathbf{r}_{1}-\boldsymbol {\sigma }_{ij}\mid
\left\{ n_{k}\right\} \right)
f_{i}(\mathbf{r}_{1},\mathbf{v}_{1};t)f_{j}(\mathbf{r}
_{1}-\boldsymbol {\sigma }_{ij},\mathbf{v}_{2};t)  \nonumber \\
&&+\alpha _{ij}^{-2}\Theta (\widehat{\boldsymbol {\sigma }}\cdot
\mathbf{g} _{12})\chi _{ij}\left(
\mathbf{r}_{1},\mathbf{r}_{1}-\boldsymbol {\sigma }_{ij}\mid
\left\{ n_{k}\right\} \right) f_{i}(\mathbf{r}_{1},\mathbf{v}
_{1}^{\prime \prime
};t)\nonumber\\
& & \times f_{j}(\mathbf{r}_{1}-\boldsymbol {\sigma
}_{ij},\mathbf{v} _{2}^{\prime \prime };t).  \label{a.5}
\end{eqnarray}
Since $\mathbf{v}_{1}^{\prime \prime }$ and
$\mathbf{v}_{2}^{\prime \prime }$ are functions of both
$\mathbf{v}_{1}$ and $\mathbf{v}_{2}$ there are velocity
correlations on the complementary hemisphere, even when they are
neglected on the precollision hemisphere.

An important consequence of (\ref{a.5}) is the relationship of $\chi_{ij}\left(
\mathbf{r}_{1},\mathbf{r}_{1}-\boldsymbol {\sigma }_{ij}\mid \left\{ n_{i}\right\} \right) $ to the pair
correlation function $g_{ij}\left( \mathbf{r}_{1},\mathbf{r}_{1}-\boldsymbol {\sigma}_{ij}\mid \left\{
n_{k}\right\} \right)$ defined by
\begin{equation}
n_{i}\left( \mathbf{r}_{1}\right) n_{j}\left(
\mathbf{r}_{1}-\boldsymbol {\sigma }_{ij}\right) g_{ij}\left(
\mathbf{r}_{1},\mathbf{r}_{1}-\boldsymbol {\sigma}_{ij}\mid
\left\{ n_{k}\right\} \right) =\int d\mathbf{v}_{1}\int
d\mathbf{v}
_{2}\;f_{ij}(\mathbf{r}_{1},\mathbf{v}_{1},\mathbf{r}_{1}-\boldsymbol
{\sigma}_{ij},\mathbf{v}_{2};t)  \label{a.6}
\end{equation}
Use of the approximation (\ref{a.5}) gives the result
\begin{eqnarray}
n_{i}\left( \mathbf{r}_{1}\right) n_{j}\left(
\mathbf{r}_{1}-\boldsymbol {\sigma}_{ij}\right) g_{ij}\left(
\mathbf{r}_{1},\mathbf{r}_{1}-\boldsymbol {\sigma}_{ij}\mid
\left\{ n_{k}\right\} \right) &=&\frac{1+\alpha _{ij}}{\alpha
_{ij}}\chi _{ij}\left( \mathbf{r}_{1}, \mathbf{r}_{1}-\boldsymbol
{\sigma}_{ij}\mid \left\{ n_{k}\right\} \right)\int d \mathbf{v}_{1}\int d\mathbf{v}_{2}\nonumber\\
& & \times \Theta (-\widehat{\boldsymbol {\sigma}}\cdot \mathbf{g
}_{12})f_{i}(\mathbf{r}_{1},\mathbf{v}_{1};t)f_{j}(\mathbf{r}_{1}-\boldsymbol
{\sigma}_{ij},\mathbf{v}_{2};t), \nonumber\\ \label{a.6a}
\end{eqnarray}
where a change of variables has been made in the integration of
the second term in (\ref{a.5})
\begin{equation}
\int d\mathbf{v}_{1}\int \;d\mathbf{v}_{2}X(\mathbf{v}_{1}^{\prime
\prime },\mathbf{ v}_{2}^{\prime \prime })=\alpha _{ij}\int
d\mathbf{v}_{1}^{\prime \prime }\int d \mathbf{v}_{2}^{\prime
\prime }\;X(\mathbf{v}_{1}^{\prime \prime },\mathbf{v}
_{2}^{\prime \prime }).  \label{a.8}
\end{equation}
For a uniform system, $g_{ij}\left(
\mathbf{r}_{1},\mathbf{r}_{2}\mid \left\{ n_{k}\right\} \right)
\rightarrow g_{ij}\left( \left| \mathbf{r}
_{1}-\mathbf{r}_{2}\right| ;\left\{ n_{k}\right\} \right)$ and
this expression reduces to
\begin{equation}
g_{ij}\left( \sigma _{ij};\left\{ n_{k}\right\} \right)
=\frac{1+\alpha _{ij}}{2\alpha _{ij}}\chi _{ij}\left( \sigma
_{ij};\left\{ n_{k}\right\} \right)  \label{a.7}
\end{equation}
Equation (\ref{a.7}) is the result quoted in the text and provides
the interpretation for $\chi _{ij}\left(
\mathbf{r}_{1},\mathbf{r}_{1}-\boldsymbol {\sigma }_{ij}\mid
\left\{ n_{k}\right\} \right)$ . For elastic collisions  $\chi
_{ij}\left( \sigma _{ij};\left\{ n_{k}\right\}\right)$ is indeed
the pair correlation function at contact. The Enskog theory in
that case takes $\chi _{ij}\left(
\mathbf{r}_{1},\mathbf{r}_{1}-\boldsymbol {\sigma }_{ij}\mid
\left\{ n_{k}\right\} \right) $ to be the pair correlation
function for an equilibrium nonuniform fluid whose densities are
equal to those for the actual nonequilibrium state being
considered. This assumption is based on the fact that structural
correlations for hard spheres are entirely due to excluded volume
effects which should be similar for equilibrium and nonequilibrium
states. It is reasonable to extend this choice for $\chi
_{ij}\left( \mathbf{r}_{1},\mathbf{r}_{1}-\boldsymbol {\sigma
}_{ij}\mid \left\{ n_{k}\right\} \right)$ to granular fluids  as
well. Its accuracy can be judged by measuring (via MD simulation)
the pair correlation given by (\ref{a.7}) with this choice on the
right side. This has been done for the one component fluid,
indicating reasonable results over a range of values for the
restitution coefficient \cite{Lutsko01}.

\section{Balance Equations and Fluxes}
\label{appB}

The macroscopic balance equations follow from the definitions (\ref{3.1})--(\ref{3.3}) and the first hierarchy
equation (\ref{2.1})
\begin{equation}
\partial _{t}n_{i}+\mathbf{\nabla }_{\mathbf{r}_{1}}\cdot \int d\mathbf{v}
_{1}\mathbf{v}_{1}f_{i}=\int d\mathbf{v}_{1}C_{i},  \label{b.1}
\end{equation}
\begin{equation}
\partial _{t}e+\mathbf{\nabla }_{\mathbf{r}_{1}}\cdot \sum_{i=1}^{s}\int d
\mathbf{v}_{1}\frac{1}{2}m_{i}v_{1}^{2}\mathbf{v}_{1}f_{i}-\sum_{i=1}^{s}
\mathbf{F}_{i}\cdot \int
d\mathbf{v}_{1}\mathbf{v}_{1}f_{i}=\sum_{i=1}^{s} \int
d\mathbf{v}_{1}\frac{1}{2}m_{i}v_{1}^{2}C_{i},  \label{b.2}
\end{equation}
\begin{equation}
\partial _{t}p_{\beta }+\partial _{r_{1\gamma }}\sum_{i=1}^{s}\int d\mathbf{v}
_{1}m_{i}v_{1\gamma }v_{1\beta }f_{i}-\sum_{i=1}^{s}n_{i}F_{i\beta
}=\sum_{i=1}^{s}\int d\mathbf{v}_{1}m_{i}v_{1\beta }C_{i}.
\label{b.3}
\end{equation}
The integrals over the collisional contribution $C_{i}$ are
analyzed below with the results
\begin{equation}
\int d\mathbf{v}_{1}C_{i}=0,  \label{b.4}
\end{equation}
\begin{equation}
\sum_{i=1}^{s}\int
d\mathbf{v}_{1}\frac{1}{2}m_{i}v_{1}^{2}C_{i}=-\nabla \cdot
\mathbf{s}^{c}-w,  \label{b.5}
\end{equation}
\begin{equation}
\sum_{i=1}^{s}\int d\mathbf{v}_{1}m_{i}v_{1\beta }C_{i}=-\partial
_{r_\gamma}t_{\gamma \beta }^{c}.  \label{b.6}
\end{equation}
Use of these expressions in (\ref{b.1})--(\ref{b.3}) gives the
balance equations (\ref{3.4})--(\ref{3.6}) of the text with
\begin{equation}
\mathbf{j}_{i}=m_i\;\int \;d\mathbf{v}_{1}\mathbf{v}_{1}f_{i}, \label{b.7}
\end{equation}
\begin{equation}
\mathbf{s}=\sum_{i=1}^{s}\int
d\mathbf{v}_{1}\frac{1}{2}m_{i}v_{1}^{2}
\mathbf{v}_{1}f_{i}+\mathbf{s}^{c},  \label{b.8}
\end{equation}
\begin{equation}
t_{\gamma \beta }=\sum_{i=1}^{s}\int
d\mathbf{v}_{1}m_{i}v_{1\gamma }v_{1\beta }f_{i}+t_{\gamma \beta
}^{c}.  \label{b.9}
\end{equation}

The terms $w$, $\mathbf{s}^{c}$, and $t_{\gamma \beta }^{c}$
arising from the collisional contribution $C_{i}$ are identified
by further analysis of the left sides of (\ref{b.5}) and
(\ref{b.6}). To do so consider the general expression for some
arbitrary function $\psi_{i}\left( \mathbf{v}_{1}\right)$
\begin{eqnarray}
\int d\mathbf{v}_{1}\psi _{i}C_{i} &=&\sum_{j=1}^{s}\sigma
_{ij}^{d-1}\int d \mathbf{v}_{1}\int d\mathbf{v}_{2}\int
d\widehat{\boldsymbol {\sigma}}\,\Theta ( \widehat{\boldsymbol
{\sigma }}\cdot \mathbf{g}_{12})(\widehat{\boldsymbol {\sigma}}
\cdot
\mathbf{g}_{12})\psi _{i}\left( \mathbf{v}_{1}\right)  \nonumber\\
&&\times \left[ \alpha _{ij}^{-2}f_{ij}(\mathbf{r}_{1},\mathbf{v}_{1}^{\prime \prime
},\mathbf{r}_{1}-\boldsymbol {\sigma }_{ij},\mathbf{v}_{2}^{\prime \prime
};t)-f_{ij}(\mathbf{r}_{1},\mathbf{v}_{1},\mathbf{r}_{1}+\boldsymbol {\sigma}_{ij},\mathbf{v}_{2};t)\right].
\label{b.10}
\end{eqnarray}
The restituting velocities are functions of the given velocities, $\mathbf{v} _{1}^{\prime \prime
}=\mathbf{v}_{1}^{\prime \prime }(\mathbf{v}_{1},\mathbf{ v}_{2}),$ $\mathbf{v}_{2}^{\prime \prime
}=\mathbf{v}_{2}^{\prime \prime }( \mathbf{v}_{1},\mathbf{v}_{2})$, defined by (\ref{2.2}). These relations can
be inverted to get
\begin{equation}
\mathbf{v}_{1}=\mathbf{v}_{1}^{\prime \prime }-\mu _{ji}\left(
1+\alpha _{ij}\right) (\widehat{\boldsymbol {\sigma}}\cdot
\mathbf{g}_{12}^{\prime \prime})\widehat{\boldsymbol {\sigma
}},\hspace{0.3in}\mathbf{v}_{2}= \mathbf{v}_{2}^{\prime \prime
}+\mu _{ij}\left( 1+\alpha _{ij}\right) (\widehat{\boldsymbol
{\sigma}}\cdot \mathbf{g}_{12}^{\prime \prime
})\widehat{\boldsymbol {\sigma }}.  \label{b.11}
\end{equation}
Therefore, in the first term of (\ref{b.10}) it is possible to
change integration variables from $d\mathbf{v}_{1}d\mathbf{v}_{2}$
to $d\mathbf{v} _{1}^{\prime \prime }d\mathbf{v}_{2}^{\prime
\prime }$, with a Jacobian $\alpha _{ij}$ to get
\begin{eqnarray}
& &\int d\mathbf{v}_{1}\int d\mathbf{v}_{2}\int
d\widehat{\boldsymbol {\sigma }}\,\Theta ( \widehat{\boldsymbol
{\sigma }}\cdot \mathbf{g}_{12})(\widehat{\boldsymbol {\sigma}}
\cdot \mathbf{g}_{12})\psi_{i}\left( \mathbf{v}_{1}\right) \alpha
_{ij}^{-2}f_{ij}(\mathbf{r}_{1},\mathbf{v}_{1}^{\prime \prime },
\mathbf{r}_{1}-\boldsymbol {\sigma
}_{ij},\mathbf{v}_{2}^{\prime \prime };t)\nonumber\\
&=&\int d\mathbf{v}_{1}^{\prime \prime }\int
d\mathbf{v}_{2}^{\prime \prime }\int d \widehat{\boldsymbol
{\sigma }}\,\Theta (-\widehat{\boldsymbol {\sigma }}\cdot
\mathbf{g} _{12}^{\prime \prime })(-\widehat{\boldsymbol {\sigma
}}\cdot \mathbf{g} _{12}^{\prime \prime })\psi _{i}\left(
\mathbf{v}_{1}\left( \mathbf{v} _{1}^{\prime \prime
},\mathbf{v}_{2}^{\prime \prime }\right) \right)
f_{ij}(\mathbf{r}_{1},\mathbf{v}_{1}^{\prime \prime
},\mathbf{r}_{1}- \boldsymbol {\sigma
}_{ij},\mathbf{v}_{2}^{\prime \prime };t)\nonumber\\
&=&\int d\mathbf{v}_{1}\int d\mathbf{v}_{2}\int
d\widehat{\boldsymbol {\sigma }}\,\Theta ( \widehat{\boldsymbol
{\sigma }}\cdot \mathbf{g}_{12})(\widehat{\boldsymbol {\sigma}}
\cdot \mathbf{g}_{12})\psi _{i}\left( \mathbf{v}_{1}^{\prime
}\left( \mathbf{ v}_{1},\mathbf{v}_{2}\right) \right)
f_{ij}(\mathbf{r}_{1},\mathbf{v} _{1},\mathbf{r}_{1}+\boldsymbol
{\sigma}_{ij},\mathbf{v}_{2};t), \label{b.12}
\end{eqnarray}
where use has been made of $(\widehat{\boldsymbol {\sigma }}\cdot
\mathbf{g} _{12})=-\alpha _{ij}(\widehat{\boldsymbol {\sigma
}}\cdot \mathbf{g}_{12}^{\prime \prime })$. In the last line the
dummy variables $\left( \mathbf{v} _{1}^{\prime \prime
},\mathbf{v}_{2}^{\prime \prime }\right) $ have been relabelled
$\left( \mathbf{v}_{1},\mathbf{v}_{2}\right) $, and a change of
integration from $\widehat{\boldsymbol {\sigma}}$ to
$-\widehat{\boldsymbol {\sigma}}$ has been performed. Accordingly
$\mathbf{v}_{1}\left( \mathbf{v}_{1}^{\prime \prime },
\mathbf{v}_{2}^{\prime \prime }\right) $ has been relabelled
$\mathbf{v} _{1}^{\prime }\left(
\mathbf{v}_{1},\mathbf{v}_{2}\right)$ with (\ref{b.11}) becoming
in this notation
\begin{equation}
\mathbf{v}_{1}^{\prime}=\mathbf{v}_{1}-\mu _{ji}\left(1+\alpha
_{ij}\right) (\widehat{\boldsymbol {\sigma }}\cdot
\mathbf{g}_{12})\widehat{\boldsymbol {\sigma
}},\hspace{0.3in}\mathbf{v}_{2}^{\prime }=\mathbf{v}_{2}+\mu
_{ij}\left( 1+\alpha _{ij}\right) (\widehat{\boldsymbol {\sigma
}}\cdot \mathbf{g}_{12})\widehat{\boldsymbol {\sigma}}.
\label{b.13}
\end{equation}
This is the direct scattering law, which differs from the
restituting scattering law (\ref{2.2}) for $\alpha_{ij} \neq 1$.
With this transformation the integral (\ref{b.10}) is
\begin{eqnarray}
\int d\mathbf{v}_{1}\psi _{i}C_{i} &=&\sum_{j=1}^{s}\sigma
_{ij}^{d-1}\int d \mathbf{v}_{1}\int d\mathbf{v}_{2}\int
d\widehat{\boldsymbol {\sigma }}\,\Theta ( \widehat{\boldsymbol
{\sigma }}\cdot \mathbf{g}_{12})(\widehat{\boldsymbol {\sigma}}
\cdot \mathbf{g}_{12})  \notag \\
&&\times \left( \psi _{i}\left( \mathbf{v}_{1}^{\prime }\right)
-\psi _{i}\left( \mathbf{v}_{1}\right) \right)
f_{ij}(\mathbf{r}_{1},\mathbf{ v}_{1},\mathbf{r}_{1}+\boldsymbol
{\sigma}_{ij},\mathbf{v}_{2};t). \label{b.14}
\end{eqnarray}
The special choice $\psi _{i}\left( \mathbf{v}_{1}\right) =1$
proves (\ref {b.4}) above.

Next, consider the sum of (\ref{b.14}) over all species
\begin{eqnarray}
\sum_{i=1}^{s}\int d\mathbf{v}_{1}\psi _{i}C_{i}
&=&\sum_{i,j=1}^{s}\sigma _{ij}^{d-1}\int d\mathbf{v}_{1}\int
d\mathbf{v}_{2}\int d\widehat{\boldsymbol {\sigma}}\, \Theta
(\widehat{\boldsymbol {\sigma}}\cdot \mathbf{g}_{12})
(\widehat{\boldsymbol {\sigma}}\cdot \mathbf{g}_{12}) \nonumber\\
& & \times\left[ \psi _{i}\left( \mathbf{v}_{1}^{\prime }\right)
-\psi _{i}\left( \mathbf{v}_{1}\right) \right]
f_{ij}^{(2)}(\mathbf{r}_{1},\mathbf{
v}_{1},\mathbf{r}_{1}+\boldsymbol {\sigma}_{ij},\mathbf{v}_{2};t)
\nonumber\\
&=&\frac{1}{2}\sum_{i,j=1}^{k}\sigma _{ij}^{d-1}\int
d\mathbf{v}_{1}\int d\mathbf{v} _{2}\int d \widehat{\boldsymbol
{\sigma}}\;\Theta (\widehat{\boldsymbol {\sigma}}\cdot
\mathbf{g}_{12})(\widehat{\boldsymbol {\sigma}}\cdot
\mathbf{g}_{12})  \nonumber\\
&&\times \left\{ \left[ \psi _{i}\left( \mathbf{v}_{1}^{\prime
}\right) -\psi _{i}\left( \mathbf{v}_{1}\right) \right]
f_{ij}(\mathbf{r}_{1},\mathbf{ v}_{1},\mathbf{r}_{1}+\boldsymbol
{\sigma}_{ij},\mathbf{v}_{2};t)\right.\nonumber\\
& & \left.+\left[ \psi _{j}\left( \mathbf{v}_{2}^{\prime }\right)
-\psi _{j}\left( \mathbf{v} _{2}\right) \right]
f_{ji}(\mathbf{r}_{1},\mathbf{v}_{2},\mathbf{r} _{1}-\boldsymbol
{\sigma}_{ij},\mathbf{v}_{1};t)\right\} . \label{b.15}
\end{eqnarray}
The second equality is obtained from the first by taking half the
sum of the first plus an equivalent form obtained by interchanging
$\mathbf{v}_{1}$ and $\mathbf{v}_{2}$, interchanging $i$ and $j$,
and changing $\widehat{\boldsymbol {\sigma}}$ to
$-\widehat{\boldsymbol {\sigma}}$. To simplify this further, note
the relation $
f_{ji}(\mathbf{r}_{1},\mathbf{v}_{2},\mathbf{r}_{1}-\boldsymbol
{\sigma}_{ij}, \mathbf{v}_{1};t)=f_{ij}(\mathbf{r}_{1}-\boldsymbol
{\sigma}_{ij},\mathbf{v} _{1},\mathbf{r}_{1},\mathbf{v}_{2};t)$
and arrange terms as
\begin{eqnarray}
\sum_{i=1}^{s}\int d\mathbf{v}_{1}\psi _{i}C_{i} &=&\frac{1}{2}
\sum_{i,j=1}^{s}\sigma _{ij}^{d-1}\int d\mathbf{v}_{1}\int
d\mathbf{v}_{2}\int d \widehat{\boldsymbol {\sigma}}\,\Theta
(\widehat{\boldsymbol {\sigma}}\cdot \mathbf{g}
_{12})(\widehat{\boldsymbol
{\sigma}}\cdot \mathbf{g}_{12})  \notag \\
&\times& \left\{ \left[ \psi _{i}\left( \mathbf{v}_{1}^{\prime
}\right) +\psi _{j}\left( \mathbf{v}_{2}^{\prime }\right) -\psi
_{i}\left( \mathbf{v} _{1}\right) -\psi _{j}\left(
\mathbf{v}_{2}\right) \right] f_{ij}(
\mathbf{r}_{1},\mathbf{v}_{1},\mathbf{r}_{1}+\boldsymbol
{\sigma}_{ij},\mathbf{v}
_{2};t)\right.  \notag \\
&+&\left.\left[ \psi _{i}\left( \mathbf{v}_{1}^{\prime }\right)
-\psi _{i}\left( \mathbf{v}_{1}\right) \right] \left[
f_{ij}(\mathbf{r}_{1},\mathbf{v} _{1},\mathbf{r}_{1}+\boldsymbol
{\sigma}_{ij},\mathbf{v}_{2};t)-f_{ij}(\mathbf{r }_{1}-\boldsymbol
{\sigma}_{ij},\mathbf{v}_{1},\mathbf{r}_{1},\mathbf{v}_{2};t)\right]\right\}.\nonumber\\
\label{b.16}
\end{eqnarray}
The first term of the integrand represents a collisional effect
due to scattering with a change in the velocities. The second term
is a collisional effect due to the spatial difference of the
colliding pair. This second effect is called ``collisional
transfer''. It can be written as a divergence through the identity
\begin{eqnarray}
f_{ij}(\mathbf{r}_{1},\mathbf{v}_{1},\mathbf{r} _{1}+\boldsymbol
{\sigma}_{ij},\mathbf{v}_{2};t)&-&
f_{ij}(\mathbf{r}_{1}-\boldsymbol
{\sigma}_{ij},\mathbf{v}_{1},\mathbf{r}_{1},
\mathbf{v}_{2};t)\nonumber\\
& =& \int_{0}^{1}dx\frac{\partial}{\partial
x}f_{ij}(\mathbf{r}_{1}-x\boldsymbol {\sigma}_{ij},
\mathbf{v}_{1},\mathbf{r}_{1}+\left( 1-x\right) \boldsymbol
{\sigma}_{ij},\mathbf{v} _{2};t)\nonumber\\
&=&\nabla _{\mathbf{r}_{1}}\cdot \boldsymbol {\sigma}_{ij}
\int_{0}^{1}dx\;f_{ij}(\mathbf{r}_{1}-x\boldsymbol
{\sigma}_{ij},\mathbf{v}_{1}, \mathbf{r}_{1}+\left( 1-x\right)
\boldsymbol {\sigma}_{ij},\mathbf{v}_{2};t).\nonumber\\
\label{b.17}
\end{eqnarray}
Using the identity (\ref{b.17}), Eq.\ (\ref{b.16}) can be finally
written as
\begin{eqnarray}
\sum_{i=1}^{s}\int d\mathbf{v}_{1}\psi _{i}C_{i} &=&\frac{1}{2}
\sum_{i,j=1}^{s}\sigma _{ij}^{d-1}\int d\mathbf{v}_{1}\int
d\mathbf{v}_{2}\int d \widehat{\boldsymbol {\sigma}}\,\Theta
(\widehat{\boldsymbol {\sigma}}\cdot \mathbf{g}
_{12})(\widehat{\boldsymbol
{\sigma}}\cdot \mathbf{g}_{12})  \notag \\
&\times& \left\{ \left[ \psi _{i}\left( \mathbf{v}_{1}^{\prime
}\right) +\psi _{j}\left( \mathbf{v}_{2}^{\prime }\right) -\psi
_{i}\left( \mathbf{v} _{1}\right) -\psi _{j}\left(
\mathbf{v}_{2}\right) \right] f_{ij}(
\mathbf{r}_{1},\mathbf{v}_{1},\mathbf{r}_{1}+\boldsymbol
{\sigma}_{ij},\mathbf{v}
_{2};t)\right.  \notag \\
&+&\left.\nabla _{\mathbf{r}_{1}}\cdot \boldsymbol {\sigma}_{ij}\left[ \psi _{i}\left( \mathbf{v}_{1}^{\prime
}\right) -\psi _{i}\left( \mathbf{v}_{1}\right) \right] \int_{0}^{1}dx\;f_{ij}(\mathbf{r}_{1}-x\boldsymbol
{\sigma}_{ij},\mathbf{v}_{1}, \mathbf{r}_{1}+\left( 1-x\right) \boldsymbol
{\sigma}_{ij},\mathbf{v}_{2};t)\right\}.\nonumber\\ \label{b.17.1}
\end{eqnarray}

Now, apply this result to the case $\psi
_{i}=m_{i}\mathbf{v}_{1}$. Since the total momentum is conserved
in all pair collisions, $\psi _{i}\left( \mathbf{v}_{1}^{\prime
}\right) +\psi _{j}\left( \mathbf{v}_{2}^{\prime }\right) -\psi
_{i}\left( \mathbf{v}_{1}\right) -\psi _{j}\left( \mathbf{v}
_{2}\right) =0$ for this case and (\ref{b.17.1}) gives (\ref{b.6})
with
\begin{eqnarray}
t_{\gamma \beta }^{c} &\equiv &-\frac{1}{2}\sum_{i,j=1}^{s}\sigma
_{ij}^{d}\int d\mathbf{v}_{1}\int d\mathbf{v}_{2}\int
d\widehat{\boldsymbol {\sigma}} \,\Theta (\widehat{\boldsymbol
{\sigma}}\cdot \mathbf{g}_{12})(\widehat{\boldsymbol
{\sigma}}\cdot \mathbf{g}_{12})\widehat{\sigma }_{\gamma} \notag
\\
&&\times \left( m_{i}v_{1\beta}^{\prime}-m_{i}v_{1\beta }\right)
\int_{0}^{1}dx\,f_{ij}(\mathbf{r}_{1}-x\boldsymbol
{\sigma}_{ij},\mathbf{v}_{1}, \mathbf{r}_{1}+\left( 1-x\right)
\boldsymbol {\sigma}_{ij},\mathbf{v}_{2};t)  \notag
\\
&=&\frac{1}{2}\sum_{i,j=1}^{s}m_{i}\mu _{ji}\left( 1+\alpha _{ij}\right) \sigma _{ij}^{d}\int
d\mathbf{v}_{1}\int d\mathbf{v}_{2}\int d\widehat{\boldsymbol {\sigma}} \,\Theta (\widehat{\boldsymbol
{\sigma}}\cdot \mathbf{g}_{12})(\widehat{\boldsymbol {\sigma} }\cdot
\mathbf{g}_{12})^{2}\widehat{\sigma}_{\gamma }
\widehat{\sigma}_{\beta}  \notag \\
&&\times \int_{0}^{1}dx\;f_{ij}(\mathbf{r}_{1}-x\boldsymbol
{\sigma}_{ij},\mathbf{ v}_{1},\mathbf{r}_{1}+\left( 1-x\right)
\boldsymbol {\sigma}_{ij},\mathbf{v}_{2};t). \label{b.18}
\end{eqnarray}
The analysis leading to (\ref{b.6}) follows from (\ref{b.16}) in a
similar way with $\psi _{i}=m_{i}v_{1}^{2}/2$. However, since
energy is not conserved in pair collisions the first term on the
right side does not vanish. Instead, it represents the collisions
energy loss $w$
\begin{eqnarray}
w &=&-\sum_{i,j=1}^{s}\frac{1}{4}\sigma _{ij}^{d-1}\int
d\mathbf{v}_{1}\int d \mathbf{v}_{2}\int d\widehat{\boldsymbol
{\sigma}}\;\Theta (\widehat{\boldsymbol {\sigma}} \cdot
\mathbf{g}_{12})(\widehat{\boldsymbol {\sigma}}\cdot \mathbf{g}
_{12})  \notag \\
&&\left( m_{i}v_{1}^{\prime 2}+m_{j}v_{2}^{\prime
2}-m_{i}v_{1}^{2}-m_{j}v_{2}^{2}\right)
f_{ij}(\mathbf{r}_{1},\mathbf{v
}_{1},\mathbf{r}_{1}+\boldsymbol {\sigma}_{ij},\mathbf{v}_{2};t)  \notag \\
&=&\frac{1}{4}\sum_{i,j=1}^{s}\left( 1-\alpha _{ij}^{2}\right)
m_{i}\mu _{ji}\sigma _{ij}^{d-1}\int d\mathbf{v}_{1}\int
d\mathbf{v}_{2}\int d\widehat{\boldsymbol {\sigma}}\; \Theta
(\widehat{\boldsymbol {\sigma}}\cdot \mathbf{g}_{12})(\widehat{
\boldsymbol {\sigma}}\cdot \mathbf{g}_{12})^{3}  \notag \\
&&\times
f_{ij}(\mathbf{r}_{1},\mathbf{v}_{1},\mathbf{r}_{1}+\boldsymbol
{\sigma}_{ij},\mathbf{v}_{2};t).  \label{b.19}
\end{eqnarray}
The second term on the right side of (\ref{b.17.1}) gives the
collisional transfer contribution to the flux
\begin{eqnarray}
\nabla \cdot \mathbf{s}^{c} &=&-\nabla _{\mathbf{r}_{1}}\cdot
\sum_{i,j=1}^{s} \frac{1}{4}m_{i}\sigma _{ij}^{d}\int
d\mathbf{v}_{1}\int d\mathbf{v}_{2}\int d \widehat{\boldsymbol
{\sigma}}\,\Theta (\widehat{\boldsymbol {\sigma}}\cdot \mathbf{g}
_{12})(\widehat{\boldsymbol {\sigma}}\cdot
\mathbf{g}_{12})\widehat{\boldsymbol {\sigma}}  \notag \\
&&\left( v_{1}^{\prime 2}-v_{1}^{2}\right) \int_{0}^{1}dx\;f_{ij}(
\mathbf{r}_{1}-x\boldsymbol
{\sigma}_{ij},\mathbf{v}_{1},\mathbf{r}_{1}+\left(
1-x\right) \boldsymbol {\sigma}_{ij},\mathbf{v}_{2};t)  \notag \\
&=&\nabla _{\mathbf{r}_{1}}\cdot \sum_{i,j=1}^{s}\frac{1}{4}\left(
1+\alpha _{ij}\right) m_{i}\mu _{ji}\sigma _{ij}^{d}\int
d\mathbf{v}_{1}\int d\mathbf{v}_{2}\int d\widehat{\boldsymbol
{\sigma}}\;\Theta (\widehat{\boldsymbol {\sigma}}\cdot
\mathbf{g}_{12})(\widehat{\boldsymbol {\sigma}}\cdot
\mathbf{g}_{12})^{2} \widehat{\boldsymbol {\sigma}} \nonumber\\
&&\times \left[\mu_{ji} \left( 1-\alpha _{ij}\right)
(\widehat{\boldsymbol {\sigma}}\cdot \mathbf{g}_{12})+2
\widehat{\boldsymbol {\sigma}}\cdot \left( \mu
_{ij}\mathbf{v}_{1}+\mu _{ji}\mathbf{v}
_{2}\right) \right] \notag \\
&&\times \int_{0}^{1}dx\;f_{ij}(\mathbf{r}_{1}-x\boldsymbol
{\sigma}_{ij},\mathbf{v}_{1},\mathbf{r}_{1}+\left( 1-x\right)
\boldsymbol {\sigma}_{ij},\mathbf{v}_{2};t). \label{b.20}
\end{eqnarray}
This confirms (\ref{b.5}) and identifies $\mathbf{s}^{c}$, which
has the equivalent form (obtained by taking half the sum of forms
with $i$ and $j$ interchanged)
\begin{eqnarray}
\mathbf{s}^{c} &=&\sum_{i,j=1}^{s}\frac{1}{8}\left( 1+\alpha
_{ij}\right) m_{i}\mu _{ji}\sigma _{ij}^{d}\int
d\mathbf{v}_{1}\int d\mathbf{v}_{2}\int d \widehat{\boldsymbol
{\sigma}}\,\Theta (\widehat{\boldsymbol {\sigma}}\cdot \mathbf{g}
_{12})(\widehat{\boldsymbol {\sigma}}\cdot \mathbf{g}_{12})^{2}  \notag \\
&&\times \widehat{\boldsymbol {\sigma}}\left[ \left( 1-\alpha
_{ij}\right) \left( \mu _{ji}-\mu
_{ij}\right)(\widehat{\boldsymbol {\sigma}}\cdot
\mathbf{g}_{12})+4\widehat{\boldsymbol {\sigma}}\cdot \left( \mu
_{ij}\mathbf{v}
_{1}+\mu _{ji}\mathbf{v}_{2}\right) \right]  \notag \\
&&\times \int_{0}^{1}dx\;f_{ij}(\mathbf{r}_{1}-x\boldsymbol
{\sigma}_{ij},\mathbf{v}_{1},\mathbf{r}_{1}+\left( 1-x\right)
\boldsymbol {\sigma}_{ij},\mathbf{v}_{2};t). \label{b.21}
\end{eqnarray}

\section{Chapman--Enskog Solution}
\label{appC}

As described in the text a normal solution to the kinetic equation
is a non-local functional of the hydrodynamic fields
$f_{i}(\mathbf{ v}_{1}\mid \left\{ y_{\beta }\left( t\right)
\right\} )$. This is equivalent to a function of the fields at a
point and all their derivatives at that point
\begin{equation}
f_{i}(\mathbf{r}_{1},\mathbf{v}_{1}\mid \left\{ y_{\beta }\left(
t\right) \right\} )=f_{i}(\mathbf{v}_{1};\left\{ y_{\beta }\left(
\mathbf{r} _{1},t\right) \right\} ;\left\{ \partial
_{\mathbf{r}_{1}}y_{\beta }\left( \mathbf{r}_{1},t\right)
;..\right\} ).  \label{c.1}
\end{equation}
If the gradients are small, this function can be expanded in the
appropriate dimensionless small parameter
\begin{equation}
f_{i}(\mathbf{v}_{1}\mid \left\{ y_{\beta }\left( t\right)
\right\} )=f_{i}^{(0)}(\mathbf{v}_{1};\left\{ y_{\beta }\left(
\mathbf{r} _{1},t\right) \right\}
)+f_{i}^{(1)}(\mathbf{v}_{1};\left\{ y_{\beta }\left(
\mathbf{r}_{1},t\right) \right\} ;\left\{ \partial
_{\mathbf{r}_{1}}y_{\beta }\left( \mathbf{r}_{1},t\right) \right\}
)+\cdots \label{c.2}
\end{equation}
where $f_{i}^{(0)}$ is a function of the fields alone, $f_{i}^{(1)}$ is a function of the fields and linear in
their gradients, and so on. Thus the kinetic equation can be solved perturbatively by requiring that
contributions from common order in this gradient expansion vanish.

To perform this ordering it is necessary to expand the collision
operators of (\ref{2.6})
\begin{eqnarray}
J_{ij}\left[ \mathbf{r}_{1},\mathbf{v}_{1}\mid f(t)\right] &\equiv
&\sigma _{ij}^{d-1}\int d\mathbf{v}_{2}\int d\widehat{\boldsymbol
{\sigma}}\Theta ( \widehat{\boldsymbol {\sigma}}\cdot
\mathbf{g}_{12})(\widehat{\boldsymbol {\sigma }}
\cdot \mathbf{g}_{12})  \nonumber\\
&&\times \left[ \alpha _{ij}^{-2}\chi _{ij}\left( \mathbf{r}_{1},
\mathbf{r}_{1}-{\boldsymbol{\sigma}}_{ij}\mid \left\{
n_{i}\right\} \right) f_{i}( \mathbf{r}_{1},\mathbf{v}_{1}^{\prime
\prime };t)f_{j}(\mathbf{r}_{1}-
{\boldsymbol{\sigma}}_{ij},\mathbf{v}_{2}^{\prime \prime
};t)\right.
\nonumber\\
&&\left. -\chi_{ij}\left(
\mathbf{r}_{1},\mathbf{r}_{1}+{\boldsymbol{\sigma}}_{ij} \mid
\left\{ n_{i}\right\} \right) f_{i}(\mathbf{r}_{1},\mathbf{
v}_{1};t)f_{j}(\mathbf{r}_{1}+{\boldsymbol{\sigma}}_{ij},\mathbf{v}_{2};t)\right]
.  \label{c.3}
\end{eqnarray}
For the purposes here it is sufficient to go up through first
order. The distribution functions evaluated at $\mathbf{r}_{1}\pm
\boldsymbol{\sigma}_{ij}$ become
\begin{eqnarray}
& & f_{i}(\mathbf{v}_{1};\left\{ y_{\beta }\left(
\mathbf{r}_{1}\pm {\boldsymbol{\sigma}}_{ij},t\right) \right\}
;\left\{ \partial _{\mathbf{r}_{1}}y_{\beta }\left(
\mathbf{r}_{1}\pm {\boldsymbol{\sigma}}_{ij},t\right)
;\cdots\right\} ) \rightarrow \left( 1\pm
{\boldsymbol{\sigma}}_{ij}\cdot \nabla _{\mathbf{r} _{1}}\right)
f_{i}^{(0)}(\mathbf{v}_{1};\left\{ y_{\beta }\left( \mathbf{r}
_{1},t\right) \right\} ) \nonumber\\
&&+f_{i}^{(1)}(\mathbf{v}_{1};\left\{ y_{\beta }\left( \mathbf{r}
_{1},t\right) \right\} ;\left\{
\partial _{\mathbf{r}_{1}}y_{\beta }\left( \mathbf{r}_{1},t\right)
\right\} )\nonumber\\
&=&f_{i}^{(0)}(\mathbf{v}_{1};\left\{ y_{\beta }\left( \mathbf{r}
_{1},t\right) \right\} )\pm \left( \partial _{y_{\beta
}}f_{i}^{(0)}(\mathbf{ v}_{1};\left\{ y_{\beta }\left(
\mathbf{r}_{1},t\right) \right\} )\right)
{\boldsymbol{\sigma}}_{ij}\cdot \nabla _{\mathbf{r}_{1}}y_{\beta
}\left( \mathbf{r
}_{1},t\right)  \nonumber\\
&&+f_{i}^{(1)}(\mathbf{v}_{1};\left\{ y_{\beta }\left( \mathbf{r}
_{1},t\right) \right\} ;\left\{ \partial _{\mathbf{r}_{1}}y_{\beta
}\left( \mathbf{r}_{1},t\right) \right\} ).  \label{c4}
\end{eqnarray}
The functional expansion of $\chi _{ij}\left(
\mathbf{r}_{1},\mathbf{r} _{1}\pm \mathbf{\sigma }_{ij}\mid
\left\{ n_{i}\right\} \right) $ to this order is obtained by a
functional expansion of all species densities about their values
at $\mathbf{r}_{1}$
\begin{equation}
\chi _{ij}\left( \mathbf{r}_{1},\mathbf{r}_{1}\pm \mathbf{\sigma
}_{ij}\mid \left\{ n_{i}\left( t\right) \right\} \right) =\chi
_{ij}^{(0)}\left({\sigma}_{ij}\mid \left\{ n_{k}\left(
\mathbf{r}_{1},t\right) \right\} \right)
\end{equation}
\begin{equation}
+\sum_{\ell =1}^{s}\int d\mathbf{r}^{\prime }\frac{\delta \chi
_{ij}\left( \mathbf{r}_{1},\mathbf{r}_{1}\pm
{\boldsymbol{\sigma}}_{ij}\mid \left\{ n_{i}\right\} \right)
}{\delta n_{\ell }(\mathbf{r}^{\prime },t)}\mid _{\delta
n=0}\left( n_{\ell }(\mathbf{r}^{\prime })-n_{\ell }(\mathbf{r}
_{1})\right) +..  \nonumber\\
\end{equation}
\begin{equation}
\rightarrow \chi _{ij}^{(0)}\left( \sigma _{ij};\left\{
n_{k}\left( \mathbf{r}_{1},t\right) \right\} \right) +\sum_{\ell
=1}^{s}\left( \nabla _{\mathbf{r} _{1}}n_{\ell
}(\mathbf{r}_{1};t)\right) \cdot \int d\mathbf{r}^{\prime }\left(
\mathbf{r}^{\prime }-\mathbf{r}_{1}\right) \frac{\delta \chi
_{ij}\left( \mathbf{r}_{1},\mathbf{r}_{1}\pm
{\boldsymbol{\sigma}}_{ij}\mid \left\{ n_{i}\right\} \right)
}{\delta n_{\ell }(\mathbf{r}^{\prime },t)} \mid _{\delta n=0}.
\label{c.6}
\end{equation}
The arrow denotes the leading terms of a Taylor series for $\left(
n_{\ell }( \mathbf{r}^{\prime })-n_{\ell }(\mathbf{r}_{1})\right)$
. The integral can be simplified by noting at $\delta n=0$ the
functional integral has translational invariance
\begin{equation}
\frac{\delta \chi _{ij}\left( \mathbf{r}_{1},\mathbf{r}_{1}\pm
{\boldsymbol{\sigma}}_{ij}\mid \left\{ n_{i}\right\} \right)
}{\delta n_{\ell }(\mathbf{r} ^{\prime },t)}\mid _{\delta
n=0}=F_{ij\ell}\left( \mathbf{r}_{1}-\mathbf{r}^{\prime },
\mathbf{r}_{1}\pm {\boldsymbol{\sigma}}_{ij}-\mathbf{r}^{\prime
}\right) \label{c.6aa}
\end{equation}
so
\begin{equation}
\int d\mathbf{r}^{\prime }\left( \mathbf{r}^{\prime
}-\mathbf{r}_{1}\right) \frac{\delta \chi _{ij}\left(
\mathbf{r}_{1},\mathbf{r}_{1}\pm {\boldsymbol{\sigma}}_{ij}\mid
\left\{ n_{i}\right\} \right) }{\delta n_{\ell }(\mathbf{r}
^{\prime },t)}\mid _{\delta n=0}=\int d\mathbf{r}^{\prime }\left(
\mathbf{r}^{\prime }\pm
\frac{1}{2}{\boldsymbol{\sigma}}_{ij}\right) \frac{\delta \chi
_{ij}\left( \mp \frac{1}{2}{\boldsymbol{\sigma}}_{ij},\pm
\frac{1}{2}{\boldsymbol{\sigma}}_{ij}\mid \left\{ n_{i}\right\}
\right) }{\delta n_{\ell }(\mathbf{r} ^{\prime },t)}\mid _{\delta
n=0}
\end{equation}
\begin{equation}
=\pm \frac{1}{2}{\boldsymbol{\sigma}}_{ij}\frac{\partial \ln \chi
_{ij}^{(0)}\left( \sigma _{ij};\left\{ n_{\ell }\left(
\mathbf{r}_{1}\right) \right\} \right) }{\partial n_{\ell }\left(
\mathbf{r}_{1}\right) }+\int d \mathbf{r}^{\prime
}\mathbf{r}^{\prime }\frac{\delta \chi _{ij}\left( \mp
\frac{1}{2}{\boldsymbol{\sigma}}_{ij},\pm
\frac{1}{2}{\boldsymbol{\sigma}}_{ij}\mid \left\{ n_{i}\right\}
\right) }{\delta n_{\ell }(\mathbf{r}^{\prime },t)} \mid _{\delta
n=0} \label{c.6b}
\end{equation}
The expansion for $\chi _{ij}\left(
\mathbf{r}_{1},\mathbf{r}_{1}\pm {\boldsymbol{\sigma}}_{ij}\mid
\left\{ n_{i}\left( t\right) \right\} \right) $ becomes
\begin{eqnarray}
\chi _{ij}\left( \mathbf{r}_{1},\mathbf{r}_{1}\pm
{\boldsymbol{\sigma}}_{ij}\mid \left\{ n_{i}\left( t\right)
\right\} \right)&=& \chi _{ij}^{(0)}\left( \sigma _{ij};\left\{
n_{k}\left( \mathbf{r} _{1},t\right) \right\} \right) \left[ 1\pm
\frac{1}{2}{\boldsymbol{\sigma}}_{ij}\cdot \sum_{\ell
=1}^{s}\nabla _{\mathbf{r}_{1}}\ln n_{\ell }(\mathbf{r}
_{1};t)\right.
\nonumber\\
&\times&\left.  \left( n_{\ell }\left( \mathbf{r}_{1}\right)
\frac{\partial \ln \chi _{ij}^{(0)}\left( \sigma _{ij};\left\{
n_{\ell }\left( \mathbf{r} _{1}\right) \right\} \right) }{\partial
n_{\ell }\left( \mathbf{r} _{1}\right) }+I_{ij\ell }(\sigma
_{ij};\left\{ n_{k}\left( \mathbf{r} _{1},t\right) \right\}
)\right) \right].\nonumber\\
\end{eqnarray}
The last line defines $I_{ij\ell }(\sigma _{ij};\left\{
n_{k}\left( \mathbf{r }_{1},t\right) \right\} )$ as
\begin{equation}
I_{ij\ell }(\sigma _{ij};\left\{ n_{k}\right\} )\equiv
\frac{2n_{\ell }( \mathbf{r}_{1},t)}{\chi _{ij}^{(0)}\left( \sigma
_{ij};\left\{ n_{k}\right\} \right) \sigma _{ij}}\int
d\mathbf{r}^{\prime }\left(
\widehat{\boldsymbol{\sigma}}_{ij}\cdot \mathbf{r}^{\prime
}\right) \frac{\delta \chi _{ij}\left(
-\frac{1}{2}{\boldsymbol{\sigma}}_{ij},+\frac{1}{2}{\boldsymbol{\sigma}}_{ij}\mid
\left\{ n_{i}\right\} \right) }{\delta n_{\ell }(\mathbf{r}
^{\prime },t)}\mid _{\delta n=0}.  \label{c.7}
\end{equation}

These results give the expansion of $J_{ij}\left[
\mathbf{r}_{1},\mathbf{v} _{1}\mid f_i\right]$ to first order in
the gradients
\begin{equation}
\sum_{j=1}^{s}J_{ij}\left[ \mathbf{r}_{1},\mathbf{v}_{1}\mid
f\right] \rightarrow \sum_{j=1}^{s}J_{ij}^{(0)}\left[
\mathbf{v}_{1}\mid f_{i}^{(0)},f_{j}^{(0)}\right] -\left(
Lf^{(1)}\right) _{i}
\end{equation}
\begin{equation}
-\sum_{j=1}^{s}\mathcal{K}_{ij,\gamma }[\mathbf{v}_{1}\mid
\partial
_{y_{\beta }}f_{j}^{(0)}(t)]\partial _{\gamma }y_{\beta }\left(
\mathbf{r}_{1},t\right)
\end{equation}
\begin{equation}
-\sum_{j,\ell =1}^{s}\left( n_{\ell }\frac{\partial \ln \chi
_{ij}^{(0)}\left( \sigma _{ij};\left\{ n_{k}\right\} \right)
}{\partial n_{\ell }}+I_{ij\ell }(\sigma _{ij};\left\{
n_{k}\right\} )\right) \mathcal{K}_{ij,\gamma }[\mathbf{v}_{1}\mid
f_{j}^{(0)}(t)]\partial _{\gamma }\ln n_{\ell }(\mathbf{r}_{1};t)
\label{c.8}
\end{equation}
with the definitions $\partial_{\gamma}X\equiv
\partial X/\partial r_{\gamma}$,
\begin{eqnarray}
J_{ij}^{(0)}\left[ \mathbf{v}_{1}\mid g_i,f_j\right] &\equiv &\chi
_{ij}^{(0)}\left( \sigma _{ij};\left\{ n_{i}\left(
\mathbf{r}_{1},t\right) \right\} \right) \sigma _{ij}^{d-1}\int
d\mathbf{v}_{2}\int d\widehat{\boldsymbol {\sigma }}\Theta
(\widehat{\boldsymbol {\sigma}}\cdot \mathbf{g}_{12})(\widehat{
\boldsymbol {\sigma }}\cdot \mathbf{g}_{12})  \nonumber\\
&&\times \left[ \alpha _{ij}^{-2}g_{i}(\mathbf{V}_{1}^{\prime \prime })f_{j}( \mathbf{V}_{2}^{\prime \prime
})-g_{i}(\mathbf{V}_{1})f_{j}(\mathbf{V}_{2}) \right], \label{c.10}
\end{eqnarray}
\begin{eqnarray}
\mathcal{K}_{ij,\gamma }[X_{j}] &=&\sigma _{ij}^{d}\chi
_{ij}^{(0)}\int d \mathbf{v}_{2}\int d\widehat{\boldsymbol {\sigma
}}\Theta (\widehat{\boldsymbol {\sigma}} \cdot
\mathbf{g}_{12})(\widehat{\boldsymbol {\sigma }}\cdot
\mathbf{g}_{12})
\widehat{\sigma }_\gamma  \nonumber\\
&&\times \left[ \alpha _{ij}^{-2}f_{i}^{(0)}(\mathbf{r}_{1},\mathbf{V} _{1}^{\prime \prime
};t)X_{j}(\mathbf{r}_{1},\mathbf{V}_{2}^{\prime \prime
})+f_{i}^{(0)}(\mathbf{r}_{1},\mathbf{V}_{1};t)X_{j}(\mathbf{r}_{1},\mathbf{V }_{2})\right] .  \label{c.12}
\end{eqnarray}
Finally, $L$ is a linear operator defined over $s$ dimensional
vectors $\left\{ X_{i}\right\}$ whose components are labelled by
the species
\begin{equation}
\left( LX\right) _{i}=-\sum_{j=1}^{s}\left( J_{ij}^{(0)}\left[
\mathbf{v} _{1}\mid X_{i},f_{j}^{(0)}\right] +J_{ij}^{(0)}\left[
\mathbf{v}_{1}\mid f_{i}^{(0)},X_{j}\right] \right) . \label{c.11}
\end{equation}

It remains to choose the magnitude of the external forces $\mathbf{F}_{i}$ to consistently order this expansion.
To be specific, and for comparison with Ref.\ \cite{deHaro83} it is assumed here to be of first order in the
gradients.

\subsection{Zeroth Order}

At lowest order all gradients of the hydrodynamic fields are
neglected, and (\ref{2.5}) becomes
\begin{equation}
\partial _{t}^{(0)}f_{i}^{(0)}(\mathbf{v}_{1};\left\{ y_{\beta }\left(
\mathbf{r}_{1},t\right) \right\} =\sum_{j=1}^{s}J_{ij}^{(0)}\left[
\mathbf{v} _{1}\mid f_{i}^{(0)}(\left\{ y_{\beta }\left(
\mathbf{r}_{1},t\right) \right\} ),f_{j}^{(0)}(\left\{ y_{\beta
}\left( \mathbf{r}_{1},t\right) \right\} )\right] \;. \label{c.13}
\end{equation}
The notation $\partial _{t}^{(0)}$ for the time derivative means
that the balance equations are to be used to zeroth order in the
gradients
\begin{eqnarray}
\partial _{t}^{(0)}f_{i}^{(0)}(V_{1};\left\{ y_{\beta }\left( \mathbf{r}
_{1},t\right) \right\} &=&\left( \partial _{y_{\beta
}}f_{i}^{(0)}(V_{1};\left\{ y_{\beta }\left(
\mathbf{r}_{1},t\right) \right\} )\right) \partial
_{t}^{(0)}y_{\beta }\left( \mathbf{r}_{1},t\right) \nonumber\\
&=&-\zeta ^{(0)}\left( \left\{ y_{\beta }\left(
\mathbf{r}_{1},t\right) \right\} \right) T\partial
_{T}f_{i}^{(0)}(V_{1};\left\{ y_{\beta }\left(
\mathbf{r}_{1},t\right) \right\} ).  \label{c.14}
\end{eqnarray}
Use of (\ref{c.14}) in (\ref{c.13}) gives the zeroth
order equation (\ref {5.1}) of the text.

\subsection{First Order}

The kinetic equation for contributions of first order in the gradients is
\begin{eqnarray}
\partial _{t}^{(0)}f_{i}^{(1)}+\left( Lf^{(1)}\right) _{i} &=&-\left(
\partial _{t}^{(1)}+\mathbf{v}_{1}\cdot \nabla_{\mathbf{r}_{1}}
+m_{i}^{-1}\mathbf{F}_{i}\cdot \nabla_{\mathbf{v} _{1}}\right)
f_{i}^{(0)}-\sum_{j=1}^{s}\mathcal{K}_{ij,\gamma }[\mathbf{v}
_{1}\mid \partial _{y_{\beta }}f_j^{(0)}]\partial
_{\gamma}y_{\beta
} \nonumber\\
& & -\frac{1}{2}\sum_{j,\ell =1}^{s}\;\mathcal{K}_{ij,\gamma
}\left[\mathbf{v}_{1}|\left( n_{\ell }\frac{\partial \ln \chi
_{ij}^{(0)}}{\partial n_{\ell }}+I_{ij\ell
}\right)f_j^{(0)}\right]\partial _{\gamma}\ln n_{\ell },
\label{c.15}
\end{eqnarray}
where $\nabla_{\bf r} \equiv \partial/\partial {\bf r}$ and
$\nabla_{\bf V} \equiv \partial/\partial {\bf V}$. The first term
on the right side of (\ref{c.15}) can be expressed explicitly in
terms of the gradients, where now $\partial _{t}^{(1)}$ means that
the balance equations are to be used with only terms of first
order in the gradients
\begin{eqnarray}
\left( \partial_{t}^{(1)}+\mathbf{v}_{1}\cdot \mathbf{\nabla
}_{\mathbf{r}_{1}}\mathbf{+}m_{i}^{-1}\mathbf{F}_{i}\cdot
\mathbf{\nabla }_{\mathbf{V} _{1}}\right) f_{i}^{(0)}
&=&m_{i}^{-1}\mathbf{F}_{i}\cdot \mathbf{\nabla }_{
\mathbf{V}_{1}}f_{i}^{(0)}+\left( \partial _{y_{\beta
}}f_i^{(0)}\right) \left( D_{t}^{(1)}+\mathbf{V} _{1}\cdot
\mathbf{\nabla
}_{\mathbf{r}_{1}}\right) y_{\beta } \nonumber\\
&=&\left( m_{i}^{-1}\mathbf{F}_{i}-\rho
^{-1}\sum_{j=1}^{s}n_{j}\mathbf{F} _{j}\right) \cdot
\mathbf{\nabla
}_{\mathbf{V}_{1}}f_{i}^{(0)}\nonumber\\
& & -\left( \mathbf{\nabla }_{\mathbf{V}_{1}}f_{i}^{(0)}\right)
\cdot \left( -\rho ^{-1} \mathbf{\nabla
}_{\mathbf{r}_{1}}p+\mathbf{V}_{1}\cdot \mathbf{\nabla }_{
\mathbf{r}_{1}}\mathbf{U}\right)  \nonumber\\
&&-\left( \partial _{T}f_{i}^{(0)}\right) \left( \left( T\zeta
_{U}+\frac{2}{nd}p\right) \mathbf{\nabla }\cdot
\mathbf{U}-\mathbf{V}_{1}\cdot \mathbf{\nabla
}_{\mathbf{r}_{1}}T\right)  \nonumber\\  &&+\sum_{j=1}^{s}\left(
\partial _{n_{j}}f_{i}^{(0)}\right) \left(
-n_{j}\nabla \cdot \mathbf{U}+\mathbf{V}_{1}\cdot \mathbf{\nabla
}_{\mathbf{r }_{1}}n_{j}\right) .  \label{c.17}
\end{eqnarray}
In (\ref{c.17}), $D_t^{(1)}\equiv \partial_t^{(1)}+{\bf
U}\cdot \nabla$ and use has been made of the fact that
$f_{i}^{(0)}$ depends on $\mathbf{U}$ only through the combination
$\mathbf{V}_{1}=\mathbf{v}_{1}-\mathbf{U}$, so that
\begin{equation}
\mathbf{\nabla }_{\mathbf{U}}f_{i}^{(0)}=-\mathbf{\nabla
}_{\mathbf{V} _{1}}f_{i}^{(0)}.
\end{equation}
The pressure gradient can be expressed in terms of the temperature
and density gradients
\begin{equation}
\nabla_{\mathbf{r}_{1}}p=\frac{p}{T}\mathbf{\nabla }_{\mathbf{r}
_{1}}T+\sum_{j=1}^{s}\left( \partial _{n_{j}}p\right)
\mathbf{\nabla }_{\mathbf{r}_{1}}n_{j},  \label{c.18}
\end{equation}
to give
\begin{eqnarray}
\left( \partial _{t}^{(1)}+\mathbf{v}_{1}\cdot \mathbf{\nabla
}_{\mathbf{r} _{1}}\mathbf{+}m_{i}^{-1}\mathbf{F}_{i}\cdot
\mathbf{\nabla }_{\mathbf{v} _{1}}\right)
f_{i}^{(0)}&=&m_{i}^{-1}\mathbf{F}_{i}\cdot \mathbf{\nabla }_{
\mathbf{V}_{1}}f_{i}^{(0)}-\rho ^{-1}\sum_{j=1}^{s}
 n_{j}\mathbf{\nabla }_{\mathbf{V}_{1}}f_{i}^{(0)}\cdot\mathbf{F}_{j}
\nonumber\\
 &&+\left( \frac{p}{\rho T}\mathbf{\nabla
}_{\mathbf{v}_{1}}f_{i}^{(0)}+ \mathbf{V}_{1}\partial
_{T}f_{i}^{(0)}\right) \cdot \mathbf{\nabla }_{ \mathbf{r}_{1}}T
\nonumber\\
& & +\sum_{j=1}^{s}\left( \rho ^{-1} \partial _{n_{j}}p
\mathbf{\nabla
}_{\mathbf{v}_{1}}f_{i}^{(0)}+\mathbf{V}_{1}\partial
_{n_{j}}f_{i}^{(0)}\right) \cdot \mathbf{\nabla}_{\mathbf{r}
_{1}}n_{j}  \nonumber\\
&&-\left( \left( \zeta _{U}+\frac{2}{n T d}p\right) T\partial
_{T}f_{i}^{(0)}+\sum_{j=1}^{s}n_{j}\partial
_{n_{j}}f_{i}^{(0)}\right) \nabla \cdot
\mathbf{U}\nonumber\\
& & -V_{1\gamma}(\partial _{V_{1\eta }}f_{i}^{(0)})\partial
_{\gamma }U_{\eta }. \label{c.19}
\end{eqnarray}
Equations (\ref{c.15}) for the first order distributions,
$f_{i}^{(1)}$, now become
\begin{eqnarray}
\partial _{t}^{(0)}f_{i}^{(1)}+\left( Lf^{(1)}\right) _{i} &=&\mathbf{A}
_{i}\left( \mathbf{V}_1;\left\{ n_{i}\right\} \right) \cdot \nabla
\ln T+\sum_{j=1}^{s}\mathbf{B}_{i}^j\left( \mathbf{V}_1;\left\{
n_{i}\right\} \right) \cdot \nabla \ln n_{j} \nonumber\\
&&+C_{i,\gamma \eta }\left( \mathbf{V}_1;\left\{ n_{i}\right\}
\right) \frac{1}{2}\left( \partial _{\gamma }U_{\eta }+\partial
_{\eta }U_{\gamma }-\frac{2}{d}\delta
_{\gamma \eta }\nabla \cdot \mathbf{U}\right) \nonumber\\
&&+D_{i}\left( \mathbf{V}_1;\left\{ n_{i}\right\} \right) \nabla
\cdot \mathbf{U}+\sum_{j=1}^{s}\mathbf{E}_{i}^j\left(
\mathbf{V}_1;\left\{ n_{i}\right\} \right) \cdot \mathbf{F}_{j}.
\label{c.20}
\end{eqnarray}
The functions of velocity on the right side of (\ref{c.20})
are identified as
\begin{equation}
A_{i,\gamma}\left( \mathbf{V}\right)=\frac{1}{2} V_{\gamma}\nabla
_{\mathbf{V}}\cdot \left( \mathbf{V}f_{i}^{(0)}\right)
-\frac{p}{\rho }\partial_{V_{\gamma}}
f_{i}^{(0)}+\frac{1}{2}\sum_{j=1}^{k}\mathcal{ K}_{ij,\gamma
}\left[\nabla_{\mathbf{V}}\cdot \left( \mathbf{V}
f_{j}^{(0)}\right) \right] ,  \label{c.20a}
\end{equation}
\begin{eqnarray}
B_{i,\gamma}^{j}\left( \mathbf{V}\right) &=& -V_\gamma n_{j}\partial _{n_{j}}f_{i}^{(0)}-\rho^{-1}
(\partial_{V_\gamma}f_{i}^{(0)})n_{j}(\partial _{n_{j}}p)
\nonumber\\
 &&-\sum_{\ell
=1}^{s}\mathcal{K}_{i\ell,\gamma}\left[\left( n_{j}\partial
_{n_{j}}+\frac{1}{2}\left( n_{\ell }\frac{\partial \ln \chi
_{i\ell }^{(0)}}{\partial n_{j}}+I_{i\ell j}\right) \right)
f_{\ell }^{(0)} \right],  \label{c.20b}
\end{eqnarray}
\begin{eqnarray}
\label{c.20.c} C_{i,\gamma \beta }\left(
\mathbf{V}\right)&=&\frac{1}{2} \left(V_\gamma
\partial_{V_\beta}f_{i}^{(0)}+V_\beta
\partial_{V_\gamma}f_{i}^{(0)}-\frac{2}{d}\delta_{\beta
\gamma}{\bf V}\cdot \nabla_{\bf V}f_{i}^{(0)}\right)\nonumber\\
& & +\frac{1}{2}\sum_{j
=1}^{s}\left(\mathcal{K}_{ij,\gamma}[\partial_{V_\beta}f_{j}^{(0)}]+
\mathcal{K}_{ij,\beta}[\partial_{V_\gamma}f_{j}^{(0)}]-\frac{2}{d}
\delta_{\beta
\gamma}\mathcal{K}_{ij,\lambda}[\partial_{V_\lambda}f_{j}^{(0)}]\right),
\end{eqnarray}
\begin{eqnarray}
D_{i}({\bf V})&=&\frac{1}{d}\mathbf{V}\cdot \nabla _{\mathbf{V}} f_{i}^{(0)}-\frac{1}{2} \left( \zeta
_{U}+\frac{2}{n T d}p\right) \nabla _{\mathbf{V}}\cdot \left( \mathbf{V}
f_{i}^{(0)}\right)\nonumber\\
& &  +\sum_{j=1}^s\left( n_{j}\partial _{n_{j}}f_{i}^{(0)}+\frac{1
}{d}\mathcal{K}_{ij,\gamma}\left[\partial
_{V_{\gamma}}f_{j}^{(0)}\right]\right),   \label{c.20d}
\end{eqnarray}
\begin{equation}
\mathbf{E}_{i}^{j}({\bf V})=-\left(\nabla_{\mathbf{V}} f_{i}^{(0)}\right)\frac{1}{m_j}\left(\delta
_{ij}-\frac{n_{j}m_{j}}{\rho}\right).  \label{c.20e}
\end{equation}
Upon deriving (\ref{c.20a})--(\ref{c.20e}), use has been
made of the relations
\begin{equation}
T\partial _{T}f_{i}^{(0)}=-\frac{1}{2}\mathbf{\nabla
}_{\mathbf{V}}\cdot \left( \mathbf{V}f_{i}^{(0)}\right)
,\hspace{0.3in}\partial _{U_{\beta }}f_{i}^{(0)}=-\partial
_{V_{\beta }}f_{i}^{(0)}.
\end{equation}
The tensor derivative of the flow field $\partial _{\gamma
}U_{\eta }$ has been expressed in terms of its independent trace
and traceless parts, using the spherical symmetry of
$f_{i}^{(0)}$, e.g.
\begin{eqnarray}
\left(V_{\gamma }\partial _{V_{\eta}}f_{i}^{(0)}\right)\partial
_{\gamma }U_{\eta} &=&\widehat{V}_{\gamma }\widehat{V}_{\eta
}(V\partial _{V})f_{i}^{(0)}\left( \partial _{\gamma
}U_{\eta}\right) =\widehat{V}_{\gamma }\widehat{V}_{\beta
}(V\partial _{V}f_{i}^{(0)})\frac{1}{2}\left(
\partial _{\gamma}U_{\eta}+\partial _{\eta}U_{\gamma}\right)
\nonumber\\
&=&\widehat{V}_{\gamma }\widehat{V}_{\eta }(V\partial
_{V}f_{i}^{(0)})\frac{1}{2}\left( \partial _{\gamma}U_{\eta
}+\partial _{\eta}U_{\gamma
}-\frac{2}{d}\delta _{\gamma \eta }\nabla \cdot \mathbf{U}\right) \nonumber\\
& & +\frac{1}{d}(V\partial _{V}f_{i}^{(0)})\nabla \cdot
\mathbf{U}, \label{c.21}
\end{eqnarray}
and a similar analysis of the contribution from
$\sum_{j=1}^{s}\mathcal{K} _{ij,\gamma }[\mathbf{v}\mid
\partial _{V_{\beta }}f_j^{(0)}].$ Equation (\ref{c.20}) is an
inhomogeneous, linear integral equation, where the inhomogeneity (the right side) is a linear combination of the
the external force and the gradients of the hydrodynamic fields. The coefficients of these fields are specified
functions of the velocity $ \mathbf{V}$. Since by definition $f_{i}^{(1)}$ is proportional to the external force
and the gradients of the hydrodynamic fields, it must have the form
\begin{eqnarray}
f_{i}^{(1)} &\rightarrow &\boldsymbol{\mathcal{A}}_{i}\left(
\mathbf{V}\right)\cdot  \nabla \ln
T+\sum_{j=1}^{s}\boldsymbol{\mathcal{B}}_{i}^j\left(
\mathbf{V}\right) \cdot \nabla \ln n_{j}  \nonumber\\
&&+\mathcal{C}_{i,\gamma \eta }\left( \mathbf{V} \right)
\frac{1}{2}\left( \partial _{\gamma }U_{\eta }+\partial _{\eta
}U_{\gamma }-\frac{2}{d}\delta _{\gamma \eta }\nabla \cdot
\mathbf{U} \right)  \nonumber\\
&&+\mathcal{D}_{i}\left( \mathbf{V} \right) \nabla \cdot
\mathbf{U}+\sum_{j=1}^{s}\boldsymbol{\mathcal{E}}_{i}^j\left(
\mathbf{V} \right) \cdot \mathbf{F}_{j}. \label{c.22}
\end{eqnarray}
The unknown functions of the peculiar velocity,
$\boldsymbol{\mathcal{A}}_{i},\boldsymbol{\mathcal{B}}
_{i}^j,\mathcal{C}_{i,\gamma \eta },\mathcal{D}_{i},$ and
$\boldsymbol{\mathcal{E}}_{i}^j$ are determined by solving Eq.
(\ref{c.20}). By dimensional analysis,
$\boldsymbol{\mathcal{A}}_{i}\left(
\mathbf{V}\right)=v_0^{-d}\ell^{1-d}\boldsymbol{\mathcal{A}}_{i}^*\left(
\mathbf{V}^*\right)$, $\boldsymbol{\mathcal{B}}_{i}^j\left(
\mathbf{V}\right)=v_0^{-d}\ell^{1-d}\boldsymbol{\mathcal{B}}_{i}^{j*}\left(
\mathbf{V}^*\right)$, $\mathcal{C}_{i,\gamma \eta}\left(
\mathbf{V}\right)=v_0^{-(d+1)}\ell^{1-d}\mathcal{C}_{i,\gamma
\eta}^*\left( \mathbf{V}^*\right)$, $\mathcal{D}_{i}\left(
\mathbf{V}\right)=v_0^{-(d+1)}\ell^{1-d}\mathcal{D}_{i}^*\left(
\mathbf{V}^*\right)$, and $\boldsymbol{\mathcal{E}}_{i}^j\left(
\mathbf{V}\right)=m^{-1}v_0^{-(d+2)}\ell^{1-d}\boldsymbol{\mathcal{E}}_{i}^{j*}\left(
\mathbf{V}^*\right)$, where $\ell$ is an effective mean free path
and $\boldsymbol{\mathcal{A}}_{i}^*\left( \mathbf{V}^*\right)$,
$\boldsymbol{\mathcal{B}}_{i}^{j*}\left( \mathbf{V}^*\right)$,
$\mathcal{C}_{i,\gamma \eta}^*\left( \mathbf{V}^*\right)$,
$\mathcal{D}_{i}^*\left( \mathbf{V}^*\right)$, and
$\boldsymbol{\mathcal{E}}_{i}^{j*}\left( \mathbf{V}^*\right)$ are
dimensionless functions of the reduced velocity ${\bf V}^*={\bf
V}/v_0$, $v_0=\sqrt{2T/m}$ being a thermal speed. Consequently,
\begin{equation}
\label{c.23}
\partial_t^{(0)}\boldsymbol{\mathcal{A}}_{i}\left(
\mathbf{V}\right)=-\zeta^{(0)}T\partial_T\boldsymbol{\mathcal{A}}_{i}\left(
\mathbf{V}\right)=\frac{1}{2}\zeta^{(0)} \nabla _{\mathbf{V}
}\cdot \left( \mathbf{V}\boldsymbol{\mathcal{A}}_{i}\left(
\mathbf{V}\right)\right),
\end{equation}
\begin{equation}
\label{c.23.1}
\partial_t^{(0)}\boldsymbol{\mathcal{B}}_{i}^j\left(
\mathbf{V}\right)=-\zeta^{(0)}T\partial_T\boldsymbol{\mathcal{B}}_{i}^j\left(
\mathbf{V}\right)=\frac{1}{2}\zeta^{(0)} \nabla _{\mathbf{V}
}\cdot \left( \mathbf{V}\boldsymbol{\mathcal{B}}_{i}^j\left(
\mathbf{V}\right)\right),
\end{equation}
\begin{equation}
\label{c.23.2}
\partial_t^{(0)}\mathcal{C}_{i,\gamma\eta}\left(
\mathbf{V}\right)=-\zeta^{(0)}T\partial_T\mathcal{C}_{i,\gamma\eta}\left(
\mathbf{V}\right)=\frac{1}{2}\zeta^{(0)}\mathcal{C}_{i,\gamma\eta}+
\frac{1}{2}\zeta^{(0)} \nabla _{\mathbf{V} }\cdot \left(
\mathbf{V}\mathcal{C}_{i,\gamma\eta}\left(
\mathbf{V}\right)\right),
\end{equation}
\begin{equation}
\label{c.23.3}
\partial_t^{(0)}\mathcal{D}_{i}\left(
\mathbf{V}\right)=-\zeta^{(0)}T\partial_T\mathcal{D}_{i}\left(
\mathbf{V}\right)=\frac{1}{2}\zeta^{(0)}\mathcal{D}_{i}+
\frac{1}{2}\zeta^{(0)} \nabla _{\mathbf{V} }\cdot \left(
\mathbf{V}\mathcal{D}_{i}\left( \mathbf{V}\right)\right),
\end{equation}
\begin{equation}
\label{c.23.4}
\partial_t^{(0)}\boldsymbol{\mathcal{E}}_{i}^j\left(
\mathbf{V}\right)=-\zeta^{(0)}T\partial_T\boldsymbol{\mathcal{E}}_{i}^j\left(
\mathbf{V}\right)=\zeta^{(0)}\boldsymbol{\mathcal{E}}_{i}^j\left(
\mathbf{V}\right)+\frac{1}{2}\zeta^{(0)} \nabla _{\mathbf{V}
}\cdot \left( \mathbf{V}\boldsymbol{\mathcal{E}}_{i}^j\left(
\mathbf{V}\right)\right).
\end{equation}
In addition,
\begin{equation}
\partial _{t}^{(0)}\nabla \ln T=\nabla \partial_{t}^{(0)}\ln T=-\nabla
\zeta ^{(0)}=-\frac{1}{2}\zeta ^{(0)}\nabla \ln
T-\sum_{j=1}^{s}n_{j}\frac{
\partial \zeta ^{(0)}}{\partial n_{j}}\nabla \ln n_{j}.  \label{c.24}
\end{equation}
Since the external force and gradients of the fields are all
independent, Eq. (\ref{c.20}) can be separated into independent
equations for the coefficients of each. This leads to the set of
linear, inhomogeneous integral equations
\begin{equation}
\left( \left( \mathcal{L}-\frac{1}{2}\zeta ^{(0)}\right) \boldsymbol{\mathcal{A}}\right) _{i}=\mathbf{A}_{i},
\label{c.25}
\end{equation}
\begin{equation}
\left( \mathcal{L}\boldsymbol{\mathcal{B}}^j\right)_{i}-n_{j}\frac{\partial \zeta ^{(0)}}{\partial
n_{j}}\boldsymbol{\mathcal{A}}_{i}=\mathbf{B}_{i}^{j}, \label{c.26}
\end{equation}
\begin{equation}
\left( \left( \mathcal{L}+\frac{1}{2}\zeta ^{(0)}\right) \mathcal{C}_{\gamma \eta }\right) _{i}=C_{i,\gamma
\eta}, \label{c.27}
\end{equation}
\begin{equation}
\left( \left( \mathcal{L}+\frac{1}{2}\zeta ^{(0)}\right) \mathcal{D}\right) _{i}=D_{i},  \label{c.28}
\end{equation}
\begin{equation}
\left(
\left(\mathcal{L}+\zeta^{(0)}\right)\boldsymbol{\mathcal{E}}^j\right)_{i}=\mathbf{E}_{i}^{j}.
\label{c.29}
\end{equation}
The linear operator $\mathcal{L}$ is
\begin{equation}
\left( \mathcal{L}X\right) _{i}=\frac{1}{2}\zeta ^{(0)}\nabla
_{\mathbf{V} }\cdot \left( \mathbf{V}X_{i}\right) +\left(
LX\right) _{i}. \label{c.30}
\end{equation}
Notice that (\ref{c.25}) can be used in (\ref{c.26}) to give the
equivalent representation for the latter
\begin{equation}
\left( \mathcal{L}\left( \boldsymbol{\mathcal{B}}^{j}-\left(
2n_{j}\partial _{n_{j}}\ln \zeta ^{(0)}\right)
\boldsymbol{\mathcal{A}}\right) \right) _{i}=\mathbf{B}
_{i}^{j}-\left( 2n_{j}\partial _{n_{j}}\ln \zeta ^{(0)}\right)
\mathbf{A}_i. \label{c.31}
\end{equation}

This completes the CE solution up through first order in the gradients and first order in the external force.
Once (\ref{5.1}) has been solved for $f_{i}^{(0)}$ the integral equations for $\boldsymbol{\mathcal{A}}_{i}$,
 $\boldsymbol{\mathcal{B}}_{i}^j$, $\mathcal{C}_{i,\gamma \eta }$, $\mathcal{D}_{i},$ and $
\boldsymbol{\mathcal{E}}_{i}^j$ can be solved for $f_{i}^{(1)}$.
Then, the cooling rate, heat flux, and pressure tensor can be
calculated as linear functions of the gradients and the external
force, and the explicit forms for the transport coefficients
identified.

\section{An Eigenvalue Problem for $\mathcal{L}$}
\label{CMH.App.1}

To simplify and interpret the linear integral equations defining
the first order solutions $\left\{ f_{i}^{(1)}\right\} $ it is
useful to identify a special set of eigenvalues and eigenfunctions
for the operator $\mathcal{L}$. Consider the equation for
$f_{i}^{(0)}$
\begin{equation}
-\zeta ^{(0)}T
\partial_T\;f_{i}^{(0)}=\sum_{j=1}^{s}J_{ij}^{(0)}\left[
\mathbf{r} _{1},\mathbf{v}_{1}\mid f_{i}^{(0)},f_{j}^{(0)}\right]
. \label{d.1}
\end{equation}
Since the temperature occurs through the form (\ref{5.7}), the
temperature derivatives can be expressed as velocity derivatives
\begin{equation}
\frac{1}{2}\zeta ^{(0)}\nabla _{\mathbf{V}}\cdot \left( \mathbf{V} f_{i}^{(0)}\right)
=\sum_{j=1}^{s}J_{ij}^{(0)}\left[ \mathbf{r}_{1},\mathbf{v }_{1}\mid f_{i}^{(0)},f_{j}^{(0)}\right]. \label{d.2}
\end{equation}
Noting that $\zeta ^{(0)}\propto \sqrt{T}$, the derivative of this
equation with respect to $T$ gives directly
\begin{equation}
\left( \mathcal{L}T\mathbf{\partial }_{T}f^{(0)}\right)
_{i}=\frac{1}{2} \zeta ^{(0)}T\mathbf{\partial }_{T}f_{i}^{(0)}.
\label{d.3}
\end{equation}
where use has been made of (\ref{c.23}), i.e. $T\mathbf{\partial }
_{T}f_{i}^{(0)}=-\nabla _{\mathbf{V}}\cdot \left( \mathbf{V}
f_{i}^{(0)}\right) /2$. An equivalent dimensionless form is
\begin{equation}
\left( \mathcal{L}\nabla _{\mathbf{V}}\cdot \left( \mathbf{V}f^{(0)}\right) \right) _{i}=\frac{1}{2}\zeta
^{(0)}\nabla _{\mathbf{V}}\cdot \left( \mathbf{ V}f_{i}^{(0)}\right).  \label{d.3a}
\end{equation}
In a similar way, differentiation of (\ref{d.2}) with respect to
each component of the flow velocity gives
\begin{equation}
\left( \mathcal{L}\mathbf{\partial }_{U_{\gamma }}f^{(0)}\right)
_{i}=-\frac{ 1}{2}\zeta ^{(0)}\mathbf{\partial }_{U_{\gamma
}}f_{i}^{(0)}. \label{d.4}
\end{equation}

Finally, differentiate (\ref{d.2}) with respect to each of the
species densities and noting that the density dependence of all
quantities occurs only through the $f^{(0)}$'s and the $\chi
_{ij}$'s, one gets
\begin{eqnarray}
\left( \mathcal{L}\partial _{n_{\ell }}f^{(0)}\right) _{i}
&=&\left(
\partial _{n_{\ell }}\sum_{j=1}^{s}J_{ij}^{(0)}\left[ f_{i}^{(0)},f_{j}^{(0)}
\right] \right) \mid _{f^{(0)}}-\frac{1}{2}\left( \partial
_{n_{\ell }}\zeta ^{(0)}\right) \nabla _{\mathbf{V}}\cdot \left(
\mathbf{V}f_{i}^{(0)}\right)
\notag \\
&=&\frac{1}{2}\left( \partial _{n_{\ell }}\zeta ^{(0)}\mid
_{f^{(0)}}-\partial _{n_{\ell }}\zeta ^{(0)}\right) \nabla
_{\mathbf{V}
}\cdot \left( \mathbf{V}f_{i}^{(0)}\right)  \notag \\
&=&-\partial _{n_{\ell }}\ln \zeta ^{(0)}\mid _{\chi
_{ij}^{(0)}}\frac{1}{2} \zeta ^{(0)}\nabla _{\mathbf{V}}\cdot
\left( \mathbf{V}f_{i}^{(0)}\right). \label{d.5}
\end{eqnarray}
This last form can be simplified by taking into account
(\ref{d.3a}) to get
\begin{equation}
\left( \mathcal{L}\left( \partial _{n_{\ell }}f^{(0)}+\partial
_{n_{\ell }}\ln \zeta ^{(0)}\mid _{\chi _{ij}^{(0)}}\nabla
_{\mathbf{V}}\cdot \left( \mathbf{V}f^{(0)}\right) \right) \right)
_{i}=0 . \label{d.7}
\end{equation}

In summary, there are $s+d+1$ eigenvalues and eigenfunctions of
the operator $\mathcal{L}$. Equations (\ref{d.3a}) and (\ref{d.4})
identify these for the eigenvalue $\zeta ^{(0)}/2$ and the
$d$--fold degenerate value $-\zeta ^{(0)}/2$, respectively.
Equation (\ref{d.7}) identifies the eigenfunctions for the $s$--
fold degenerate eigenvalue $0$. In dimensionless form this
eigenvalue problem is written as
\begin{equation}
\left( \mathcal{L}\Psi ^{(m)}\right) _{i}=\lambda ^{(m)}\Psi
_{i}^{(m)}. \label{d.8}
\end{equation}
The eigenvectors are
\begin{eqnarray}
\Psi _{i}^{(\ell)} &=&n_{\ell }\partial _{n_{\ell
}}f_i^{(0)}-2n_{\ell }\partial _{n_{\ell }}\ln \zeta ^{(0)}\mid
_{\chi _{ij}^{(0)}}\Psi
_{i}^{(s+1)},\hspace{0.3in}\ell =1,..,s,  \notag \\
\Psi _{i}^{(s+1)} &=&-\frac{1}{2}\nabla _{\mathbf{V}}\cdot \left(
\mathbf{V} f_{i}^{(0)}\right) ,\hspace{0.3in}\Psi _{i}^{(\gamma
)}=-v_0\partial _{V_{\gamma}}f_{i}^{(0)},  \label{d.9}
\end{eqnarray}
with the corresponding eigenvalues
\begin{equation}
\lambda ^{(m)}\Leftrightarrow \left( 0,..,0,\frac{1}{2}\zeta
^{(0)},-\frac{1}{2}\zeta ^{(0)},-\frac{1}{2}\zeta
^{(0)},-\frac{1}{2}\zeta ^{(0)}\right) . \label{d.10}
\end{equation}
These eigenvalues are the same as those of the linearized
hydrodynamic equations in the long wavelength limit. This provides
a direct link between hydrodynamics and the spectrum of the
linearized Enskog operator. In addition to this physical
interpretation, the eigenvalues and eigenfunctions allow a
practical formulation of the integral equations, as follows.

\subsection{Biorthogonal set}

Define a scalar product by
\begin{equation}
\left( a,b\right) =\sum_{i=1}^{s}\int d\mathbf{V}a_{i}^{\dagger
}\left( \mathbf{V}\right) b_{i}\left( \mathbf{V}\right),
\label{d.11}
\end{equation}
where the dagger denotes complex conjugation. A biorthogonal basis
set is then defined by the eigenfunctions $\Psi_i ^{(m)}$ above,
and
\begin{equation}
\psi _{\alpha i}=\left( \frac{\delta _{i1}}{n_{i}},..\frac{\delta
_{is}}{ n_{i}},\left( \frac{2m_{i}}{dm}V^{\ast 2}-1\right)
,\frac{m_{i}}{\rho} \mathbf{V}^{\ast }\right), \label{d.12}
\end{equation}
where ${\bf V}^*={\bf V}/v_0$. The orthonormality condition
\begin{equation}
\sum_{i=1}^{s}\int d\mathbf{V}^*\psi _{\alpha i}\left(
\mathbf{V}^{\ast }\right) \Psi _{\beta i}\left( \mathbf{V}^{\ast
}\right) =\delta _{\alpha \beta }  \label{d.13}
\end{equation}
is easily verified. An associated projection operator is given by
\begin{equation}
\left( \mathcal{P}X\right) _{i}=\sum_{\alpha }\Psi_{\alpha
i}\left( \mathbf{ V}^{\ast}\right) \sum_{j}\int d\mathbf{V}^*\psi
_{\alpha j}\left( \mathbf{V} ^{\ast }\right) X_{j}\left(
\mathbf{V}^{\ast }\right) .  \label{d.14}
\end{equation}
It follows from (\ref{d.14}) that $\mathcal{P}^{2}=\mathcal{P}$.
The corresponding orthogonal projection is
\begin{equation}
\mathcal{Q}=1-\mathcal{P}.  \label{d.15}
\end{equation}

Consider the quantity $\left( \mathcal{PL}X\right) _{i}$
\begin{eqnarray}
\left( \mathcal{PL}X\right)_{i} &=&\sum_{\alpha }\Psi _{\alpha
i}\left( \mathbf{V}^{\ast }\right) \sum_{j}\int d\mathbf{V}^*\psi
_{\alpha j}\left(
\mathbf{V}^*\right) \left( \mathcal{L}X\right) _{j}  \notag \\
&=&\Psi _{i}^{(s+1)}\left( \mathbf{V}^{\ast }\right)
\sum_{j=1}^{s}\frac{2m_{j}}{d m}\int d\mathbf{V}^{\ast }V^{\ast
2}\left( \mathcal{L}X\right) _{j}. \label{d.16}
\end{eqnarray}
Only the projection onto $\Psi_{i}^{(s+1)}$ contributes due to
conservation of species number and momentum. It follows then that
\begin{equation}
\left( \mathcal{PL}\boldsymbol{\mathcal{A}}\right) _{i}=0,\hspace{0.3in} \left(
\mathcal{PL}\boldsymbol{\mathcal{B}}^j\right) _{i}=0,\hspace{0.3in}
\left(\mathcal{PL}\boldsymbol{\mathcal{E}}^j\right) _{i}=0,\hspace{ 0.3in}\left( \mathcal{PLC}\right)_{i}=0.
\label{d.17}
\end{equation}
The terms with $ \boldsymbol{\mathcal{A}}, \boldsymbol{\mathcal{B}}^j$, and $ \boldsymbol{\mathcal{E}}^{j}$
vanish from symmetry since these all vectors; the last equality follows because $ \mathcal{C}$ has zero trace.
Next, note that
\begin{equation}
\zeta _{U}\nabla \cdot \mathbf{U}=\zeta ^{(1)}=\frac{2}{dnT}
\sum_{i=1}^{s}\int
d\mathbf{v}_{1}\frac{1}{2}m_{i}v_{1}^{2}\left(\mathcal{L}
\mathcal{D}\right)_i -\frac{2}{dnT}\sum_{i,j=1}^{s}\int
d\mathbf{v}_{1} \frac{1}{2}m_{i}v_{1}^{2}\mathcal{K}_{ij,\eta
}\left[\partial _{V_\eta}f_{j}^{(0)}\right],  \label{d.18}
\end{equation}
or equivalently
\begin{eqnarray}
\Psi _{i}^{s+1}\zeta _{U} &=&\Psi _{2i}\sum_{j=1}^{k}\int
d\mathbf{v} _{1}\psi _{j}^{s+1}\mathcal{L}_{j}\left[
\mathcal{D}\right] -\Psi _{i}^{s+1}
\frac{2m}{dnT}\sum_{j=1}^{k}\int d\mathbf{v}_{1}\frac{1}{2mdv_{0}}
m_{j}v_{1}^{2}\sum_{p}\mathcal{K}_{jp,\eta }[\partial _{V_{\eta
}^{\ast
}}f_{j}^{(0)}]  \notag \\
&=&\Psi _{2i}\sum_{j=1}^{k}\int d\mathbf{v}_{1}\psi
_{2j}\mathcal{L}_{j} \left[ \mathcal{D}\right]
-\mathcal{P}\frac{1}{dv_{0}}\mathcal{K}_{\eta }
\left[ \partial _{V_{\eta }^{\ast }}f^{(0)}\right]  \notag \\
&=&\mathcal{PL}_{j}\left[ \mathcal{D}\right]
+\mathcal{P}\frac{1}{dv_{0}} \mathcal{K}_{\eta }\left[ \partial
_{V_{\eta }^{\ast }}f^{(0)}\right] . \label{d.19}
\end{eqnarray}
This can be used to eliminate the explicit occurrence of the transport coefficient $\zeta _{U}$ in the integral
equation (\ref{c.28}). Finally, two additional identities are needed for the proofs of Appendix E
\begin{equation}
-\partial _{V_{\gamma }^{\ast }}f_{i}^{(0)}\frac{p-nT}{v_{0}^{2}\rho }=
\mathcal{P}\sum_{j=1}^{k}\frac{1}{2v_{0}}\mathcal{K}_{ij,\gamma }[\mathbf{ \nabla }_{\mathbf{V}^{\ast }}\cdot
\left( \mathbf{V}^{\ast }f_{j}^{(0)}\right) ],  \label{d.20}
\end{equation}
\begin{equation}
\left( \nabla _{\mathbf{V\ast }}f_{i}^{(0)}\right) \left( v_{0}^{2}\rho \right) ^{-1}n_{j}\partial
_{n_{j}}\left( p-nT\right) =-\mathcal{P} \sum_{\ell =1}^{k}\frac{1}{v_{0}}\mathcal{K}_{i\ell }[\left(
n_{j}\partial _{n_{j}}+\frac{1}{2}I_{i\ell j}\right) f_{\ell }^{(0)}]. \label{d.21}
\end{equation}

\section{Solubility Conditions and Uniqueness}
The results of Appendix D allow proof that the integral equations
have solutions and that they are unique. These equations have the
generic form
\begin{equation}
\left( \mathcal{L}-\lambda \right) \mathcal{X}=X,  \label{e.1}
\end{equation}
where $\lambda $ is one of the eigenvalues (\ref{d.10}). Let
relation $ \mathcal{X}$ be a solution to (\ref{e.1}). Then adding
any solution to the corresponding homogeneous integral equation
also gives a solution
\begin{equation}
\mathcal{X}^{\prime }=\mathcal{X}+c\Psi ,  \label{e.2}
\end{equation}
where $\left( \mathcal{L}-\lambda \right) \Psi =0$. However, the property
\begin{equation}
\left( \mathcal{P}f^{(1)}\right) _{i}=\sum_{\alpha }\Psi _{\alpha
i}\left( \mathbf{V}\right) \sum_{j}\int d\mathbf{v}\psi _{\alpha
j}\left( \mathbf{V} \right) f_{j}^{(1)}\left( \mathbf{V}\right) =0
\label{e.3}
\end{equation}
follows from the fact that the average densities, temperature, and flow velocity are given exactly by the first
order term $f_i^{(0)}$, so that contributions to these averages from all higher order terms must vanish.
Equivalently, (\ref{e.3}) implies
\begin{equation}
\mathcal{P}\left(
\begin{array}{c}
\boldsymbol{\mathcal{A}} \\
\boldsymbol{\mathcal{B}}^{j} \\
\mathcal{C} \\
\mathcal{D} \\
\boldsymbol{\mathcal{E}}^{j}
\end{array}
\right) =0.  \label{e.4}
\end{equation}
Consequently, the solution to (\ref{e.1}) with the condition
(\ref{e.4}) is unique.

To show that solutions exist the integral equations are written in
the equivalent form
\begin{equation}
\left( \mathcal{Q}\left( \mathcal{L}-\frac{1}{2}\zeta ^{\ast }\right)\boldsymbol{\mathcal{A}}\right)
_{i}=\mathcal{Q}\mathbf{A}_{i}, \label{e.5}
\end{equation}
\begin{equation}
\left( \mathcal{QL}\left( \boldsymbol{\mathcal{B}}^{j}-2n_{j}\frac{\partial \ln \zeta ^{(0)}}{\partial
n_{j}}\boldsymbol{\mathcal{A}}\right) \right) _{i}=\mathcal{Q}\left( \mathbf{B}_{i}^{j}-2n_{j}\frac{\partial \ln
\zeta ^{\ast }}{\partial n_{j}} \mathbf{A}_i\right),  \label{e.6}
\end{equation}
\begin{equation}
\left( \mathcal{Q}\left( \mathcal{L}+\frac{1}{2}\zeta ^{(0)}\right) \mathcal{ C}_{\gamma \eta }\right)
_{i}=\mathcal{Q}C_{i,\gamma \eta}, \label{e.7}
\end{equation}
\begin{equation}
\left( \mathcal{Q}\left( \mathcal{L}+\frac{1}{2}\zeta ^{\ast }\right) \mathcal{D}\right) _{i}=\mathcal{Q}D_{i},
\label{e.8}
\end{equation}
\begin{equation}
\left( \mathcal{Q}\left( \mathcal{L}+\zeta ^{\ast}\right)\boldsymbol{\mathcal{E}}^j\right)
_{i}=\mathcal{Q}\mathbf{E}_{i}^{j}. \label{e.9}
\end{equation}
These equations are the same as (\ref{c.25})--(\ref{c.29}). The appearance of the factors of $\mathcal{Q}$
simply represent a convenient rearrangement of those equations, using the identities of Appendix D. They show
that the relevant linear operator is $\mathcal{Q}\left( \mathcal{L}-\lambda \right) $ where $\lambda $ is one of
the eigenvalues (\ref{d.10}). The orthogonal projection $\mathcal{Q}$ identifies the left eigenfunctions with
zero eigenvalue as being those of the biorthogonal set $\psi _{\alpha i}$ in (\ref {d.12}). According to the
Fredholm alternative \cite{Dunford67}, solutions to these equations exist if and only if the inhomogeneity is
orthogonal to the null space of the left eigenfunctions. Here, all the inhomogeneities on the right sides of
(\ref{e.5})--(\ref{e.9}) appear explicitly orthogonal to this null space. Hence, solutions exist and are unique.

\section{Details of the Constitutive Equations}
\label{app4}

The cooling rate, and fluxes of mass, momentum, and energy are
given exactly as explicit integrals of solutions to the kinetic
equation. Once the CE solution is obtained, approximately to first
order in the gradients, these expressions give the cooling rate
and fluxes in the form of the constitutive equations
(\ref{4.5})--(\ref{4.8}). The objective of this Appendix is to
simplify these expressions to the extent possible without making
any approximations. This is accomplished in most cases by
performing solid angle integrations using the results
\begin{equation}
\int d\widehat{\boldsymbol {\sigma }}\Theta (\widehat{\boldsymbol {\sigma}}\cdot \mathbf{
g})(\widehat{\boldsymbol {\sigma }}\cdot \mathbf{g})^{n}=\pi ^{\left( d-1\right) /2}\frac{\Gamma \left(
\frac{n+1}{2}\right) }{\Gamma \left( \frac{n+d}{2} \right) }g^{n}\equiv B_{n}g^{n},  \label{f.1}
\end{equation}
\begin{equation}
\int d\widehat{\boldsymbol {\sigma }}\Theta (\widehat{\boldsymbol {\sigma}}\cdot \mathbf{
g})(\widehat{\boldsymbol {\sigma }}\cdot \mathbf{g})^{n}\widehat{\boldsymbol {\sigma }}
=B_{n+1}g^{n}\widehat{\mathbf{g}},  \label{f.2}
\end{equation}
\begin{equation}
\int d\widehat{\boldsymbol {\sigma }}\Theta (\widehat{\boldsymbol
{\sigma}}\cdot \mathbf{ g})(\widehat{\boldsymbol {\sigma }}\cdot
\mathbf{g})^{n}\widehat{\sigma }_{k} \widehat{\sigma }_{\ell
}=\frac{B_{n}}{n+d}g^{n}\left( n\widehat{g}_{k} \widehat{g}_{\ell
}+\delta _{k\ell }\right) ,  \label{f.3}
\end{equation}
\begin{equation}
\int d\widehat{\boldsymbol {\sigma }}\Theta (\widehat{\boldsymbol
{\sigma}}\cdot \mathbf{ g})(\widehat{\boldsymbol {\sigma }}\cdot
\mathbf{g})^{n}\widehat{\sigma }_{k} \widehat{\sigma }_{\ell
}\widehat{\sigma }_{m}=g^{n}\frac{B_{n+1}}{n+d+1} \left[ \left(
n-1\right) \widehat{g}_{k}\widehat{g}_{\ell }\widehat{g}_{m}+
\widehat{g}_{m}\delta _{k\ell }+\widehat{g}_{k}\delta _{m\ell
}+\widehat{g} _{\ell }\delta _{km}\right] ,  \label{f3a}
\end{equation}
where $d$ is the dimension $\left( d\geq 2\right) $, and $\Gamma
\left( x\right) $ is the usual Gamma function
\begin{equation}
\Gamma \left( x+1\right) =x\Gamma \left( x\right)
,\hspace{0.3in}\Gamma \left( \frac{1}{2}\right) =\sqrt{\pi
},\hspace{0.3in}\Gamma \left( 1\right) =1.  \label{f.4}
\end{equation}
In addition, for the sake of convenience, henceforth we will use
the notation ${\bf g}_{12}\equiv {\bf g}$.

To get the collisional transfer contributions to the fluxes, one
has to consider the following expansion
\begin{eqnarray}
\int_{0}^{1}dx& & f_{ij}(\mathbf{r}_{1}-x\boldsymbol {\sigma }_{ij},\mathbf{v}_{1}, \mathbf{r}_{1}+\left(
1-x\right) \boldsymbol {\sigma }_{ij},\mathbf{v} _{2};t)=\int_{0}^{1}dx\chi _{ij}\left(
\mathbf{r}_{1}-x\boldsymbol {\sigma }_{ij}, \mathbf{r}_{1}+\left( 1-x\right) \boldsymbol {\sigma }_{ij}\mid
\left\{ n_{i}\right\} \right)
\nonumber\\
& & \times f_{i}(\mathbf{r}_{1}-x\boldsymbol {\sigma
}_{ij},\mathbf{v}_{1};t)f_{j}( \mathbf{r}_{1}+\left( 1-x\right)
\boldsymbol {\sigma}_{ij},\mathbf{v}_{2};t)\nonumber\\
& &\rightarrow\chi _{ij}^{(0)}\left( \sigma _{ij};\left\{
n_{i}\right\} \right)
f_{i}^{(0)}(\mathbf{v}_{1};t)f_{j}^{(0)}(\mathbf{v}_{2};t)\nonumber\\
 &&+\frac{1}{2}\chi _{ij}^{(0)}\left( \sigma _{ij};\left\{ n_{i}\right\}
\right) \left[f_{i}^{(0)}(\mathbf{v}_{1};t)\partial
_{y_\beta}f_{j}^{(0)}(\mathbf{v}_{2};t)-f_{j}^{(0)}(\mathbf{v}_{2};t)\partial
_{y_\beta}f_i^{(0)}( \mathbf{v}_{1};t)\right] \boldsymbol {\sigma
}_{ij} \mathbf{\cdot \nabla }y_\beta \nonumber\\
 &&+\chi _{ij}^{(0)}\left[f_{i}^{(0)}(\mathbf{v}_{1};t)f_{j}^{(1)}(\mathbf{r}_{1},\mathbf{v}
_{2};t)+f_{i}^{(1)}(\mathbf{v}_{1};t)f_{j}^{(0)}(\mathbf{r}_{1},\mathbf{v}
_{2};t)\right]\nonumber\\
& &
+f_{i}^{(0)}(\mathbf{v}_{1};t)f_{j}^{(1)}(\mathbf{r}_{1},\mathbf{v}
_{2};t)\delta \chi _{ij},  \label{f.4a}
\end{eqnarray}
where $\delta \chi _{ij}$ is defined by
\begin{equation}
\delta \chi _{ij}= \sum_{\ell =1}^{s}\left( \nabla
_{\mathbf{r}_{1}}n_{\ell }( \mathbf{r}_{1};t)\right) \cdot
\int_{0}^{1}dx\int d\mathbf{r}^{\prime }\left( \mathbf{r}^{\prime
}-\mathbf{r}_{1}\right) \frac{\delta \chi _{ij}\left(
\mathbf{r}_{1}-x\boldsymbol {\sigma }_{ij},\mathbf{r}_{1}+\left(
1-x\right) \boldsymbol {\sigma }_{ij}\mid \left\{ n_{i}\right\}
\right) }{\delta n_{\ell }(\mathbf{r}^{\prime },t)}\mid _{\delta
n=0} \label{f.4b}
\end{equation}
The functional derivative is evaluated at $\delta n=0$ and so it
depends only on differences of pairs of coordinates, as in
(\ref{c.6aa}). A change of variables then makes the dependence on
$x$ explicit
\begin{eqnarray}
\int d\mathbf{r}^{\prime }\left( \mathbf{r}^{\prime
}-\mathbf{r}_{1}\right) & & \frac{\delta \chi _{ij}\left(
\mathbf{r}_{1}-x\boldsymbol {\sigma }_{ij},\mathbf{r }_{1}+\left(
1-x\right) \boldsymbol {\sigma }_{ij}\mid \left\{ n_{i}\right\}
\right) }{\delta n_{\ell }(\mathbf{r}^{\prime },t)}\mid _{\delta
n=0}=\int d \mathbf{r}^{\prime \prime }\left( \mathbf{r}^{\prime
\prime }+\frac{1}{2} \boldsymbol {\sigma }_{ij}-x\boldsymbol
{\sigma}_{ij}\right) \nonumber\\
& & \times \frac{\delta \chi _{ij}\left(
-\frac{1}{2}\mathbf{\sigma }_{ij},\frac{1}{2}\boldsymbol {\sigma
}_{ij}\mid \left\{ n_{i}\right\} \right) }{\delta n_{\ell
}(\mathbf{r} ^{\prime \prime },t)}\mid _{\delta n=0}\nonumber\\
&=&-\left( x-\frac{1}{2}\right)\boldsymbol {\sigma
}_{ij}\frac{\partial \chi _{ij}^{(0)}\left(\sigma_{ij}\mid \left\{
n_{i}\right\} \right) }{\partial n_{\ell }(\mathbf{r}^{\prime
\prime },t)}\mid _{\delta n=0}+\int d\mathbf{r}^{\prime \prime
}\mathbf{r}^{\prime \prime }\frac{ \delta \chi _{ij}\left(
-\frac{1}{2}\mathbf{\sigma }_{ij},\frac{1}{2}\mathbf{ \sigma
}_{ij}\mid \left\{ n_{i}\right\} \right) }{\delta n_{\ell
}(\mathbf{r} ^{\prime \prime },t)}\mid
_{\delta n=0}  \nonumber\\
 &=&-\frac{\chi _{ij}^{(0)}}{n_{\ell
}}\boldsymbol {\sigma }_{ij}\left[ \left( x-\frac{1 }{2}\right)
n_{\ell }\frac{\partial \ln \chi _{ij}^{(0)}(\sigma _{ij};\left\{
n_{k}\right\} )}{\partial n_{\ell }}-\frac{1}{2}I_{ij\ell }(\sigma
_{ij};\left\{ n_{k}\right\} )\right] ,  \label{f.4c}
\end{eqnarray}
where $I_{ij\ell }(\sigma _{ij};\left\{ n_{k}\right\} )$ is
defined in (\ref {c.6aa}). Finally, then
\begin{equation}
\delta \chi _{ij}=\frac{1}{2}\chi _{ij}^{(0)}\sum_{\ell
=1}^{s}\left( \nabla _{\mathbf{r}_{1}}\ln n_{\ell
}(\mathbf{r}_{1};t)\right) \cdot \boldsymbol {\sigma }_{ij}\left(
I_{ij\ell }(\sigma _{ij};\left\{ n_{k}\right\} )\right).
\label{f.5}
\end{equation}

\subsection{Cooling rate}

Since $\zeta $ is a scalar, the only gradient contributions are
proportional to $\nabla \cdot \mathbf{U}$, and (\ref{3.37}) to
first order in the gradients becomes
\begin{equation}
\zeta =\zeta ^{(0)}+\zeta _{U}\nabla \cdot \mathbf{U}, \label{f.6}
\end{equation}
with
\begin{equation}
\zeta ^{(0)} =\frac{1}{2dnT}\sum_{i,j=1}^{s}\left( 1-\alpha
_{ij}^{2}\right) m_{i}\mu _{ji}\chi_{ij}^{(0)}\sigma
_{ij}^{d-1}\int d\mathbf{v}_{1}\int d \mathbf{v}_{2}\int
d\widehat{\boldsymbol {\sigma}}\Theta (\widehat{\boldsymbol
{\sigma}}\cdot \mathbf{g})(\widehat{ \boldsymbol {\sigma }}\cdot
\mathbf{g})^{3}f_{i}^{(0)}({\bf V}_{1})f_{j}^{(0)}({\bf V}_{2}),
\end{equation}
\begin{eqnarray}
\zeta _{U} &=&\frac{1}{2dnT}\sum_{i,j=1}^{s}\left( 1-\alpha
_{ij}^{2}\right) m_{i}\mu _{ji}\chi_{ij}^{(0)}\sigma
_{ij}^{d-1}\int d\mathbf{v}_{1}\int d\mathbf{v}_{2}\int d
\widehat{\boldsymbol {\sigma }} \Theta (\widehat{\boldsymbol
{\sigma}}\cdot \mathbf{g})(\widehat{ \boldsymbol {\sigma }}\cdot
\mathbf{g})^{3}\nonumber\\
& & \times \left[\frac{1}{d}f_{j}^{(0)}({\bf V}_{2}){\boldsymbol
{\sigma }}_{ij}\cdot \nabla_{{\bf V}_1}f_{i}^{(0)}({\bf
V}_{1})+2f_{i}^{(0)}({\bf V}_{1})\mathcal{D}_{j}({\bf
V}_{2})\right].
\end{eqnarray}
Performing the solid angle integrals gives
\begin{equation}
\zeta ^{(0)}=\frac{B_{3}}{2dnT}\sum_{i,j=1}^{s}\left( 1-\alpha
_{ij}^{2}\right) \frac{m_{i}m_{j}}{m_{i}+m_{j}}\chi
_{ij}^{(0)}\sigma _{ij}^{d-1}\int d\mathbf{v}_{1}\int d\mathbf{v}
_{2}g^{3}f_{i}^{(0)}({\bf V}_{1})f_{j}^{(0)}({\bf V}_{2}),
\label{f.7}
\end{equation}
\begin{eqnarray}
\label{f.8} \zeta _{U}
&=&\frac{B_{4}}{2d^2nT}\sum_{i,j=1}^{s}\left( 1-\alpha
_{ij}^{2}\right) \frac{m_{i}m_{j}}{m_{i}+m_{j}}\chi
_{ij}^{(0)}\sigma _{ij}^{d}\int d\mathbf{v}_{1}\int
d\mathbf{v}_{2}\;g^2f_{j}^{(0)}({\bf V}_{2})\left({\bf
g}\cdot\nabla_{{\bf
V}_1}f_{i}^{(0)}({\bf V}_{1})\right)\nonumber\\
& & +\frac{B_{3}}{d n T}\sum_{i,j=1}^{s}\left( 1-\alpha
_{ij}^{2}\right) \frac{m_{i}m_{j}}{m_{i}+m_{j}}\chi
_{ij}^{(0)}\sigma _{ij}^{d-1}\int d\mathbf{v}_{1}\int
d\mathbf{v}_{2}\; g^3f_{i}^{(0)}({\bf V}_{1})\mathcal{D}_{j}({\bf
V}_{2}).
\end{eqnarray}
Finally, an integration by parts in the first term of the velocity
integrals gives the result quoted in the text
\begin{eqnarray}
\label{f.9}
 \zeta _{U}
&=&-\frac{d+2}{2dnT}B_{4}\sum_{i,j=1}^{s}\left( 1-\alpha
_{ij}^{2}\right) \frac{m_{i}m_{j}}{m_{i}+m_{j}}\chi
_{ij}^{(0)}\sigma _{ij}^{d} n_{i}n_{j} \left(
\frac{T_{i}^{(0)}}{m_{i} }+\frac{T_{j}^{(0)}}{m_{j}}\right)
\nonumber\\
&&+\frac{B_{3}}{d n T}\sum_{i,j=1}^{s}\left( 1-\alpha
_{ij}^{2}\right) \frac{m_{i}m_{j}}{m_{i}+m_{j}}\chi
_{ij}^{(0)}\sigma _{ij}^{d-1}\int d\mathbf{v}_{1}\int
d\mathbf{v}_{2}\;g^3f_{i}^{(0)}({\bf V}_{1})\mathcal{D}_{j}({\bf
V}_{2}),
\end{eqnarray}
where the species temperatures are defined by
\begin{equation}
\frac{d}{2}n_{i}T_{i}^{(0)}=\int d\mathbf{v}\frac{1}{2}
m_{i}V^{2}f_{i}^{(0)}({\bf V}).  \label{f.10}
\end{equation}
In the case of mechanically equivalent particles, Eq.\ (\ref{f.9}) reduces to previous results obtained for a
monocomponent gas \cite{Garzo99a,Lutsko05}.

\subsection{Mass Flux}

The mass fluxes are determined from the definition of (\ref{3.27})
\begin{eqnarray}
\mathbf{j}_{0i}\left( \mathbf{r}_{1},t\right) &\rightarrow
&m_i\,\int
d\mathbf{v}\mathbf{V}f_{i}^{(1)}(\mathbf{r}_{1},\mathbf{v};t)
\nonumber\\
 &=&\frac{1}{d}\int d\mathbf{v}m_i\mathbf{V}\cdot
\left(\boldsymbol{\mathcal{A}}_{i}\left( \mathbf{V}\right) \nabla
\ln T+\sum_{j=1}^{s}\left( \boldsymbol{\mathcal{B}}_{i}^{j}\left(
\mathbf{V}\right) \nabla \ln
n_{j}+\boldsymbol{\mathcal{E}}_{i}^{j}\left( \mathbf{V}\right)
\mathbf{F}_{j}\right) \right),  \label{f.39}
\end{eqnarray}
where the contribution from $f_{i}^{(0)}$ vanishes. The transport coefficients according to (\ref{4.6}) are
identified as
\begin{equation}
D_{i}^{T}=-\frac{m_i}{\rho d}\int d\mathbf{v}\mathbf{V}\cdot \boldsymbol{\mathcal{A}}_{i}\left(
\mathbf{V}\right), \label{f.40}
\end{equation}
\begin{equation}
D_{ij}=-\frac{\rho }{m_{j}n_{j}d}\int d\mathbf{v}\mathbf{V}\cdot \boldsymbol{\mathcal{B}}_{i}^{j}\left(
\mathbf{V}\right), \label{f.41}
\end{equation}
\begin{equation}
D_{ij}^{F}=-\frac{m_i}{d}\int d\mathbf{v}\mathbf{V}\cdot
\boldsymbol{\mathcal{E}}_{i}^{j}\left( \mathbf{V}\right).
\label{f.42}
\end{equation}

\subsection{Energy Flux}

\bigskip

The energy flux to first order in the gradients is obtained from Eqs.\ (\ref{3.29}) and (\ref{3.29a}) as
\begin{equation}
\mathbf{q}=\mathbf{q}^{k}+\mathbf{q}^{c}.  \label{f.22}
\end{equation}
The kinetic contribution is
\begin{eqnarray}
\mathbf{q}^{k} &=&\sum_{i=1}^{s}\int
d\mathbf{v}_1\frac{1}{2}m_iV_1^{2}
\mathbf{V}_{1}f_{i}^{(1)}(\mathbf{r}_{1},\mathbf{v}_{1};t)
\nonumber\\
 &=&\frac{1}{d}\sum_{i=1}^{s}\int
d\mathbf{v}_{1}\frac{1}{2}m_{i}V_{1}^{2} \mathbf{V}_{1}\cdot
\boldsymbol{\mathcal{A}}_{i}\left(\mathbf{V}_1\right) \nabla
\ln T \nonumber\\
 &&\left. +\frac{1}{d}\sum_{i,j=1}^{s}\int
d\mathbf{v}_{1}\frac{1}{2}m_{i}V_{1}^{2} \mathbf{V}_{1}\cdot\left(
\boldsymbol{\mathcal{B}}_{i}^j\left( \mathbf{V}_1\right) \nabla
\ln n_{j}+\boldsymbol{\mathcal{E}}_{i}^j\left( \mathbf{V}_1\right)
\mathbf{F} _{j}\right)\right..  \label{f23}
\end{eqnarray}
The contributions proportional to derivatives of the flow velocity
vanish from symmetry. The collisional transfer contribution is
\begin{eqnarray}
{\bf q}^{c}&=&\sum_{i,j=1}^{s}\frac{1}{8}\left( 1+\alpha
_{ij}\right) m_{j}\mu _{ij}\sigma _{ij}^{d}\chi _{ij}^{(0)}\int
d\mathbf{v}_{1}\int d\mathbf{v}_{2}\int d\widehat{\boldsymbol
{\sigma }}\,\Theta (\widehat{\boldsymbol {\sigma}}\cdot
\mathbf{g})(\widehat{\boldsymbol {\sigma }}\cdot
\mathbf{g})^{2}\widehat{\boldsymbol {\sigma}}  \nonumber \\
&&\times \left[ -\left( 1-\alpha _{ij}\right) \left( \mu _{ij}-\mu
_{ji}\right)(\widehat{\boldsymbol {\sigma}}\cdot \mathbf{g})+4
(\mathbf{G}_{ij}\cdot \widehat{\boldsymbol {\sigma}})\right]
\left[ f_{i}^{(0)}({\bf V}_{1})f_{j}^{(1)}({\bf
V}_{2})+f_{i}^{(1)}({\bf V}_{1})f_{j}^{(0)}({\bf V}_{2})
\right.  \nonumber \\
&&\left. -\frac{1}{2}f_{j}^{(0)}({\bf V}_{2})\partial _{y_{\beta
}}f_{i}^{(0)}({\bf V}_{1})\boldsymbol {\sigma}_{ij}\cdot \nabla
y_\beta+ \frac{1}{2}f_{i}^{(0)}({\bf V}_{1})\partial _{y_{\beta
}}f_{j}^{(0)}({\bf V}_{2})\boldsymbol {\sigma}_{ij}\cdot \nabla
y_\beta\right],
 \label{f.24}
\end{eqnarray}
where $\mathbf{G}_{ij}\equiv \mu _{ij}\mathbf{V}_{1}+\mu
_{ji}\mathbf{V}_{2}$ . The contribution from $\delta \chi _{ij}$
in (\ref{f.5}) vanishes from symmetry. The angular integrals can
be performed to get
\begin{eqnarray}
q_{\gamma }^{c}&=&\sum_{i,j=1}^{s}\frac{1}{8}\left( 1+\alpha
_{ij}\right) m_{j}\mu _{ij}\sigma _{ij}^{d}\chi _{ij}^{(0)}\int
d\mathbf{v}_{1}\int d\mathbf{v} _{2}\left\{ -B_{4}\left( 1-\alpha
_{ij}\right)\left( \mu _{ij}-\mu _{ji}\right)
g^{2}g_{\gamma}\right.\nonumber\\
& & +\frac{4B_{2}}{2+d}\left[2\left( \mathbf{G} _{ij}\cdot
\mathbf{g}\right) g_{\gamma }+g^{2}G_{ij,\gamma} \right]\left(
f_{i}^{(0)}({\bf V}_{1})f_{j}^{(1)}({\bf V}_{2})+f_{i}^{(1)}({\bf
V}_{1})f_{j}^{(0)}({\bf V}_{2})
\right)  \nonumber \\
&&+\frac{B_{3}}{3+d}\left[ -\left( 1-\alpha _{ij}\right) \left(
\mu _{ij}-\mu _{ji}\right) g^{3}\left( 3\widehat{g}_{\gamma }
\widehat{g}_{\eta}+\delta _{\gamma \eta }\right)
+4g^{2}\left(\left( \mathbf{G} _{ij}\cdot
\widehat{\mathbf{g}}\right)\left( \widehat{g}_{\gamma }\widehat{g}
_{\eta }+\delta _{\gamma \eta }\right)\right.\right. \nonumber\\
& & \left.\left.\left. +\widehat{g}_{\gamma }G_{ij,\eta }+
\widehat{g}_{\eta }G_{ij,\gamma }\right) \right] \frac{1}{2}\left[
f_{i}^{(0)}({\bf V}_{1})\partial _{y_{\beta }}f_{j}^{(0)}({\bf
V}_{2})-f_{j}^{(0)}({\bf V}_{2})\partial _{y_{\beta
}}f_{i}^{(0)}({\bf V}_{1})\right] \partial _\eta y_{\beta
}\right\}.\nonumber\\\label{f.25}
\end{eqnarray}
Interchanging the labels $i,j$ and $\mathbf{v}_{1},\mathbf{v}_{2}$ it is seen that the contributions from
$f_{i}^{(1)}$ and $f_{j}^{(1)}$ are the same. For the same reason the contributions from $\partial
_{y_{\beta}}f_{i}^{(0)}$ and $-\partial _{y_{\beta }}f_{j}^{(0)}$ are the same. The first terms of the integrand
give velocity moments of $f_{j}^{(1)}$ of degree one and three, which are proportional to the (partial) mass and
kinetic energy fluxes. Finally, the only contributions from $\partial _{y_{\beta}}f_{i}^{(0)}$ are those that
are scalar functions of the velocities, i.e. those proportional to temperature and species density gradients.
The final result is therefore
\begin{eqnarray}
{\bf q}^{c}&=&\sum_{i,j=1}^{s}\frac{1}{8}\left( 1+\alpha
_{ij}\right) m_{j}\mu _{ij}\sigma _{ij}^{d}\chi _{ij}^{(0)}\left[
2B_{4}\left(1-\alpha _{ij}\right)\left(\mu _{ij}-\mu _{ji}\right)
n_{i}\left( \frac{ 2}{m_{j}}{\bf q}_{j}^{k}+(d+2)\frac{
T_{i}^{(0)}}{m_{i}m_j}
{\bf j}_{0j}^{(1)}\right) \right.  \nonumber\\
&&\left. +\frac{8B_{2}}{2+d}n_{i}\left( \frac{2\mu _{ji}}{m_{j}}{\bf
q}_{j}^{k}-(d+2)\frac{T_{i}^{(0)}}{m_{i}m_j}\left( 2\mu _{ij}-\mu _{ji}\right)  {\bf j}_{0j}^{(1)}\right)
+C_{ij}^{T}\nabla
\ln T+\sum_{p=1}^{s}C_{ijp}^{T}\nabla \ln n_{p}\right],\nonumber\\
\label{f.26}
\end{eqnarray}
where ${\bf j}_{0i}^{(1)}$ is defined by Eq.\ (\ref{f.39}) and the partial kinetic energy flux is
\begin{equation}
{\bf q}_{i}^{k}=\int d\mathbf{v}\;\frac{m_{i}}{2}V^2{\bf V}f_{i}^{(1)}({\bf V}). \label{f.27}
\end{equation}
The constants $C_{ij}^{T}$ and $C_{ijp}^{T}$ are
\begin{equation}
C_{ij}^{T}=\frac{B_{3}}{d}\int d\mathbf{v}_{1}\int
d\mathbf{v}_{2}\left[ -\left( 1-\alpha _{ij}\right) \left( \mu
_{ij}-\mu _{ji}\right)g^{3}+4g\left({\bf
g}\cdot\mathbf{G}_{ij}\right)\right] f_{i}^{(0)}({\bf
V}_{1})T\partial _{T}f_{j}^{(0)}({\bf V}_{2}),  \label{f.28}
\end{equation}
\begin{equation}
C_{ijp}^{T}=\frac{B_{3}}{d}\int d\mathbf{v}_{1}\int
d\mathbf{v}_{2}\left[ -\left( 1-\alpha _{ij}\right) \left( \mu
_{ij}-\mu _{ji}\right)g^{3}+4g\left({\bf
g}\cdot\mathbf{G}_{ij}\right)\right] f_{i}^{(0)}({\bf
V}_{1})n_{p}\partial _{n_{p}}f_{j}^{(0)}({\bf V}_{2}).
\label{f.29}
\end{equation}
The expression of $C_{ij}^{T}$ can be simplified when one takes
into account the relation
\begin{equation}
\label{f.29.1} T\partial _{T}f_{j}^{(0)}({\bf
V})=-\frac{1}{2}\mathbf{\nabla }_{\mathbf{V}}\cdot \left(
\mathbf{V}f_{j}^{(0)}({\bf V})\right),
\end{equation}
and integrates by parts in (\ref{f.28}). The result is
\begin{eqnarray}
\label{f.29.2} C_{ij}^{T}&=&-\frac{2B_{3}}{d}\int
d\mathbf{v}_{1}\int d\mathbf{v}_{2}f_{i}^{(0)}({\bf
V}_{1})f_{j}^{(0)}({\bf V}_{2})\left\{gG_{ij}^2+g^{-1}\left({\bf
g}\cdot\mathbf{G}_{ij}\right)^2+(1+\mu_{ji})g\left({\bf
g}\cdot\mathbf{G}_{ij}\right)
\right.\nonumber\\
& &
\left.+\mu_{ji}\mu_{ij}g^3+\frac{3}{4}(1-\alpha_{ij})(\mu_{ji}-\mu_{ij})\left[g\left({\bf
g}\cdot\mathbf{G}_{ij}\right)+g^3\right]\right\}.
\end{eqnarray}
On the other hand, no significant further simplification of Eq.\
(\ref{f.29}) is possible until $f_{i}^{(0)}$ is specified in
detail.

The heat flux is seen to have the form (\ref{4.7}),
\begin{equation}
\mathbf{q}\left( \mathbf{r},t\right) \rightarrow -\lambda
\mathbf{\nabla } T-\sum_{i,j=1}^{s}\left( T^2 D_{q,ij}\nabla \ln
n_{j}+L_{ij}\mathbf{F}_{j} \right) , \label{f.30}
\end{equation}
so the transport coefficients now can be identified
\begin{equation}
\lambda =\lambda ^{k}+\lambda ^{c},\hspace{0.3in}
D_{q,ij}=D_{q,ij}^{k}+D_{q,ij}^{c},\hspace{0.3in}
L_{ij}=L_{ij}^{k}+L_{ij}^{c}.  \label{f.31}
\end{equation}
The kinetic parts are,
\begin{equation}
\lambda
^{k}=\sum_{i=1}^s\lambda_i^k=-\frac{1}{dT}\sum_{i=1}^{s}\int
d\mathbf{v}\frac{m_i}{2} V^{2} \mathbf{V}\cdot
\boldsymbol{\mathcal{A}}_{i}\left( \mathbf{V}\right) ,
\label{f.32}
\end{equation}
\begin{equation}
D_{q,ij}^{k}=-\frac{1}{dT^2}\int d\mathbf{v}\frac{m_i}{2}V^{2}
\mathbf{V}\cdot \boldsymbol{\mathcal{B}}_{i}^j\left(
\mathbf{V}\right) , \label{fr.34}
\end{equation}
\begin{equation}
L_{ij}^{k}=-\frac{1}{d}\int d\mathbf{v}\frac{m_i}{2}V^{2}
\mathbf{V}\cdot \boldsymbol{\mathcal{E}}_{i}^j\left(
\mathbf{V}\right) , \label{fr.35}
\end{equation}
while the collisional transport parts are given by Eqs.\
(\ref{f.36})--(\ref{f.38}).

\subsection{Momentum Flux}

The momentum flux to first order in the gradients is obtained from
(\ref {3.30})--(\ref{3.32})
\begin{equation}
P_{\gamma \lambda}\equiv P_{\gamma \lambda }^{k}+P_{\gamma
\lambda}^{c}, \label{f.11}
\end{equation}
where
\begin{eqnarray}
P_{\gamma \lambda}^{k} &\rightarrow &\sum_{i=1}^{s}\int
d\mathbf{v} _{1}m_{i}V_{1\gamma}V_{1\lambda}\left(
f_{i}^{(0)}({\bf V}_{1})+f_{i}^{(1)}({\bf V}_{1})\right)  \nonumber\\
&=&\delta _{\gamma \lambda}n T+\sum_{i=1}^{s}\int d\mathbf{v}
_{1}m_{i}V_{1\lambda}V_{1\gamma}f_{i}^{(1)}({\bf V}_{1}),
\label{f.12}
\end{eqnarray}
\begin{eqnarray}
P_{\gamma \lambda}^{c} &=&\frac{1}{2}\sum_{i,j=1}^{s}m_{j}\mu
_{ij}\left( 1+\alpha _{ij}\right) \sigma _{ij}^{d}\chi
_{ij}^{(0)}\int d\mathbf{v}_{1}\int d \mathbf{v}_{2}\int
d\widehat{\boldsymbol {\sigma}}\,\Theta (\widehat{\boldsymbol
{\sigma}}\cdot \mathbf{g})(\widehat{\boldsymbol {\sigma }}\cdot
\mathbf{g})^{2}\widehat{\sigma }_{\lambda}\widehat{\sigma }_{\gamma }  \nonumber\\
&&\times \left[ f_{i}^{(0)}({\bf V}_{1})f_{j}^{(0)}({\bf V}_{2})
+2f_{i}^{(0)}({\bf V}_{1})f_{j}^{(1)}({\bf V}_{2})
\right.  \nonumber\\
&&\left. -\frac{1}{2}\left(f_{j}^{(0)}({\bf V}_{2}) \partial
_{y_{\beta }}f_{i}^{(0)}({\bf V}_{1})-f_{i}^{(0)}({\bf
V}_{1})\partial _{y_{\beta }}f_{j}^{(0)}({\bf V}_{2})\right)
\boldsymbol{\sigma }_{ij}\cdot \nabla y_{\beta }\right] .
\label{f.13}
\end{eqnarray}
A contribution to (\ref{f.13}) proportional to the density
gradients from the expansion of $\chi
_{ij}(\mathbf{r}_{1}-x\boldsymbol{\sigma }_{ij},\mathbf{r}
_{1}+\left( 1-x\right) \boldsymbol{\sigma }_{ij})$ vanishes from
symmetry. For similar reasons, the only gradients contributing to
both (\ref{f.12}) and ( \ref{f.13}) are those from the flow field.
The terms proportional to $ \mathcal{D}_{i}$ in (\ref{c.22}) also
do not contribute due to the orthogonality condition (\ref{e.4}).
The solid angle integrations can be performed with the results
\begin{equation}
P_{\gamma \lambda}^{k}\rightarrow \delta _{\gamma \lambda
}nT+\frac{1}{2} \sum_{i=1}^{s}\int d\mathbf{v}_{1}m_{i}V_{1\lambda
}V_{1\gamma }\mathcal{C}_{i,\beta \mu}\left( \mathbf{V}_{1}\right)
\left( \partial _{\beta}U_{\mu }+\partial _{\mu }U_{\beta
}-\frac{2}{d}\delta _{\beta \mu }\nabla \cdot \mathbf{U}\right),
\label{f.14}
\end{equation}
\begin{eqnarray}
P_{\gamma \lambda}^{c} &=&\delta _{\gamma \lambda
}\frac{B_{2}}{2d} \sum_{i,j=1}^{s}m_{j}\mu _{ij}\left( 1+\alpha
_{ij}\right) \sigma _{ij}^{d}\chi _{ij}^{(0)}n_{i}n_{j}\left(
\frac{T_{i}^{(0)}}{m_{i}}+\frac{
T_{j}^{(0)}}{m_{j}}\right)  \nonumber \\
&&+\frac{B_{2}}{d+2}\sum_{i,j=1}^{s}\mu _{ij}\left( 1+\alpha
_{ij}\right) \chi _{ij}^{(0)}n_{i}\sigma _{ij}^{d}\int
d\mathbf{v}_{2}m_{j}V_{2\gamma }V_{2\lambda }\mathcal{C}_{j,\beta
\mu }\left( \mathbf{V}_{2}\right) \left(
\partial _{\beta}U_{\mu }+\partial _{\mu }U_{\beta }-\frac{2}{d}\delta
_{\beta \mu }\nabla \cdot \mathbf{U}\right)  \nonumber \\
&&+\frac{1}{2}\frac{B_{3}}{3+d}\sum_{i,j=1}^{s}m_{j}\mu
_{ij}\left( 1+\alpha _{ij}\right) \chi _{ij}^{(0)}\sigma
_{ij}^{d}\int d\mathbf{v}_{1}\int d\mathbf{v}
_{2}f_{j}^{(0)}(V_{2})\left(\partial _{V_{1\ell
}}f_{i}^{(0)}(V_{1})\right)
\nonumber \\
&& \times g^{2}\left( \widehat{g}_{\lambda }\widehat{g}_{\gamma }
\widehat{g}_{\mu}+\widehat{g}_{\mu}\delta _{\lambda\gamma
}+\widehat{g}_{\gamma }\delta _{\mu\lambda
}+\widehat{g}_{\lambda}\delta _{\gamma \mu}\right)
\partial_{\mu}U_{\ell } .  \label{f.15}
\end{eqnarray}
An integration by parts in the velocity integral, and use of fluid
symmetry gives, finally
\begin{eqnarray}
P_{\gamma \lambda}^{c} &=&\delta _{\gamma \lambda
}\frac{B_{2}}{2d} \sum_{i,j=1}^{s}m_{j}\mu _{ij}\left( 1+\alpha
_{ij}\right) \sigma _{ij}^{d}\chi _{ij}^{(0)}n_{i}n_{j}\left(
\frac{T_{i}^{(0)}}{m_{i}}+\frac{
T_{j}^{(0)}}{m_{j}}\right)  \nonumber \\
&&+\frac{B_{2}}{d+2}\sum_{i,j=1}^{s}\mu _{ij}\left( 1+\alpha
_{ij}\right) \chi _{ij}^{(0)}n_{i}\sigma _{ij}^{d}\int
d\mathbf{v}_{2}m_{j}V_{2\gamma }V_{2\lambda }\mathcal{C}_{j,\beta
\mu }\left( \mathbf{V}_{2}\right) \left(
\partial _{\beta}U_{\mu }+\partial _{\mu }U_{\beta }-\frac{2}{d}\delta
_{\beta \mu }\nabla \cdot \mathbf{U}\right)  \nonumber \\
&&-\frac{B_{3}\left( d+1\right) }{2d\left( d+2\right)
}\sum_{i,j=1}^{s}m_{j} \mu _{ij}\left( 1+\alpha _{ij}\right) \chi
_{ij}^{(0)}\sigma _{ij}^{d}\int d \mathbf{v}_{1}\int
d\mathbf{v}_{2}f_{i}^{(0)}({\bf V}_{1})f_{j}^{(0)}({\bf V}_{2})g
\nonumber \\
&&\times \left[ \left( \partial _\lambda U_{\gamma}+\partial
_\gamma U_{\lambda}-\frac{2}{d}\delta _{ \gamma \lambda}\nabla
\cdot \mathbf{U}\right)+\frac{d+2}{d}\delta _{\gamma
\lambda}\nabla \cdot \mathbf{U}\right] .  \label{f.16}
\end{eqnarray}

The pressure tensor therefore has the form (\ref{4.8}), and the
pressure, shear viscosity, and bulk viscosity are identified in
terms of their kinetic and collisional transfer contributions.
Their expressions are given by Eqs.\ (\ref{f.17})--(\ref{f.21}),
respectively.

\bibliography{enskog}
\bibliographystyle{apsrev}

\end{document}